\documentclass[fleqn,usenatbib]{mnras}

\usepackage{newtxtext,newtxmath}
\usepackage[T1]{fontenc}

\DeclareRobustCommand{\VAN}[3]{#2}
\let\VANthebibliography\thebibliography
\def\thebibliography{\DeclareRobustCommand{\VAN}[3]{##3}\VANthebibliography}

\usepackage{graphicx}	% Including figure files
\usepackage{amsmath}	% Advanced maths commands
\title[Crater II Streams]{La Silla-QUEST RR Lyrae Star Survey II: The Crater II Tidal Streams }

\author[Coppi, Zinn et al.]{
P.S. Coppi,$^{1,2}$\thanks{E-mail: paolo.coppi@yale.edu, robert.zinn@yale.edu}, R. Zinn,$^{1}$
C. Baltay,$^{1,2}$
D. Rabinowitz,$^{2}$ T. Girard,$^{1,3}$  R. Howard$^{2}$,  K. Ment$^{2}$, R. Rahman$^{2}$
\\
$^{1}$Department of Astronomy, Yale University,
P.O. Box 208101, New Haven, CT 06520-8101, USA\\
$^{2}$Department of Physics, Yale University, P.O. Box
208120, New Haven, CT 06520-8101, USA\\
$^{3}$Department of Physics, Southern Connecticut State University, 501 Crescent Street, New Haven, CT 06515\\
}

\date{Accepted January 31, 2024. January 30, 2024; in original May 6, 2024}

\pubyear{2024}

\begin{document}
\label{firstpage}
\pagerange{\pageref{firstpage}--\pageref{lastpage}}
\maketitle

\begin{abstract}
We describe photometry improvements in the La Silla--Quest RR Lyrae star (RRLS) survey that enable it to reach distances from the Sun ($d_{\odot}$) $\sim 140$ kpc. We report the results of surveying $\sim 300~ {\rm deg}^2$ of  sky around the large, low-surface-brightness Crater II dwarf spheroidal galaxy.  At $d_{\odot}$ $> 80$ kpc, we find a large overdensity of RRLS that extends beyond the traditional isophotal contours used for Crater II.  The majority of these RRLS (34) have a linear distribution on the sky, extending over $15^{\circ}$, that runs through Crater II and is oriented along Crater II's proper motion vector. We hypothesize that this unlikely distribution traces extended tidal streams associated with Crater II.  To test this, we search for other Crater II stellar populations that should be in the streams. Using Gaia proper motion data, we isolate $\approx$ 17 candidate stars outside of Crater II that are consistent with being luminous stars from the Crater II Red Giant Branch (RGB). Their spatial distribution is consistent with the RRLS one. The inferred streams are long, spanning a distance range $\sim 80 - 135$ kpc from the Galactic Centre.  They are oriented at a relatively small angle relative to our line-of-sight ($\sim 25^{\circ}$), which means some stream stars are likely projected onto the main body of the galaxy.  Comparing the numbers of RRLS and RGB candidate stars found in the streams to those in the main galaxy, we estimate Crater II has lost $\gtrsim 30\%$ of its stellar mass. 

\end{abstract}


%%%%%%%%%%%%%%%%%%%%%%%%%%%%%%%%%%%%%%%%%%%%%%%%%%

%%%%%%%%%%%%%%%%% BODY OF PAPER %%%%%%%%%%%%%%%%%%

\section{Introduction}
From the time of its discovery \citep{torrealba16}, the dwarf spheroidal (dSph) galaxy Crater II has been recognized as a satellite galaxy with unusual properties.  \citet{torrealba16} showed that it has a large size and low surface brightness compared to other dSph galaxies of similar luminosity ($M_{V} \sim -8$). 
 This and its moderately large distance from the Sun ($d_{\odot}$) $\sim 120$ kpc may explain why it was not discovered previously.  The measurement of a relatively low velocity dispersion by \citet{caldwell17}, later confirmed by \citet{fu19} and \citet{ji21}, indicated that Crater II also has a small mass-to-light ratio compared to similar galaxies.  \citet{caldwell17} inferred that it has a kinematically very cold dark-matter halo, which they suggested may pose a problem for the cold-dark-matter paradigm of galaxy formation.   

In most respects, the stellar populations of Crater II are not remarkable for a dSph galaxy of its luminosity.  The color-magnitude diagrams (CMD) constructed by \citet{torrealba16} and \citet{monelli18} showed that it is an old, metal-poor system with a densely populated red horizontal branch (HB).  Although there are other dSph galaxies of similar luminosity and metallicity with predominantly red HBs (e.g, the Draco I galaxy), Crater II appears to be a more extreme case where there are no blue HB stars at all \citep{monelli18}. The more extensive CMD constructed by \citet{walker19} confirmed these features, but also provided evidence for two main-sequence turnoffs.  The most populated one corresponds to an age of 12.5 Gyr and the other to 10.5 Gyr.  \citet{walker19} did not find evidence for an intermediate-age population, which suggested that star formation quenched soon after the 10.5 Gyr population.  The variable star population of Crater II has been investigated by \citet{monelli18}, \citet{joo18}, and \citet{vivas20}.  According to \citet{vivas20}, the galaxy contains 98 RR Lyrae stars (RRLS) and 7 Anomalous Cepheids (AC), and they also discovered one Dwarf Cepheid.  The characteristics of the RRLS are intermediate between Oosterhoff groups I and II \citep{monelli18, joo18, vivas20}, to which most of the Milky Way globular clusters belong.  This is also observed in other dSph galaxies with red HBs (e.g., Draco I).  The spectroscopic observations of red giants \citep{caldwell17, fu19, ji21} have confirmed that it is in the mean a very metal-poor system with a range in metallicity.  \citet{ji21}, for example, found a [Fe/H] spread from -2.8 to -1.0, with a mean of -2.16.  \citet{ji21} showed that with their mean value, Crater II lies below the luminosity-metallicity relation delineated by Milky Way satellites by an amount that may be significant and indicative of substantial mass loss from the galaxy.

The Gaia satellite has measured the proper motions of the red giants in Crater II, and a number of investigators have used these data to calculate the proper motion of the galaxy as a whole \citep{caldwell17,kallivayalil18,fu19,walker19,mcconnachie20,ji21,battaglia22, pace22}. Using these measurements and Crater II's radial velocity, several authors have computed its orbit about the Milky Way \citep{fritz18,fu19, ji21,borukhovetskays22,battaglia22, pace22}.  While the details of these calculations vary, one common result is that Crater II is likely to have passed sufficiently close to the inner regions of the Milky Way to have undergone substantial tidal stripping.  There is, however, some debate over whether tidal stripping alone can explain the cold kinematics of the Crater II stars.  Is it necessary for the dark halo of Crater II to be cored as opposed to cuspy \citep{frings17, sanders18} for stripping to be the explanation?  Is the $\Lambda CDM$ paradigm challenged by the characteristics of Crater II \citep{borukhovetskays22,errani22} or not \citep{fattahi18,amorisco19, applebaum21}. Is another form of dark matter required \citep{pozo22}?  Is modified gravity a viable solution \citep{mcgaugh16} or probably not \citep{fattahi18,ji21}?

While the orbit calculations have made a strong case for tidal stripping, which is supported by the deviation of Crater II from the luminosity-metallicity relation \citep{ji21}, until now there has been little direct evidence of this stripping. Whether stripping has occurred has thus been somewhat controversial. \citet{torrealba16}, for example, argued on the basis of the round shape of Crater II that it has not been stripped.  The distributions of RRLS and red giants inside Crater II proper do not show a clear alignment with the direction of its proper motion \citep{walker19, ji21}, and \citet{ji21} found only tentative evidence for a velocity gradient across the galaxy. 

The main difficulty in finding evidence for stripping is the low expected surface density of any features, which should be lower still than that of Crater II itself, which is already a difficult-to-find object due its low surface brightness and and large distance ($\sim 120$ kpc).  In this paper, we attempt to circumvent this issue by looking for RRLS in an extended ($20^\circ\times15^\circ$) region containing Crater II.  While RRLS constitute a rare component of a galaxy's stellar population, variability-selected RRLS with good lightcurve coverage cannot be confused with other objects and they provide robust distance estimates that eliminate confusion with potential foregrounds or backgrounds. Moreover, the density of ``field" RRLS in the halo drops rapidly with distance from the Galactic center, so that the detection of even a few extra RRLS at the approximate distance of Crater II is potentially significant.  

In this paper, we use results from the improved La Silla QUEST (LSQ) RRLS survey to show that there are in fact large streams of stars emanating from Crater II, which are aligned with the Crater II proper motion vector. The paper is organized as follows. We first summarize the characteristics of the LSQ RRLS survey, which has been enhanced recently by the addition of an ensemble photometry pipeline and improved period finding routines. We show that the survey can now find and characterize RRLS with reasonable completeness out to distances $\sim 140$ kpc.  The enhanced LSQ RRLS survey now covers a declination range $\sim -80$ to $+25$ degrees and galactic latitudes $|b| > 15$, but we focus here only on the region of sky near Crater II. At large distances $d_{\odot}> 80$ kpc, we use the survey to demonstrate that we find a very significant excess of RRLS compared to the expectations of a smooth halo model. We show that the distribution of these distant RRLS on the sky is not uniform. Besides the expected concentration located at the position of Crater II, we find that the distant RRLS outside Crater II primarily lie in ``streams" ($\sim$ linear features) stretching $\sim 15^{\circ}$ on the sky and $\sim 50$ kpc in length. Motivated by the discovery of these RRLS streams, we then
show that, in hindsight, hints of this structure are visible if one selects low proper motion (distant) stars with colors that place them in the red giant tip of the color magnitude diagram for Crater II. In the last section of the main paper, we summarize our conclusions and present a 3-D map of the stream in galactic coordinates. In the Appendix, we present lower level details on how our improved ensemble photometry pipeline works, we present some tests verifying the performance of the pipeline, we study the uniformity of the LSQ RRLS survey coverage in the Crater II region and show that it cannot be responsible for the large-scale spatial distribution of RRLS we find there, and we present more details on how we selected the red giant tip stars that we think are stream members. 

\section{The La Silla-Quest RRLS Survey}
The observations used for the La Silla – QUEST RR Lyrae survey were made with the $10~ {\rm deg}^2$
QUEST camera \citep{baltay07} attached to the 1m Schmidt telescope of the La Silla
Observatory. They were obtained over 6 years, from March 2010 to April 2016, covering a declination range $-80$ to $+25$ and  galactic latitude $|b| \gtrsim 15. $
The primary purpose of these 
observations was to search for supernovae and trans-Neptunian objects. The 
rolling cadence used, however, achieves good ``logarithmic" time coverage, probing timescales from hours to months to years,  and is suitable for a variety of variability studies. In our earlier paper on the LSQ RRL Survey,
\citet[Paper I]{zinn14}, one can find descriptions of the camera and its filter \cite[see also][]{Lsq_description_2013}. We summarize below the
improvements that we have made to our target selection,
the photometric calibration of the LSQ lightcurve data, our period finding
techniques, and our search routines for finding RRLs and other types of variables. As we demonstrate by recovering the periods and lightcurves of known distant RRLS, these improvements together with the incorporation of the full LSQ dataset allow us to probe the structure of the halo over a much larger area and out to much larger distances.
(Paper I only considered
the first two years of data in a relatively small region of sky.) In particular, at declinations ~-40 to 0, where the typical air mass of the observations is low, we can reach mean RRL  
magnitudes $V\sim21.5$ and corresponding RRL distances $\sim 120-140$ kpc. Moreover, there are several thousand square degrees of coverage where the number of observations exceeds
200. For stable RRLS in these areas, where we can use the full six years of data, this enables excellent precision in the determination of lightcurve parameters and quantities derived from lightcurve shapes such as metallicity \cite[e.g.,][]{jurcsik96}.

\subsection{Initial Target Selection}
For this paper, our lightcurve pipeline relies on the initial LSQ Sextractor catalogs, where the detection threshold was on purpose set to a low value, only 3 standard deviations above background rather than the traditional 5 standard deviations.  The disadvantage of this low
threshold, combined with the presence of bad chip cosmetics, means that majority of the 
objects in our initial search catalog are fake. One must therefore rely on an external catalog
of true sources, and the completeness of any final catalog depends on the completeness
of the external catalog. The advantage of this approach, however, is that if one already
knows the target positions and many observations are available, one can go significantly 
fainter by effectively co-adding many marginal detections.  (In a future release of our RRL catalog, we will do proper forced photometry on LSQ sources found in deep coadds.) 

For our external catalog, as in paper I, we initially used the Sloan Digital Sky Survey (SDSS) DR17 release \citep{sdss_dr17_2022}, with some point source and quality cuts applied. Besides limiting us to the Northern hemisphere, we discovered that this choice missed real RRLS at magnitudes $V \gtrsim 20.5.$ We thus switched to using PanSTARRS DR2 \citep{ps1db2020}, which allowed us to go to down to declination -30 with quite uniform coverage. Again, 
we initially considered only objects classified as stars (point sources) by the improved \citet{tachibana2018} star vs. galaxy separation algorithm. However, when we went to check how many of the known RRLS in Crater II were in this catalog, we discovered that $\sim 20 \%$ were missing.
This is higher than expected from \citet{tachibana2018} and may reflect the fact that RRLS are strongly variable sources (up to $\sim$ 1 magnitude in amplitude), 
and thus that the shape of their 
observation-averaged point spread function (PSF) does not match that of non-variable objects with the same average magnitude.  This issue (point-like vs. extended object classification for
strongly variable objects) may also  be relevant for upcoming variability surveys such as Rubin. In the end, given that we were also starting to push up against 
the limits of PanSTARRS photometry, 
we decided against applying any quality or point source cuts. Rather, we decided to extract lightcurves for {\it any} PANSTARRS object in their
in the ``mean" PS1 DR2 catalog 
that has more than one PanSTARRS detection. Comparing to the recently released DELVE DR2 catalog \citep{delvedr2_2022}, which is based on 
4m photometry and goes significantly deeper than PanSTARRS, this choice of PanSTARRS targets should contain essentially every persistent source of 
interest to LSQ in the Crater II region and removes the external target catalog as a source of incompleteness.  The $\sim 80\%$ completeness we report below based on the recovery of known RRLS in Crater II is thus caused by photometry errors, which become large at $V \sim 21$ and eventually overwhelm the distinctive 
signatures of RRLS, as well as the unavoidable observing gaps caused by bad chips and chip imperfections. 
(See Appendix A for LSQ survey coverage maps as a function of source magnitude for the Crater II region explored in this work.)

There are two other differences from Paper I that result from our target catalog choice and that may also be relevant to future variability surveys. First, because of limitations in computing resources, Paper I imposed a hard color cut, $g-r \lesssim 0.4$ (after extinction corrections), on target sources. This worked because the SDSS photometry used in paper I is typically obtained in the same driftscan for a given object, i.e., the observed fluxes in the various filter are usually simultaneous to within a few minutes -- a timescale much less than the typical variability timescales of RRLS (a few hours to a day). PanSTARRS observations in different filters, on the other hand, can be separated by days, i.e., they are {\it not} simultaneous in the case of RRLS.  Using PanSTARRS-based color data (and depending on the exact PanSTARRS exposure cadence), the color cut of Paper I would have to be expanded to $g-r \lesssim 0.9$ to capture the same fraction of RRLS, significantly lessening the usefulness of the cut.  In the end, because of the uncertainties associated with color cuts based on PanSTARRS photometry,  we decided not to apply any such cuts in determining which objects to use to extract lightcurves and classify. Similarly, paper I simplifies the analysis and classification pipeline by throwing away all objects that do not satisfy a
simple and quick-to-compute variability criterion, namely that the measured rms variability amplitude exceeds a certain (magnitude-dependent) threshold. Such a cut unfortunately throws away some moderately blended RRLS, which could still be detected, and RRLS with intrinsically lower variability amplitudes. Hence, we also do not apply a variability amplitude cut to our search catalog.  Rather, the PanSTARR colors and rms amplitude are but two of several ``features" that we combine to highlight  likely RR Lyrae lightcurve candidates for human inspection.  

\subsection{The New Photometric Calibration}
Many analyses of LSQ data \cite[e.g.,][]{cartier2015} rely on relative photometry based on {\it chip-wide} comparisons of reference stars. The absolute calibration of the photometry is also based on chip-wide comparison to external calibration stars, e.g., from the SDSS catalogs.  Unfortunately, the LSQ chips are an early design, with many cosmetic flaws such as charge traps and marginally stable regions. In particular, there are strong non-linearities and background fluctuations, particularly important for faint objects, that not only vary in time for a given chip but also as a function of position a chip, e.g., entire chunks of columns can suddenly go bad or change their non-linearity characteristics. This precludes the standard use of ``non-linearity" corrections computed only once at the time of camera commissioning.  This also limits the accuracy that can be achieved using a chip-wide calibration since parts of the same chip may be behaving quite differently at any give time. More importantly, this also means that one must be careful, whenever possible, to use comparison stars of approximately the same magnitude as the target object. As the non-linearity effects vary as a function of time, the overall instrumental correction at a given time and target magnitude may be quite different from a correction determined by comparing stars at brighter magnitudes, as sometimes happened in paper I.  In paper I, the overall impact of these photometry issues is relatively modest at the faint end of paper I's photometry ( at $V\sim 19-20$). They become quite significant, however, if one tries to go after objects as faint as those in Crater II ($V \gtrsim 21 $). To deal with them, we thus consider only target stars that have instrumental magnitudes in a given exposure within $\pm 0.5$ magnitudes of the target star instrumental magnitude for that exposure. To minimize color effects, and optimize for RRLS (which are blue), we also restrict the color of the comparison stars to be $g-r\sim 0.7.$ Finally, we restrict ourselves to comparison stars that are as close as possible in spatial position to the target star subject to the constraint that our comparison star set contains at least 3-5 ``good'' comparison stars. Here, ``good'' comparison stars are defined to be those that do not exhibit excess variability compared to the other comparison stars and whose median corrected instrumental magnitude agrees with that predicted from the external catalog. To further improve the accuracy of our relative photometry, we use the least squares fitting method of \citet{Honeycutt1992} \cite[see also][]{ubercal2008}. More details and cross-checks on the photometry pipeline, which may be useful for other experiments as it is implemented in a quite general manner, are found in the Appendix. 

\subsection{The New Period Finding Routines}
Paper I relied on a periodogram based on \cite{lafler_kinman1965} to identify the three most significant lightcurve folding periods to examine
in a given lightcurve. To restrict the number of objects to look at, Paper I required the most significant period to have a 
significance parameter $\Lambda \geq 3.0$. We compared the performance of this periodogram to that of more traditional ones like Lomb-Scargle \citet{lombscargle1976} and generally found its performance to be superior for bright sources ($V\lesssim 19$) that fell on good chips. With as few as $\sim 12$ lightcurve points, the correct period would usually be among the top three periods identified by the periodogram. As measurement errors increase, however, due to chip defects or Poisson statistics for faint sources, the LK periodogram falls apart. The correct period may still be found but the value of $\Lambda$ for that period now never exceeds 3.0. Simply dropping the threshold value of $\Lambda$ is not practical as the number of candidates to be looked at explodes. After some 
experimentation, we settled on the use of two other periodograms that can deal with significant measurement error: the Lomb-Scargle periodogram computed using the fast algorithm of \citet{press_rybicki1989} and the ``Analysis of Variance" (AOV) periodogram by \citet{schwarzenbergczernyAov1996}. We did not find other period search methods such as Supersmoother and generalized Lomb-Scargle to be worth the added computational cost -- at least for objects with the typical semi-sinusoidal shape and amplitude of RR Lyrae lightcurves. (Searches for objects with other lightcurve characteristics, such as narrow eclipses, are best done with other methods, however, such as those used for planetary transit detection.) For our application, the AOV periodogram is somewhat more sensitive than the Lomb-Scargle periodogram but it also more prone to spurious results due to bad photometry. Thus, we usually require a confirmation of a period detection in both the AOV and Lomb-Scargle periodograms to accept a candidate RRLS.

Because of aliasing, the effective time sampling (window) function of the lightcurve, and the addition of power from spurious photometry, the strongest peak in a periodogram may not represent the true period of a RRLS. The lightcurve data was automatically phase-folded on the 3 most signficant periods found in each periodogram and then fit using the RRL templates from \citet{laydentemplates1998}. To help combat bad lightcurve outliers, the templates were also fit to the phase-folded lightcurve binned and medianed over 10 phase bins/period. The template fitting step starts with the period from the periodogram as a best initial guess but then refines the period using a least squares minimization routine \cite[MINUIT, see][]{minuit94}. This compensates for the relatively sparse frequency sampling of the AOV and Lomb-Scargle routines and gives final period values that agree well with those in the literature.  We searched for periodicity in the range of $0.05$ to $2.0$ days. This range is important for identifying short period variables like Delta Scuti stars and contact binaries as well as longer period variables like Anomalous Cepheids that can easily contaminate RRLS searches and do not have the same standard candle luminosities as RRLS. Note that unlike some searches, the spurious alias periods corresponding to the typical 1-2 day cadence of observations (e.g., $P=0.25, 0.33, 0.5, 1$ day) were {\it not} excluded from consideration.

After the best RRLS candidates were selected by a combination of decision trees and human inspection, the quality of their template
fits were then assessed by eye, including checking the robustness of the best-fit to the inclusion or removal of marginal data points and making sure that an incorrect period alias was not chosen as the true period. One important further check, enabled by the LSQ supernova cadence that re-observes a source after $\sim$ two hours, was the confirmation of  true variability on short timescales.  This allows one to unambiguously distinguish, for example, an Anomalous Cepheid with a 0.9 day period from an RR Lyrae with a 0.45 day period. More importantly, this also allows one to significantly reduce contamination from quasars - a major headache for RRLS surveys that go fainter than $V\sim20,$ and one that has not been fully appreciated to date. (Note that for $V>20,$ we are looking for RRLS well out in the halo, where their sky density drops to significantly less than one per square degree. On the other hand, the density of quasars with RRL colors at $V\sim 20$ is already several per square degree and increases rapidly as one goes fainter.)  Most quasars do not show significant variability, $\gtrsim 0.1$ mag, on $\sim$ few hour timescales. RRLS do.  

\begin{figure*}
\hbox{\includegraphics[width=3.3in]{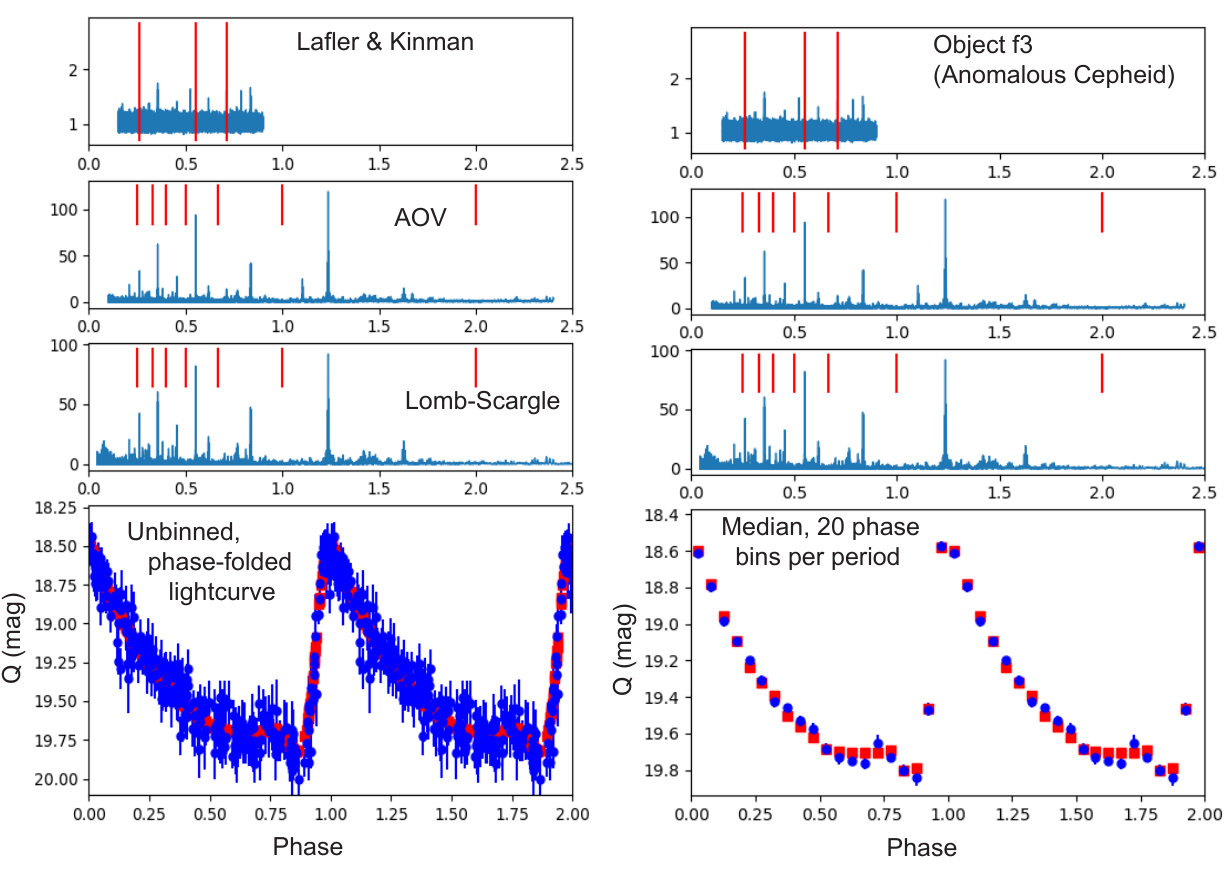}
\hfill \includegraphics[width=3.3in]{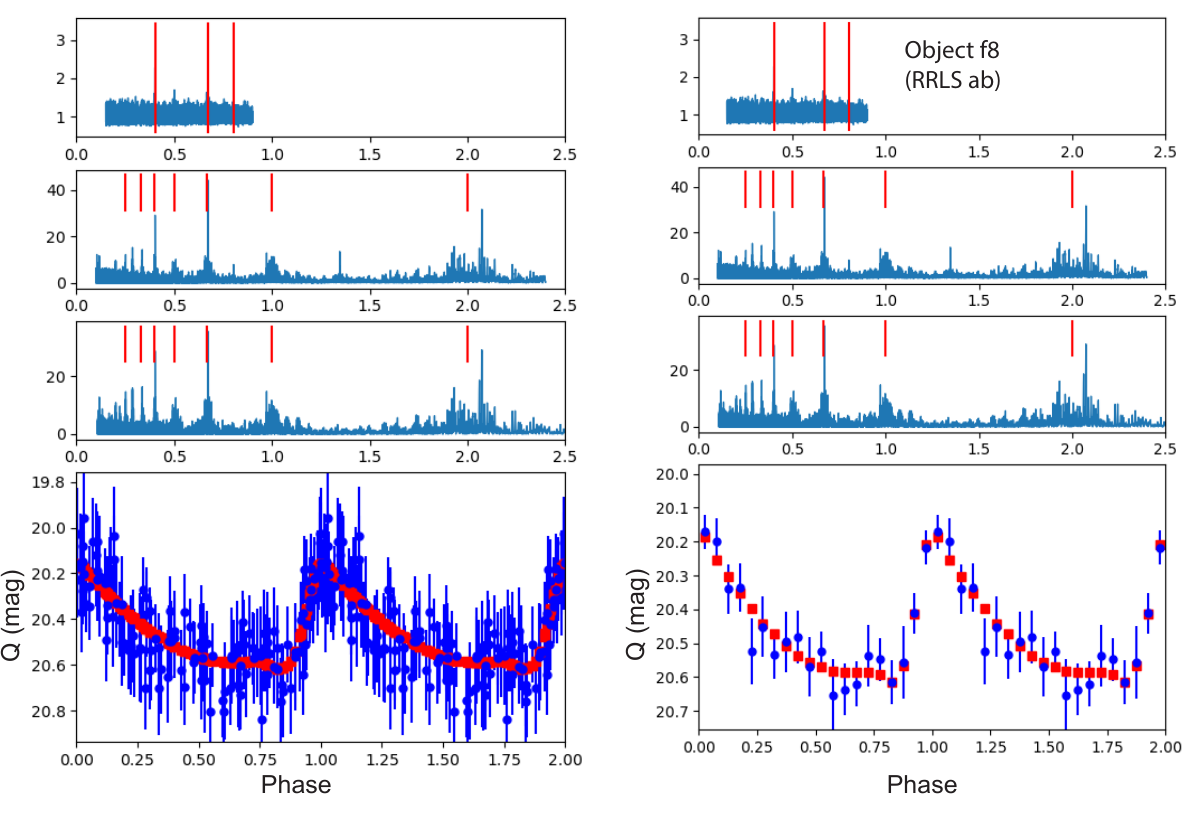}}
\hbox{
\includegraphics[width=3.3in]{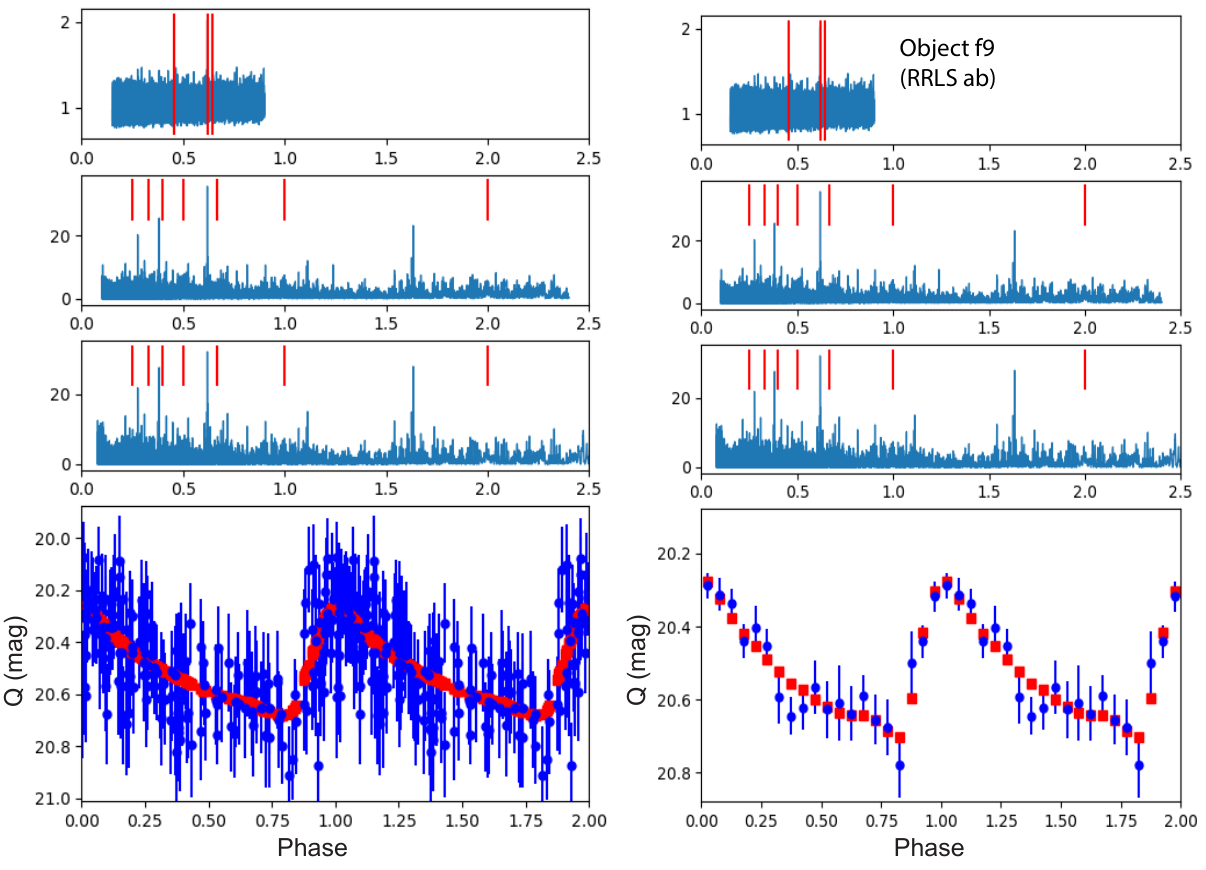}
\hfill
\includegraphics[width=3.3in]{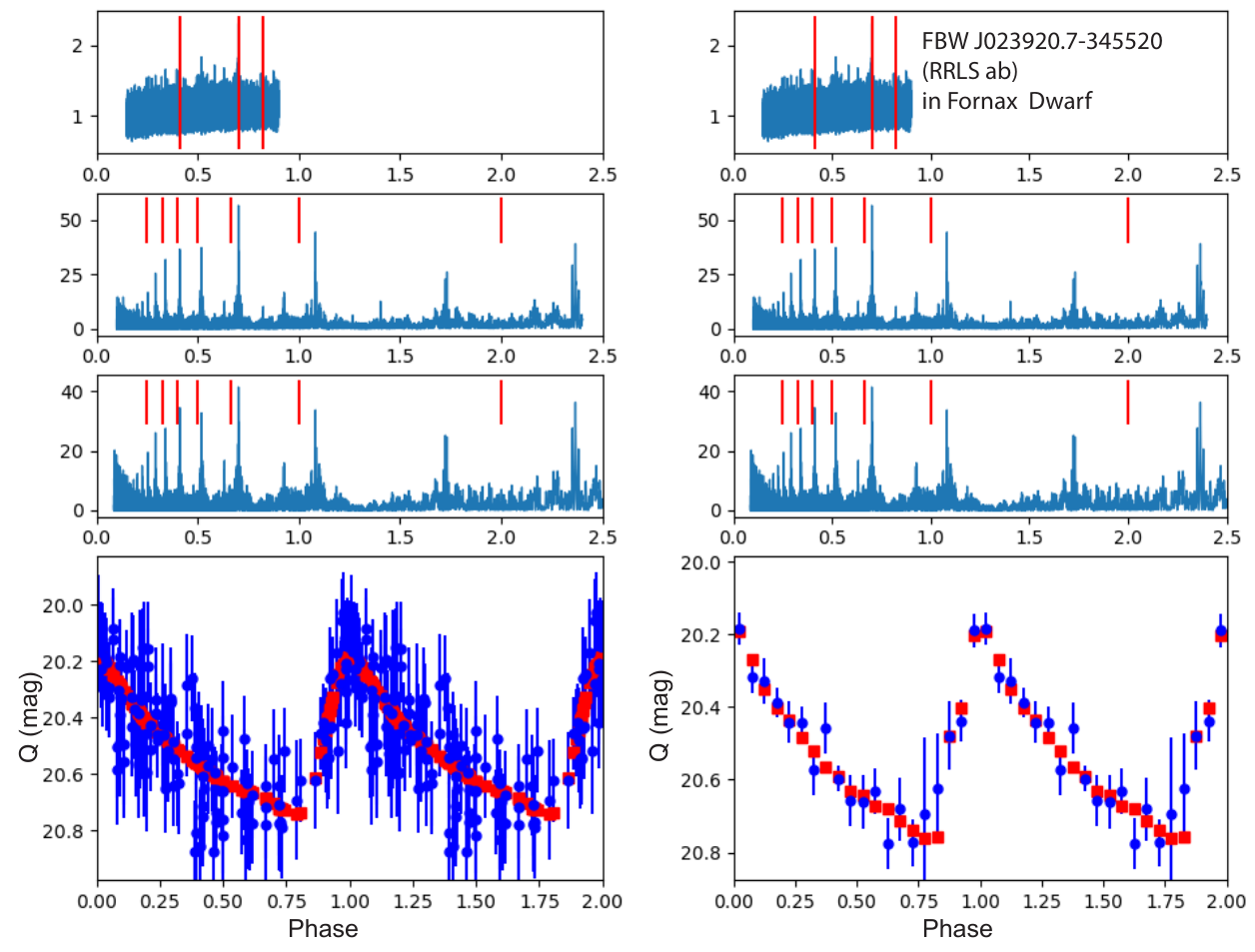}
}
\caption{Examples of the periodograms and phase folded lightcurves used to evaluate RRLS candidates. Plots are arranged in pairs, with 
the unbinned lightcurve at the bottom of the left panel and the binned, medianed lightcurve at the bottom of the right panel in a pair. 
The top 3 plots are identical in a pair and from top to bottom are the Lafler \& Kinman, AOV, and Lomb-Scargle periodograms respectively.
The the three red lines in the Lafler \& Kinman represent the 3 most significant periods as determined by the value of the $\Lambda$ parameter. The short red lines at the top of the AOV and Lomb-Scargle periodograms indicate common alias periods of the typical 1-2 night
interval between pairs of observations. (A source that does not vary much on a 1-2 day interval, like a quasar, will show up as a set
of strong peaks at these periods.) For typical LSQ data, a feature in the AOV and Lomb-Scargle periodograms that has a value exceeding
$\sim 10-15$ is significant and should be examined. All the periodicity detections for the objects shown here are thus highly significant.
To produce the binned lightcurves,
lightcurve points are grouped into 20 phase bins of width 0.05. The data point shown for each bin is then the median of the lightcurve points falling in that bin. In the folded lightcurve plots, the {\it blue circles} with errorbars are the LSQ measurements while the {\it red squares} show the value of the best-fit \citet{laydentemplates1998} RRLS template at a given phase. From left to right, plots in the upper set of panels are for objects
f3 and f8 respectively. Plots in the lower set of panels are for objects f9 and FBW J023920.7-345520, respectively. The periods used to fold
the lightcurves are those given in Table \ref{tab:fieldstars} for objects f3,f8, and f9, and 0\fd70158 for the Fornax FBW star. Lightcurves show the calibrated flux in the LSQ 
`Q' band (see Appendix, Equation \ref{eqn:qtov}). For typical RRLS, ${\rm V}\approx {\rm Q} + 0.75.$  }
\label{fig:periodograms}
\end{figure*}

To illustrate how well we can actually do in constraining the lightcurve shape and periods for faint RRL, we present in Fig. \ref{fig:periodograms}  the periodograms and template fits for an Anomalous Cepheid (object f3) and two of the faintest RRLS (objects f8 and f9) in the field near Crater II as well as for one already known RR Lyrae in the Fornax dwarf galaxy, at a distance of $\sim 140$ kpc.
In the AOV and Lomb-Scargle periodogram panels of Fig. \ref{fig:periodograms}, any line with a strength $\gtrsim 10-15$ 
typically represents a significant detection of periodicity (with Lomb-Scargle false alarm probability $\lesssim 10^{-4}$). We see
that all the periodicity detections are of very high significance because of the large number of observations (>100), especially for the Anomalous Cepheid, which is brighter, but also for the RRLS in Fornax (at $d\sim 140$ kpc). The RRLS in Fornax (at $\alpha=39.8364$ and $\delta=-34.922348$) was identified in a previous study \cite[FBWJ023920.7-345520 in][]{fornaxvariables2002}, where it had a reported lighcurve-averaged magnitude of $<\rm V> = 21.212 \pm 0.02$ and a period of 0\fd68058 based on 30 observations. Using DELVE DR2 comparison stars (the only ones available at this magnitude and declination), we instead find $<\rm V> = 21.239,$ by averaging over the best-fitting template, and a period of 0\fd70158 based on the LSQ lightcurve with 139 points. Given the strength of our period detection in all three periodograms as well as our significantly larger number of lightcurve points, we believe our period determination to be the more accurate one (assuming the period has remained constant in time, which it may not have). Indeed, this star also appears to be in the catalog of DES-selected RRLS \citep{StringerDESRRL2021}, with a period 0\fd 7016 .

The periodogram and folded lightcurve plots for the distant stream  stars f8 and f9 look similar to those for the known Fornax star, which argues for their also being real RRLS. (The stream stars have similar average magnitude as the Fornax star but somewhat lower oscillation amplitudes, hence their period detections are not as significant.) Note also that while the unbinned phase-folded lightcurve for the Anomalous Cepheid f3 looks quite ragged, the phase-binned median lightcurve is extremely well-constrained (the estimated error bars are only a few percent) and lightcurve points generally sit right on top of the template except at late phases where real RRLS/AC lightcurves often show variations in the dip just before the return to maximum light. In other words, the measurement scatter can indeed be medianed away despite potential systematic issues with the LSQ CCDs. Given enough observations, we can place very tight constraints on RRLS parameters, even for faint RRLS with $<\rm V>~ \gtrsim 20.$  Metallicity measurements based on lightcurve shape, e.g., \citet{jurcsik96},
are definitely possible with lightcurves of the quality of the Anomalous Cepheid and will be explored in a future paper.

\subsection{Search Procedure for RRLS}
Even after the application of ensemble photometry, LSQ lightcurves can still have significant outliers that will set off various variability detection triggers. Fake variability can also be induced in the lightcurves by misphotometering of extended galaxies, and at the faint magnitudes considered here ($V\gtrsim 21$), the number density of halo RRLs is dwarfed by the density of quasars, some of which do appear as real, variable, and blue point-like objects (and have contaminated some previous RRLS surveys). One must therefore be careful and conservative in dealing with LSQ data. Our attempts to use absolute cuts on the goodness-of-fit of RRLS templates, as in some previous work, failed because of LSQ's data quality, for example. The correct template was usually fit and at the right period, but the absolute value of goodness-of-fit metrics we could come up with proved too sensitive to outliers, leading to significant incompleteness. To better understand which lightcurve features are robust to LSQ data issues and then use these features to   
cut down the number of candidates to be examined by hand, we generated lightcurves for all objects in the Sextans dwarf and Crater II galaxies, where deep, well-sampled, and highly complete RRLS surveys are available based on DECAM 4m photometry, e.g., see \cite{vivas19} and \cite{vivas20}. We then examined the set of resulting LSQ lightcurves by hand to decide if LSQ could claim a detection of a given object or not. As discussed below, the fraction
of known RRLS recovered by LSQ was approximately 93\% in the Sextans dwarf  and 77\% for Crater II. (The Crater II fraction was lower because some of the lowest amplitude variables were not detected significantly above the partly systematics-induced background fluctations.)  
We then developed a series of lightcurve feature cuts (such as on clipped RMS lightcurve variability amplitude and the location and false alarm probability of the strongest Fourier peak) that reduced the candidate background significantly but still let through $\sim 95\%$ of the known RRLS in these systems. In general, what enables LSQ to succeed in pulling out and classifying persistent periodic sources is the relatively large number of observations available for a source (typically >100 in the vicinity of Crater II) and the fact that bad photometry due to uncorrelated factors (like the seeing on different nights) usually does not generate significant periodogram peaks, except at period aliases of one day. 

For southern regions, like the Crater II one, where one cannot rely on color information because of the PanSTARRS filter simultaneity issues discussed above, the number of sources passing our initial cuts is $\sim 1000$ in a $5^{\circ}x5^{\circ}$ region. These sources are then quickly scanned by eye using an interface that flashes up condensed versions of the plots in Fig. \ref{fig:periodograms}. In a region without significant halo structure ({\it not} Crater II), we typically find $\sim 50-100$ true RRLS but also up to $\sim 50$  objects that are real sources and periodic, like Delta Scuti stars and short period, contact binaries. These non-RRLS periodic sources become relatively more numerous as one goes to fainter magnitudes and can masquerade as RRLS in poorly sampled lightcurves. LSQ usually has good enough lightcurve coverage to reject them but we caution that some past surveys do not. 
The remaining background sources  seem to be mostly badly photometered (extended) galaxies, quasars, and low mass active stars 
like M stars. Experiments indicate that we may be able to remove them automatically in the future by, e.g., looking at the details of the 
AOV/Lomb-Scargle periodograms and removing objects that do not show significant intraday variability. (The last condition, intraday variability, can be problematic at faint magnitudes where the error estimates on individual measurements are not always reliable.) 

Once good periodic candidates have been identified by the scanning process, their lightcurves are examined by at least two sets of eyes to remove obviously bad outliers, to make sure the correct period has been identified, and to classify the object (e.g., RRLS Type ab vs. c vs. a close binary or Delta Scuti). For strong RRLS candidates, we then check external imaging data to make sure the accepted objects are point-like (stellar) and do not have very red colors (which galaxies often do). In the scanning by humans, we have tried to be conservative in accepting candidates, sacrificing completeness for purity. We thus expect the Crater II region sample we present below to be very clean (>95\% pure), even at the faint end. The lightcurves for the objects we consider are of comparable quality to those shown above, which represent significant, unambiguous detections of RRLS. As cross-checks on our purity, once the DELVE DR2 catalog became available, we also looked at the Sculptor and Fornax dSph galaxies, which also have well-studied RRLS populations, using the same procedure and data cuts employed here.  Except for regions that fell on obviously bad or inoperative chips, the recovery rate of known RRLS was similar to that reported here for Sextans and Crater II, $\sim 89\%$ for Sculptor and $\sim 73\%$ for the more distant Fornax. In the best studied regions of those galaxies, which we expected to be complete, we did not find any new candidates, although a few of our periods and classifications did disagree, e.g., as presented above for FBW J023920.7-345520, the star in Fornax where we suspect our period is close to the correct one. Based on checks like these, we are thus confident that the stream features we claim here are not the result of spurious RRLS detections. A further argument against a significant fraction of the RRLS in the streams being fake is that the (unlikely) failure modes we could think of, e.g., confusion with quasars or misphotometered galaxies, do not know or care about position on the sky, and in particular, the direction of Crater II's proper motion. As an example of this, we do keep a record of the possible RRLS candidates we placed in the `maybe/marginal' category. Except for the coverage holes discussed in the Appendix (which do not coincide with the likely Crater II streams), those marginal candidates, which likely include a significant fraction of fakes, are to first order isotropically distributed on the sky.

\section{RRLS in the Crater II Region}
We consider here the region of the LaSilla-Quest RRLS survey bordered by $170^{\circ} \le \alpha \le 190^{\circ}$,
$-25^{\circ} \le \delta \le -10^{\circ}$, which includes Crater II. In that region, the survey identified 78 RRLS and 7 Anomalous Cepheids (ACs) in Crater II and 587 RRLS and 2 ACs in the field. Not surprisingly, the variables in Crater II dominate the
sample at $d_{\odot}$ $\sim 120$ kpc. We show below that there is a very
significant overdensity of stars in the range in $d_{\odot} \sim 80$ to $\sim 135$ kpc, which passes
through Crater II. But first we consider the completeness of sample, the accuracy of the period
finding, and the precision of the photometry. A comparison of our results with the
previous surveys of Crater II provide estimates of these factors at the distances of the tidal
streams.
\subsection{Completeness, period finding, and photometric precision at Crater II}
The completeness of the survey depends on the brightness of the variables, the amplitudes and stability of their lightcurves, and the number and precision of the observations.
The previous surveys of Crater II \citep{monelli18, joo18, vivas20}  list 101 RRLS and 31 brighter stars in the range <V> $\sim 16.4$ to 20.7, which include the 7 ACs in Crater II and 24 field variables of different types. Our survey
identified 100\% of the ACs and the field RRLS, but was less complete for the fainter RRLS in Crater II. We have not attempted to identify double-mode type d RRLS in our survey, and these stars are included with
the single mode type c variables.  The irregular lightcurves of the type d and the small amplitudes of both types d and c make them more challenging to detect in general than the type ab variables.  Nonetheless, our survey identified, as type c, 5 of the 10 type d and 4 of the 5 type c variables catalogued by \citet{vivas20}. It found 80\% of the 86 type ab variables identified by
\citet{joo18} and \citet{vivas20}. We include among the Crater II RRLS V79 and V96, which were labelled as
`field?' by \citet{joo18} and lie beyond the boundary of the observations of \citet{vivas20}. To
within the errors, they are at the same distance as Crater II and are probably members of its
tidal streams. The examination our data for the stars that were missed revealed that some had
few observations, presumably because the star frequently fell in a gap between the chips or on
a dead chip, or because our photometry was too noisy to determine well a period.  Our overall completeness for the Crater II RRLS, which have a mean <V> of 20.95, is 77\%.  The Crater II RRLS are approaching the faint limit of our survey, and its completeness becomes worse at fainter magnitudes, where it only detects large amplitude variables such as the type ab variable in the Fornax dSph galaxy shown in Fig.~\ref{fig:periodograms}.  The completeness climbs significantly, however, at brighter magnitudes.  For example, the LSQ survey detects 93\% of the 199 RRLS catalogued by \citet{vivas19} in the Sextans dSph galaxy, which have a mean <V> of 20.32. With very few exceptions, the LSQ periods we find are in excellent agreement with those published in both \citet{vivas19} [Sextans] and \citet{vivas20} [Crater II].

Our observations of the Crater II stars are summarized in Table~\ref{tab:crater2stars}, where they are identified by the V number assigned by \citet{joo18,monelli18,vivas20}.  The columns in the table list, in order, our data for the positions of the stars (epoch 2000.0), the type of variable, the number of observations used to fit a template, the period, the amplitude, the heliocentric Julian Date of maximum light, the intensity averaged V magnitude, and the interstellar extinction in V. \footnote{From the reddening maps of \citet{schlegel98} as updated by \citet{schlafly11}.}  The actual LSQ lightcurve data for these stars can be found in the online supplementary material (in the file "rrlyrae\_lightcurves.txt").

For all but two stars, our periods agree well with the ones determined by \citet{vivas20}.
With our data, the period of 1\fd02985 for the AC V107 produces a superior lightcurve than does the
period 0\fd51347, which was determined by \citet{vivas20}. With the shorter period, the high
precision data of \citet{vivas20} produces a lightcurve with an approximately instantaneous rise
time with no g-band observations on the rising branch and almost no i-band ones. With the
longer period, their observations have large gaps in phase, so it is understandable that their
period-finding routine selected the shorter one. The longer period is adopted below in our discussion of
the ACs. For the type d RRLS V80, our data yields a period of 0\fd462295, while Vivas et al. found
a period of 0\fd44678, which with their data is much preferred over our period.  We have no explanation for this
difference in period except that V80 is double mode pulsator, which we and \citet{vivas20} observed at
different epochs.
In agreement with \citet{ngeow22}, we do not consider V108 to be an AC in Crater II. Its
luminosity is much too large for its period for it to be an AC member. It is more likely a type c
RRLS in the foreground of Crater II.

\begin{figure}
\includegraphics[width=3.3in]{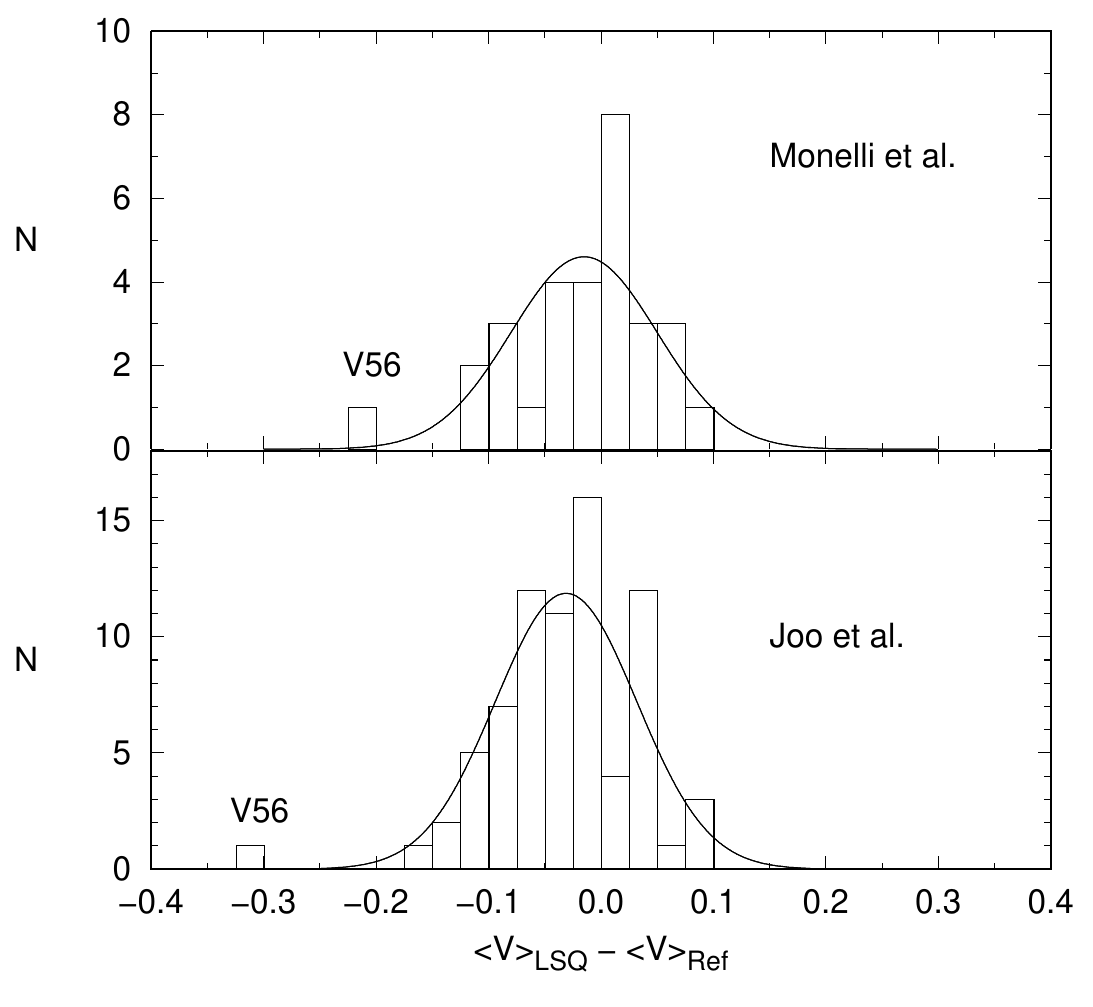}
\caption{The distributions of V magnitude differences between the LSQ measurements and the corresponding measurements of \citet{monelli18} ({\it top} panel) and \citet{joo18} ({\it bottom} panel). The {\it solid} lines show the best Gaussian fits to these distributions.}
   \label{fig:monellijoocomp}
\end{figure}

A direct comparison of our V magnitudes can be made with the ones measured by \citet{joo18} and \citet{monelli18} -- see Fig.~\ref{fig:monellijoocomp} for the distributions of the corresponding magnitude differences. 
For the 75 stars in common with \citet{joo18}, the mean difference between
their mean magnitudes and ours is -0.03, with a standard deviation ($\sigma$) of 0.06. The 30 stars in common with \citet{monelli18} instead yield a mean difference in <V> of -0.02, with $\sigma = 0.07$.  The star V56 is clearly an outlier in these comparisons because our measurement is too bright by $\sim 0.25$ mag, possibly because of its proximity to a pair of very bright stars.  Discarding this star does not significantly alter the outcome of the comparison with \citet{joo18}.  It reduces the mean offset and $\sigma$ of the relatively small sample in common with \citet{monelli18} to -0.01 and 0.05, respectively. For the V amplitudes of the RRLS lightcurves, we instead find a mean difference and spread ($\sigma$) in the measured amplitudes of -0.05, (0.12), and -0.10, (0.13) for the stars in common with \citet{joo18} and with \citet{monelli18}, respectively.  V56 is not particularly deviant in this comparison. To put these measurement differences in perspective, it is useful to compare the measurements of these authors with each other.  There are 36 stars in common between \citet{joo18} and \citet{monelli18}. 
The \citet{monelli18} measurements have a mean offset of 0.03 in <V> ($\sigma = 0.03$) and -0.03 in V amplitude ($\sigma = 0.14$) with the respect to those of \citet{joo18}.

Our accuracy and precision in <V> is what matters most in the analysis that follows. The above comparisons suggest that for the majority of variables with <V> $ \sim 20.9$, our measurements of <V> do not deviate by more than a few hundredths of a magnitude from the V scale and have $\sigma \sim 0.05$. In other words, measurement error alone {\it cannot} explain the significant spread (and trend) in distances that we find for the RRLS that make up the overdensity we discuss below. Also, our survey is able to find >~70\% of the RRLS at this magnitude, including the typically lower amplitude type c stars, as well as measure reliably their periods.  Crater II is, however, near the outer limit of our survey, and beyond its distance ($\sim 120$
kpc) our completeness and precision decline further. This limits our ability to trace the
trailing stream of Crater II beyond $\sim 140$ kpc. For RRLS closer than Crater II, our completeness rises to >~90\%.

\subsection{The Anomalous Cepheids}
We have found two stars that are probably ACs in the field surrounding Crater II.  This type of variable is rarely found in globular clusters, for only 2 are known in the Milky Way globular clusters (V19 in NGC5466, \citet{Zinn1976, McCarthy1997} and V7 in M92, \citet{Matsunaga2006, DiCriscienzo2007, Osborn2012}.  They are more common in the in the dSph galaxies, and while several have been found in the Galactic Halo, they are much rarer than the RRLS.

\begin{figure}
	\includegraphics[width=\columnwidth]{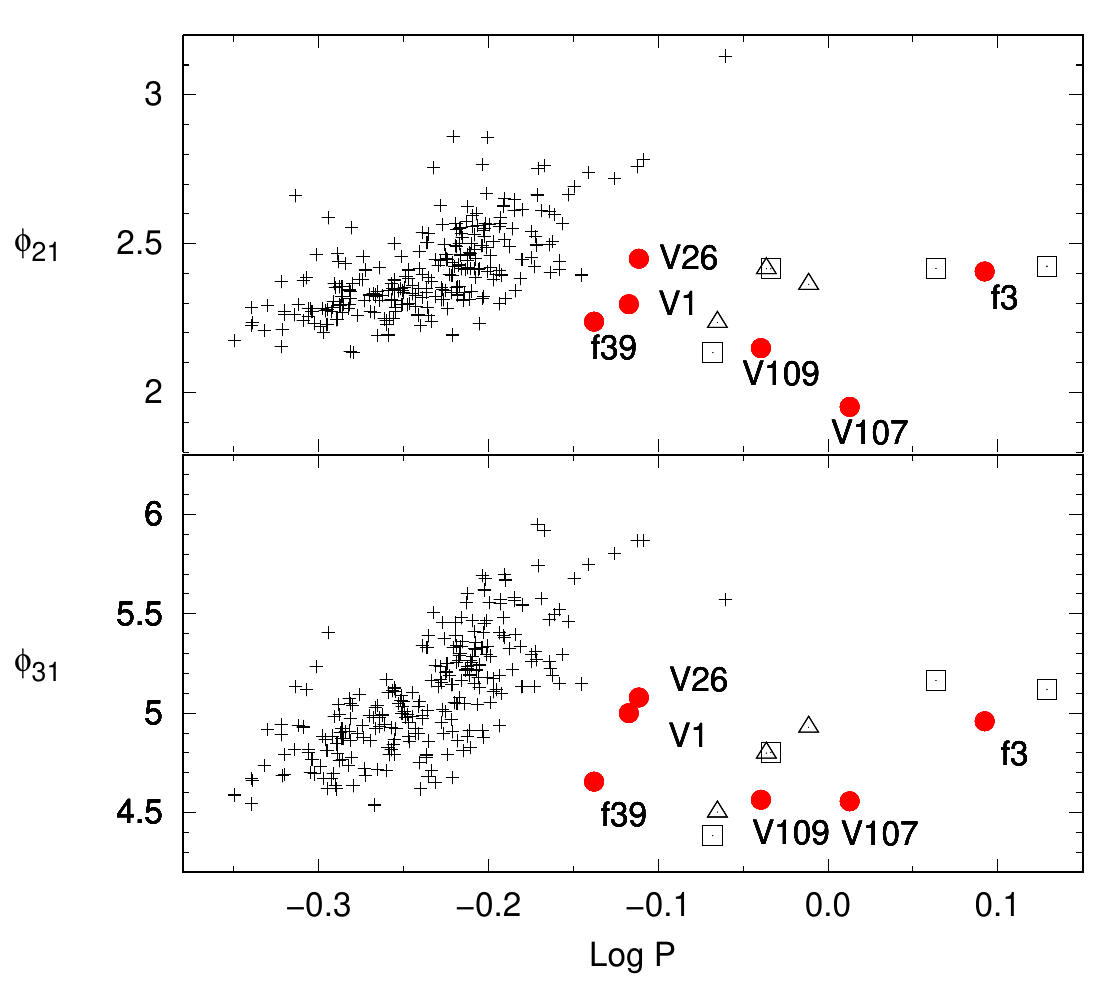}
    \caption{In plots of the phase differences $\phi_{21}$ and $\phi_{31}$ against the log of the period, the fundamental mode ACs in Crater II and the two which we have identified in the field ({\it labelled red points}),  are compared with a sample of 256 type ab RRLS ({\it small crosses}) and with ACs in the Sextans ({\it open triangles}) and Sculptor ({\it open squares}) dSph galaxies.  Even though some of the ACs overlap in period with the RRLS, they are well separated in both $\phi_{21}$ and $\phi_{31}$.}
    \label{AC_phi21_phi31}
\end{figure}

The identification of ACs is not always straightforward, because some ACs have periods in the
same range as the RRLS, and their light curves also resemble those of RRLS. As shown by
\citet{soszynski15} using i-band light curves of Magellanic Cloud stars, the combination of the periods of the stars with
the phase differences $\phi_{21}$ and $\phi_{31}$, which are obtained by Fourier decomposition of their light
curves, provides an unambiguous separation of fundamental mode ACs and type ab RRLS. Since the lightcurve shape of ACs and RRLS depends on wavelength, it is important to use lightcurve shape parameters derived from lightcurves obtained using the {\it same} filters as those used to obtain the lightcurves for the objects one wants to classify. Hence, we cannot use the i-band results of \citet{soszynski15} directly. The results we obtained using LSQ (V-band) observations are shown in Fig. \ref{AC_phi21_phi31}, where the ACs in Crater II, the two stars
that we have identified as ACs in the field (f3 and f39) are compared with 256 type ab RRLS  found in the same field as Crater II. Additionally, we plotted the positions of known ACs in the dSph galaxies Sextans (determined using LSQ photometry) and Sculptor (determined either using LSQ photometry or V band photometry from \citet{martinezvasquez2016}).  Since good quality light curves are required to determine $\phi_{21}$ and  $\phi_{31}$ reliably,
the RRLS sample contains only 18 Crater II stars. The other RRLS are brighter field stars. The Crater II-related ACs are sufficiently bright that adequate light curves were obtained for all of them. 
 Small values of $\phi_{21}$ and $\phi_{31}$ indicate that the stars have ``saw-tooth'' lightcurves, while ones with the largest values have almost sinusoidal lightcurves. Consequently, the sequences for the RRLS illustrate the well-known property that as period increases the lightcurves transform from Bailey type a to type b.  While V1 and V26 in Crater II and the field variable f39 overlap in period with the RRLS sequences, their smaller values of $\phi_{21}$ and $\phi_{31}$ indicate that their lightcurves are thus more ``saw-tooth'' in shape than those of RRLS with similar periods.  Because the Magellanic Clouds contain ACs of similar period and offsets from the RRLS sequence \citep{soszynski15}, it is most likely that these stars are ACs.  From its position in Fig. \ref{AC_phi21_phi31} there is no doubt that f3 is an AC.

\begin{figure}
	\includegraphics[width=\columnwidth]{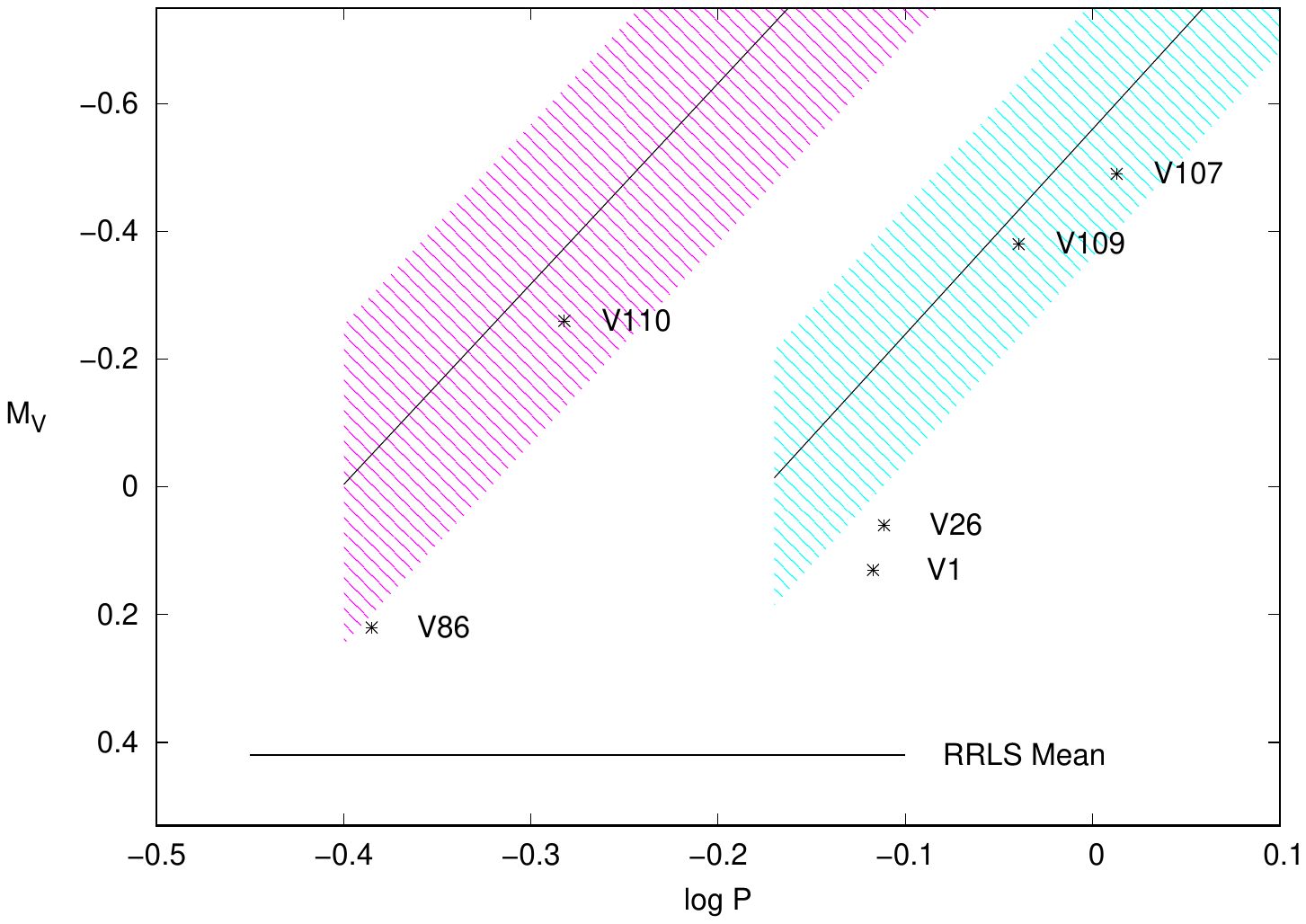}
    \caption{The ACs in Crater II are compared with the P-L relations that \citet{ripepi14} determined from observations of the ACs in the LMC.  The solid lines are the mean relations, and the shaded regions indicate the $\pm 1 \sigma$ scatter about the lines (fundamental and first overtone, cyan and magenta, respectively). The $M_{V}$ values for the stars were set by assuming $M_{V} = 0.42$ for the RRLS in Crater II.  The fundamental mode relation is used to estimate the distances of the two ACs in the Crater II tidal streams.}
    \label{fig:P_L_10_11}
\end{figure}

ACs are known to exhibit a period-luminosity relation (P-L), and a few V-band ones are available
in the literature. In Fig.~\ref{fig:P_L_10_11}, we have plotted our measurements of the Crater II ACs assuming a
distance modulus, $(m-M)_{o}$, of 20.38 for Crater II. This was derived by first estimating the
mean $M_{V}$ of the RRLS in Crater II (+0.42) from the period-metallicity relation that \citet{garofalo22}
recently derived from Gaia parallaxes of nearby RRLS variables and the mean metallicity ([Fe/H] $= -2.16$) that \citet{ji21} measured for Crater II. This value of $M_{V}$ was subtracted from
the reddening corrected mean V magnitude that \citet{joo18} measured for their sample of RRLS in
Crater II. The distance modulus used here is slightly larger than the one ($20.333 \pm 0.004$
statistical, $\pm 0.068$ systematic) \citet{vivas20} obtained from a P-L relation for RRLS in the i-band.
Adopting their value instead of the one used here would shift the AC positions in Fig.~\ref{fig:P_L_10_11} by only
0.05 mag. The solid lines in Fig.~\ref{fig:P_L_10_11} are the measured P-L relations that \citet{ripepi14} determined
from ACs in the Large Magellanic Cloud (LMC). The rms scatter of the individual stars about
these relationships is 0.20 and 0.25 mag. for the fundamental mode and the first overtone
relations, respectively \citep{ripepi14}. The shaded regions in Fig.~\ref{fig:P_L_10_11} illustrate this scatter.
\citet{vivas20} noted that the faintest ACs that they identified in Crater II (V1, V26, V86) are
clearly more luminous than RRLS, but also among the least luminous known of this type of
variable. Their values of $\phi_{21}$ and $\phi_{31}$ are consistent with their identification as ACs (Fig.~\ref{AC_phi21_phi31}). Fig.~\ref{fig:P_L_10_11}
shows that while these stars are outliers from the LMC relations, they are not extremely so.
The three brightest ACs in Crater II (V107, V109, V110) appear to be more normal ACs, because
they fall well within the scatter about the P-L relations.

The two ACs that we have identified in the field of Crater II (f3 and f39 in Table 2) have large amplitudes, saw-tooth
light curves that resemble those of type a RRLS. They are probably fundamental
mode pulsators, and we have estimated their values of $M_{V}$ from the mean P-L relation \citet{ripepi14} derived for the fundamental mode ACs in the LMC. We have also adopted 0.2 mag as the $1 \sigma$ uncertainty in their values.

\subsection{The Evidence for Stellar Streams}

\begin{figure}
	\includegraphics[width=\columnwidth]{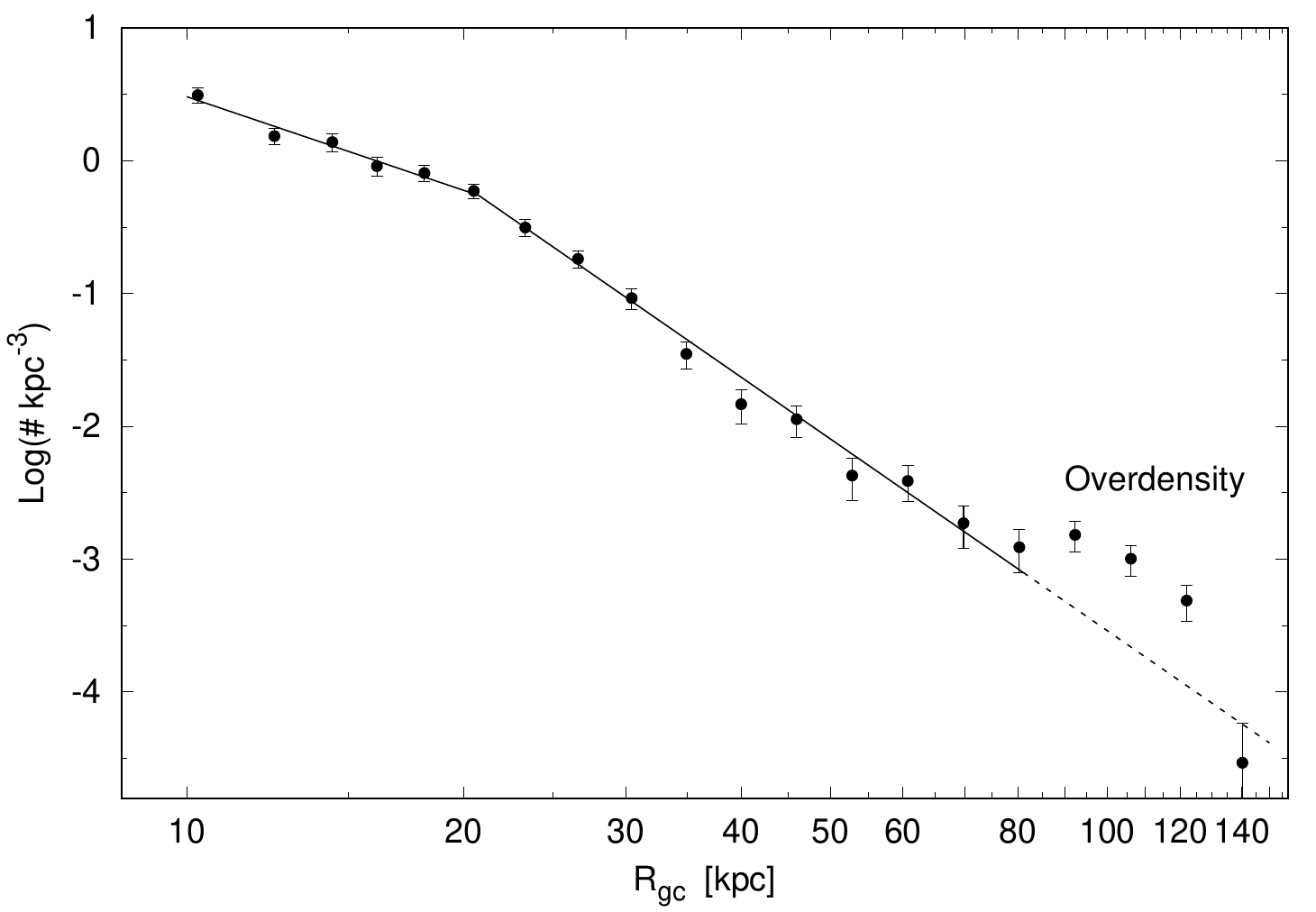}
    \caption{After {\em all} of the previously identified variables in Crater II have been removed from the sample, the number density of RRLS is plotted against galactocentric distance $R_{gc}$.  The {\it solid lines} are weighted fits of power laws to the data in the ranges 10--20 and 20--80 kpc, which have exponents of $-2.3 \pm 0.3$ and $-4.8 \pm 0.1$, respectively.  The {\it dashed line} is an extrapolation of the 20--80 line to 150 kpc.  Note the large overdensity compared to this line.}
    \label{fig:density_plot}
\end{figure}

The large overdensity in the vicinity of Crater II is evident in Fig.~\ref{fig:density_plot}, where we have plotted the
number density of RRLS in the region bordered by $170^{\circ} \le \alpha \le 190^{\circ}$ and $-25^{\circ} \le \delta \le -10^{\circ}$, as a function of galactocentric distance
($R_{gc}$). This region is nearly centered on Crater II. Before making this plot, we removed from the sample the stars identified previously
as members of Crater II \citep{joo18,monelli18, vivas20}, and in addition the stars V79 \& V96 mentioned above. The
data in Fig.~\ref{fig:density_plot} show a break in the density profile near $R_{gc} \sim 20$ kpc.  Similar breaks have been found in other directions, and they appear to be a general characteristic of the Galactic halo \citep[e.g.,][]{deason11, deason18, han22}. The power law slopes that we find for the inner and outer halo ($-2.3 \pm 0.3 ~\& -4.8 \pm 0.1$, respectively) are consistent with the many measurements in the literature for $R_{gc} < 60$ kpc (see fig. 13 in \citet{han22}).  For the
current discussion, the most noteworthy feature of Fig.~\ref{fig:density_plot} is the large excess of stars in the range $87 \lesssim R_{gc} \lesssim 132$ kpc. We found a total of 41 RRLS in this region,
whereas only 10 are expected on the basis the dashed line in Fig. \ref{fig:density_plot}. The error bars in Fig. \ref{fig:density_plot},
which were computed using Poisson statistics, suggest that this excess is highly significant. Crater II lies at $R_{gc} = 118$ kpc, which places near the middle of this overdensity in distance as well as position on
 the sky.

To examine this overdensity in more detail, we limit following discussion to the stars that lie in the same region of the sky as Fig. \ref{fig:density_plot}  and have $d_{\odot} > 80 $ kpc. The Crater II stars in this region are listed in Table~\ref{tab:crater2stars}.  The RRLS and ACs not previously identified with Crater II are listed in Table~\ref{tab:fieldstars} in order of increasing right ascension, and for ease of identification, they are given the designation `f' followed by their position in the table.  The next 9 columns list the same types of data as described above for Table~\ref{tab:crater2stars}.  The next 3 columns list the value of $d_{\odot}$, the stream longitude ($\xi$) and latitude ($\eta$), which are described in more detail below.  The final column lists whether a star belongs to the leading (L) or trailing (T) stream, or neither one (N). The LSQ lightcurve data for these stars may again be found in the online supplementary material (in the same file, "rrlyrae\_lightcurves.txt").

\begin{figure}
	\includegraphics[width=\columnwidth]{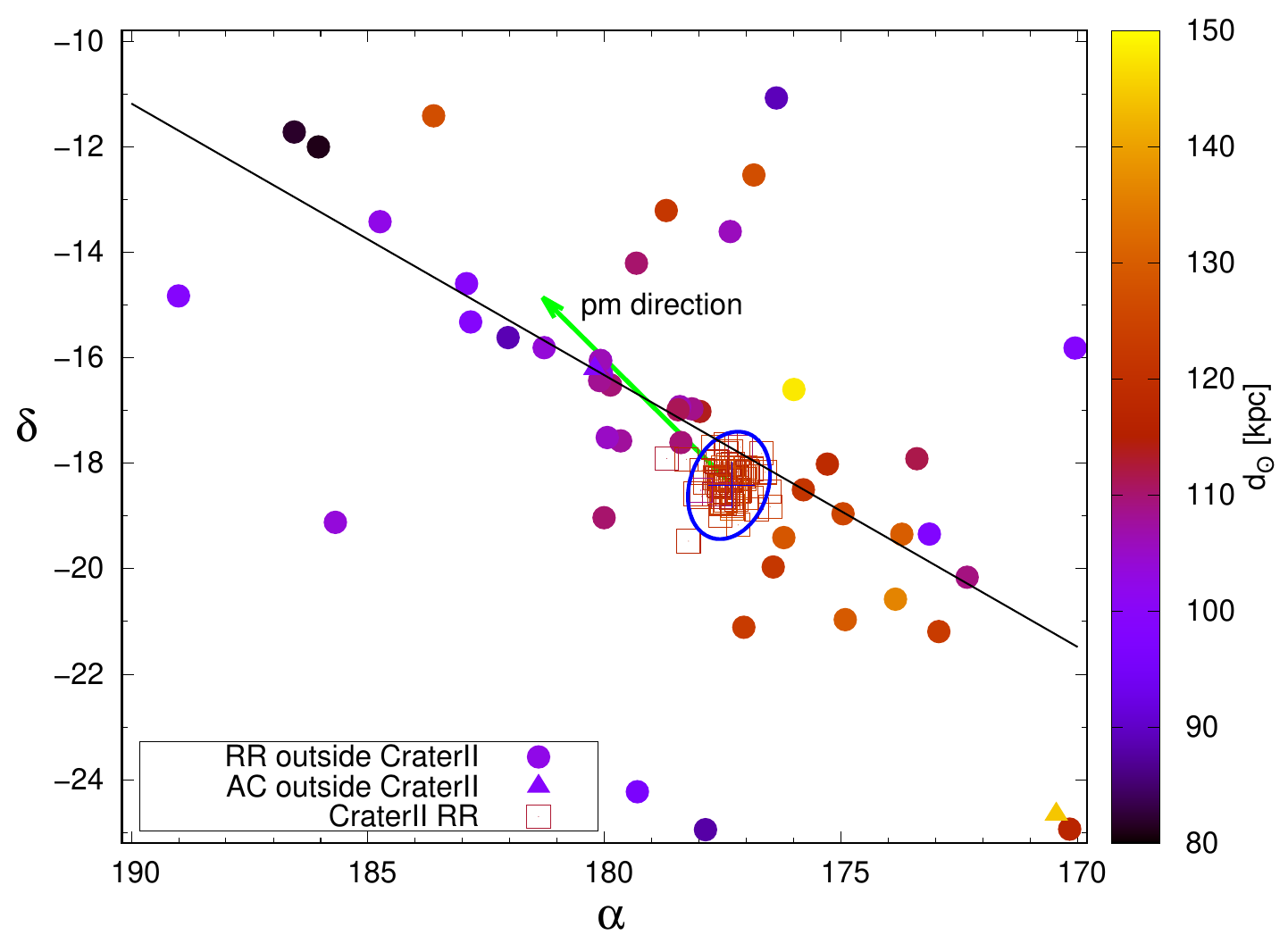}
    \caption{The plot of the sky around Crater II showing the locations of the RRLS and the ACs that were identified by our survey at $d_{\odot} > 80$ kpc.  The {\it open squares} are our sample of RRLS in Crater II.  The {\it blue ellipse} is the $3 \sigma$ bi-variate Gaussian fit that \citet{vivas20} made to their sample of Crater II stars. The {\it solid circles} and the {\it solid triangles} are the newly identified RRLS and ACs, respectively. The star symbols are color coded by $d_{\odot}$, and note that the field stars close to Crater II on the sky are also close in $d_{\odot}$. The {\it black line} is the Theil-Sen fit of a straight line to the stars not previously identified with Crater II.  Note that this line passes close to the center of Crater II and is roughly in the direction of the proper motion vector of Crater II, relative to the galactic standard of rest ({\it green arrow}, \citet{ji21}).}
    \label{fig:Sky_Map_10_11}
\end{figure}

We have plotted the distribution of these stars on the sky in Fig.~\ref{fig:Sky_Map_10_11}. The
strong concentration variables in Crater II, and the large angular size of this galaxy are clearly
evident. The solid circles are the 47 RRLS that we identified outside of the areas of Crater II that
were surveyed previously \citep{monelli18,joo18,vivas20}. These stars have $R_{gc} > 79$ kpc, and 41 lie within the range where the overdensity is so evident in Fig.~\ref{fig:density_plot}.

A simple visual inspection of Fig.~\ref{fig:Sky_Map_10_11} suggests that the stars lying outside of Crater II are not randomly distributed across the field, but have instead a linear distribution that runs through Crater II. This can be quantified as follows.  For the $\alpha$ and $\delta$ values of the 47 RRLS and the 2 ACs lying outside of Crater II, Kendall's $\tau$ statistic $= 0.4796$, which indicates that the probability that these coordinates are uncorrelated is $1.2x10^{-6}$ \citep{press92}.  We have fit a straight line to these coordinates using Theil-Sen regression, which is much more robust in the presence of outliers than the method of least-squares \citep{akritas95}.  This line (see Fig.~\ref{fig:Sky_Map_10_11}) passes within $0.6^{\circ}$ of the center of Crater II, which is an unlikely coincidence if the stars and the galaxy are unrelated. 
 In addition, this line is approximately in the same direction as the proper motion of Crater II relative to the galactic standard of rest \citep{ji21}.  The color coding of the points by $d_{\odot}$ 
 in Fig.~\ref{fig:Sky_Map_10_11} show that the majority of stars lying near Crater II are also near it in $d_{\odot}$. There are also trends of $d_{\odot}$ with position along the line that extend in either direction from Crater II (see also Fig.~\ref{fig:Dist_long_10_14}). 

 In addition to the large-scale overdensity (see Fig. \ref{fig:density_plot}), there are two small regions of the sky outside of Crater II  with remarkably high densities of RRLS. The most striking is located at $(\alpha,\delta) \sim (180\overset{\circ}{.}1,-16\overset{\circ}{.}3)$, which is $\sim 3.5^{\circ}$ from the center of Crater II and $\sim 2.6^{\circ}$ from the ellipse in Fig.~\ref{fig:Sky_Map_10_11} that roughly marks the outer boundary of the galaxy. This clump of stars is made up of 6 RRLS and one AC, which have, to within the errors, the same $d_{\odot}$.  This clump easily fits within a circular region of $0.5^{\circ}$ radius, and therefore, its RRLS density is $>7.6~{\rm deg}^{-2}$. The second is located at $(\alpha,\delta) \sim (178\overset{\circ}{.}3,-17\overset{\circ}{.}0)$, which is $\sim 0.4^{\circ}$ beyond the ellipse enclosing Crater II.  It contains 4 RRLS at the same $d_{\odot}$ to within the errors.  It too comfortably fits within a circle of $0.5^{\circ}$ radius, so its density is $> 5.1~{\rm deg}^{-2}$. In our survey, the density of RRLS with in the range $80 < d_{\odot} < 150$ kpc is $\sim 0.06~{\rm deg}^{-2}$ (see below), and unlike the RRLS in the two clumps, these halo stars are not clustered in $d_{\odot}$. It cannot be merely a coincidence that these two high density clumps of RRLS lie close to Crater II and each other both in position and $d_{\odot}$.   The densest clump is probably not a star cluster, because we do not find an excess of red giant candidates at its position (see below).  There are some candidates near the second, but since it lies the relatively close to Crater II, where the stream density is expected to be highest, this could be simply a coincidence.  Regions of higher than average density are seen, for example, in the tidal streams from the globular cluster Pal 5 \citep{odenkirchen03}.       
 
 In summary, there is a highly significant RRLS overdensity is in the direction of Crater II and at its $d_{\odot}$.  The overdensity is linear in shape, and it has small regions where the density of RRLS is exceptionally high. It extends on both sides of Crater II, and is roughly aligned in the direction of Crater's proper motion.  These features are exactly those expected of tidal streams from this galaxy. Furthermore, the discovery of tidal streams from Crater II comes as no surprise.  The orbits that have been calculated by \citep{fritz18,fu19, ji21,borukhovetskays22,battaglia22, pace22} indicate that its perigalacticon is sufficiently small that tidal stripping may have occurred.  Tidal stripping has also been discussed as a possible explanation for Crater II's large size, cold kinematics, and its deviation from the luminosity-metallicity relation for dwarf galaxies \citep{frings17, sanders18, ji21}.

\begin{figure}
	\includegraphics[width=\columnwidth]{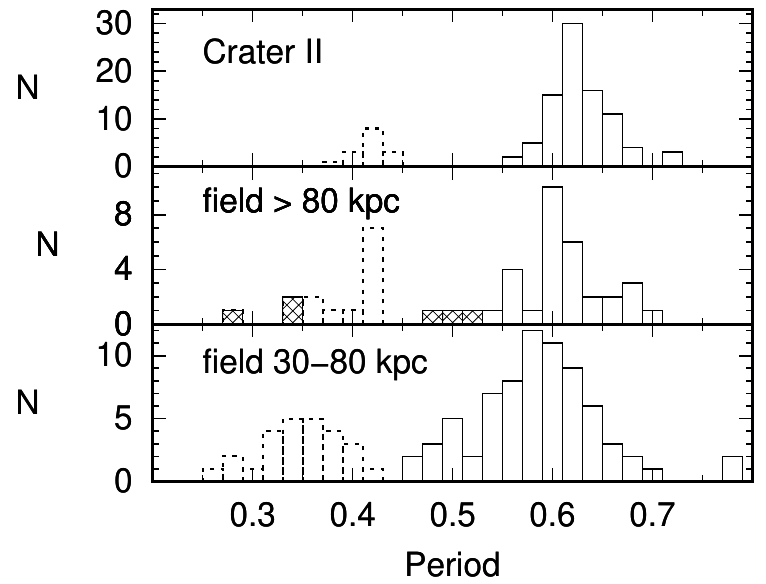}
    \caption{The period distributions of the RRLS in Crater II (top), the field surrounding Crater II (middle), and the field in the foreground of Crater II (bottom).  The histograms depicted by solid and dashed lines are the distributions of the type ab and c variables, respectively.  In the middle plot, the cross-hatched histograms  indicate the distribution of the stars that we suspect are not members of the tidal streams because their periods do not overlap with the Crater II distribution (top).}
    \label{fig:testPhisto}
\end{figure}

\subsection{The Removal of Field Interlopers}

Some of the stars plotted in Fig.~\ref{fig:Sky_Map_10_11} are expected to be unrelated to Crater II and its streams, because this figure includes a large area of the sky ($\sim 300~{\rm deg}^2$) and also a large range in $d_{\odot}$ ($ > 80$ kpc). Over the range of the stars in Fig.~\ref{fig:Sky_Map_10_11} ($79 \leq R_{gc} \leq 147$ kpc), the number density given by the dashed line in Fig.~\ref{fig:density_plot} suggests that the number of RRLS that are not part of the tidal streams is $14 \pm 4$. To obtain another estimate, we examined an equal area of the sky that is far from Crater II and other known overdensities and found 17 RRLS with $d_{\odot} > 80$ kpc ($\sim 0.06$ RRLS ${\rm deg}^{-2}$). We describe in this section how we identified 13 of the RRLS plotted in Fig.~\ref{fig:Sky_Map_10_11} as unlikely members of the tidal streams.

As noted by \citet{monelli18,joo18,vivas20}, the Crater II RRLS variables have a period distribution that is unlike the distribution of seen in the galactic halo. We can use this property to identify likely field
interlopers. The period distribution of our sample of the Crater II RRLS is shown in the top graph of Fig.~\ref{fig:testPhisto}, where one can
see that both the type ab and type c RRLS have relatively narrow distributions.  Their mean values are 0\fd62 and 0\fd42, with $\sigma =$ 0.03 and 0.02, respectively. The middle graph shows the period distributions of the
RRLS outside of Crater II with $d_{\odot} > 80$ kpc, which is the sample plotted in Fig.~\ref{fig:Sky_Map_10_11}. While
the majority of the type ab and c variables in this sample overlap in period with the Crater II
variables, there are a few that do not. The bottom graph in Fig.~\ref{fig:testPhisto} shows the period distribution
of the RRLS in the same area of the sky, but in the outer halo ($30 \le d_{\odot} \le 80$ kpc). This
distribution is much broader than the Crater II one, as expected from the previous discussions
of the RRLS in Crater II. The stars in the middle graph that are outside the period distribution of
the Crater II stars, overlap in period with stars in the bottom graph. The 6 stars 
making up the cross-hatched areas of the histogram in the middle graph deviate from the mean values of the Crater II ab and c variables by $> 3 \sigma$.  Consequently, we believe that they are more likely to be field interlopers than
members of tidal streams of Crater II.  

In Fig.~\ref{fig:Sky_Map_10_11}, there are several stars that do not to follow the the trend of the majority of non-Crater II stars between position on the sky and $d_{\odot}$.  To illustrate this in more detail, we have
used the line fitted to data in Fig.~\ref{fig:Sky_Map_10_11} by the Theil-Sen procedure to define $\xi$ as the angular distance along the line and $\eta$ as the angular distance perpendicular to the line. The tidal streams are not expected to be perfectly linear features on the sky, but the density RRLS is too low to adequately trace any curvature that may be present. Hence, $\xi$ and $\eta$ should viewed as only approximate stream coordinates.
The following equations constitute a simple rotation of ($\alpha, \delta$) into ($\xi, \eta$), with constants that
make the origin of ($\xi,\eta$) the line's closest approach to the centre of Crater II, with positive $\xi$ in the direction of the leading stream as indicated by the direction of the proper motion vector, i.e., positive $\alpha$ and $\delta$.

\begin{equation}
\xi = +0.8890\alpha + 0.4579\delta - 149.197
\end{equation}
\begin{equation}
\eta = -0.4579\alpha + 0.8890\delta + 96.942
\end{equation}

\begin{figure}
	\includegraphics[width=\columnwidth]{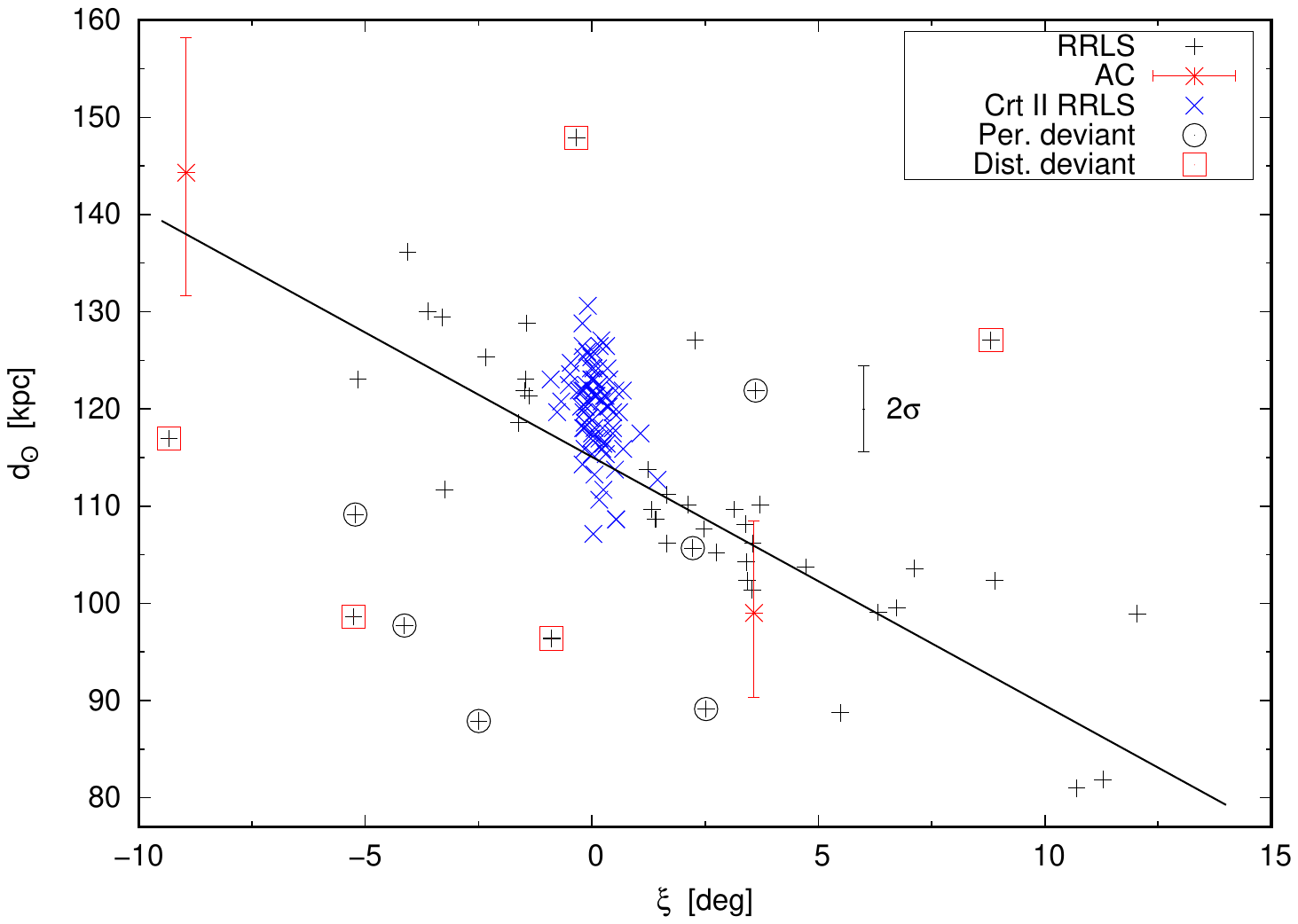}
    \caption{
    Distance from the Sun (${\rm d}_{\odot})$ is plotted against longitude along the streams ($\xi$). The
{\it blue X’s} are our sample of RRLS in Crater II, including V79 and V96. The {\it black crosses} are
the RRLS plotted in Fig.~\ref{fig:Sky_Map_10_11}.  The {\it red
asterisks} depict the ACs. Their distances and errors were estimated from the P-L relations
shown in Fig.~\ref{fig:P_L_10_11}. The points with {\it open circles} are the ``period deviant'' stars. The black errorbar is the
$\pm 1 \sigma$ scatter in the distances of our measurements of the Crater II RRLS. As discussed in the
text, this is probably an upper limit on distance errors of the RRLS in the diagram.  The line is a Theil-Sen fit to stars not in Crater II.  The stars that lie more than 4$\sigma$ in $d_{\odot}$ from this line are considered ``distance deviant'' ({\it red squares}).}
    \label{fig:Dist_long_10_14}
\end{figure}

There is a very significant correlation  between $d_{\odot}$ and $\xi$, which is illustrated in
Fig.~\ref{fig:Dist_long_10_14} where we have plotted the stars outside of Crater II and our sample of Crater II RRLS.  The line is a Theil-Sen fit to the $(\xi,d_{\odot})$ pairs of the 49 stars outside of Crater II.  One can see from this figure that majority of the stars with positive $\xi$, the leading stream, have smaller $d_{\odot}$'s than Crater II, and that it extends to $\sim 12^{\circ}$ from Crater II.  The trailing stream (negative $\xi$) ends at $\sim 5^{\circ}$ from Crater II if only the RRLS are considered, but at $\sim 9^{\circ}$ if the distant AC is included.  The apparent disappearance of the trailing stream of RRLS at $d_{\odot} > 135$ kpc may be nothing
more than inability of our survey detect RRLS fainter than V $\sim 21$ unless they have large amplitudes. At the $d_{\odot}$ of Crater II
(120 kpc), our survey can detect the majority of RRLS, and its completeness improves at shorter
distances. The apparent ending of the leading stream at $\sim 80$ kpc is unlikely to be due to
incompleteness alone. Despite these limitations, our data have detected tidal streams that
span roughly $15^{\circ}$ on the sky and 50 kpc in $d_{\odot}$, and even more if the distant AC is a stream member.

Our sample of Crater II RRLS spans $\sim 25$ kpc in $d_{\odot}$, has a mean of 120 kpc and $\sigma = 4.6$ kpc.  This scatter is  produced
by the combination of our photometric errors ($\sim 0.05$ mag, see above comparison with \citet{joo18} and \citet{monelli18}), the range in RRLS luminosity produced by the metallicity range of Crater II ($-2.8 \le $ [Fe/H] $\le -1.0$, \citet{ji21}), stellar evolution, and the back to front range in distance of
Crater II, which may not be insignificant because some stream stars may lie in the line-of-sight to the main body (see below). \citet{vivas20} found that if they divided their sample of Crater II RRLS in half by distance modulus, the
stars with the brighter modulus had a larger spatial distribution from the center of Crater II
than the fainter modulus sample, which they noted was consistent with the outskirts of Crater
II ($\gtrsim 40 \arcmin$ from the center) being more metal poor on average and more uniform in metallicity than its inner region.  This dependence of luminosity on position in Crater II may be responsible for offsets in distance in Fig.~\ref{fig:Dist_long_10_14}, because the same absolute magnitude ($M_V = 0.42$) was adopted for every RRLS. The tidally stripped RRLS, which presumably originated in the outermost regions of Crater II, may be more luminous and have less dispersion in luminosity, than the central Crater II ones.  This may explain the small scatter in $d_{\odot}$ of the 15 RRLS making up the leading stream in Fig.~\ref{fig:Dist_long_10_14} with $1^{\circ} \lesssim \xi \lesssim 5^{\circ}$, and also the $\sim 5$ kpc ($\sim 0.1$ in distance modulus) offset to larger $d_{\odot}$ of the main body with respect to the line. Nonetheless, to be conservative, in the following we adopt the $\sigma$ of the Crater II RRLS as the error in $d_{\odot}$ for the stream stars.   

In Fig.~\ref{fig:Dist_long_10_14}, we have placed circles around the 6 RRLS that deviate significantly in period from the Crater II RRLS.  Four of these 6 stars lie $> 4 \sigma$ in $d_{\odot}$ from the line in the figure, which is additional evidence that they may not belong to the streams. There are 5 more stars in the figure that lie $> 4 \sigma$ from the line ($> 20$ kpc), and red boxes have been placed around these ``distance deviant'' stars.  We have refrained from making a more stringent cut on $d_{\odot}$ because it is possible that some Crater II stars were tidally stripped during a perigalacticon passage prior to the recent one that presumably produced the more coherent streams in Fig.~\ref{fig:Dist_long_10_14}. It is more likely, in our opinion, that these  distance deviant stars are simply interlopers unrelated to Crater II.   

\begin{figure}
	\includegraphics[width=\columnwidth]{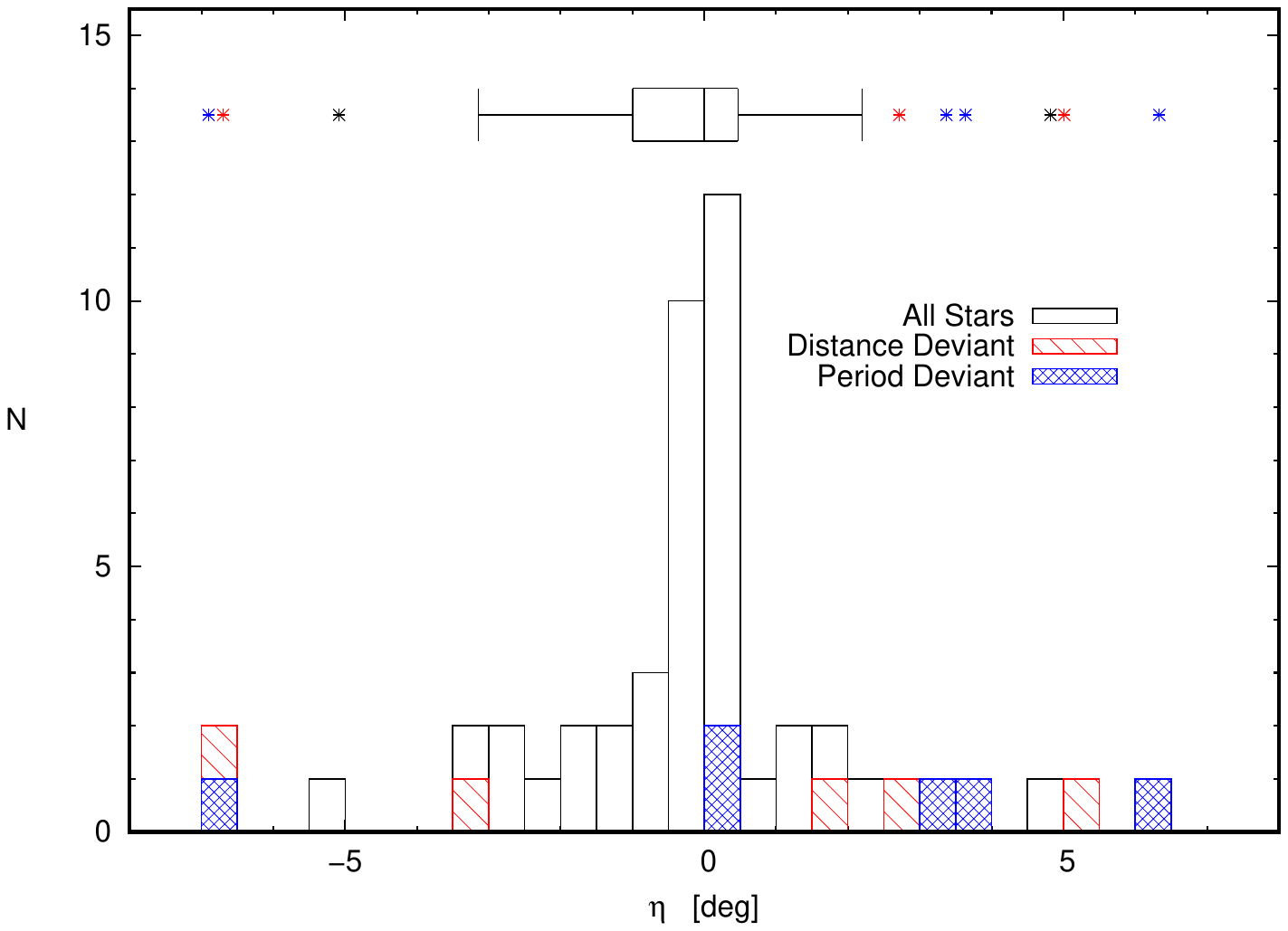}
    \caption{
    The distribution of the non-Crater II stars in $\eta$ is shown as a histogram and as a box and whisker diagram. The blue and red histograms histograms show the locations of the period and distance deviant stars, respectively.  Note for each category, the majority of the members lie in the ``outlier'' region of the box and whisker diagram.  The two additional outliers are considered to be unlikely members of the streams.}
    \label{fig:Eta_dist}
\end{figure}

The distribution in $\eta$, the angular distance on the sky perpendicular to the line in Fig.~\ref{fig:Sky_Map_10_11}, of the non-Crater II stars is shown in Fig.~\ref{fig:Eta_dist}.  The whole distribution is clearly non-Gaussian, for it has a sharp peak and a broad base.  The stars that we have identified as period deviant and distance deviant are widely spread in $\eta$ and show no significant concentration in the peak. The two period deviant stars that lie within the peak are also distance deviant as defined above; consequently, we attach no importance to their small values of $\eta$. In addition to the histogram, Fig.~\ref{fig:Eta_dist} contains the ``box and whisker'' diagram of the whole $\eta$ distribution.  As is the convention, the box spans the inter-quartile range ($IQR = Q_3 - Q_1 = 1.468$), with a vertical line at the median value of $\eta$. The whiskers, which resemble error bars in form, extend to the nearest data point, $<$ for $+\eta$ or $>$ for $-\eta$, 1.5 times the IQR from either $Q_3$ or $Q_1$.  The stars that lie outside of the whiskers are called by convention the  ``outliers'' of the distribution.  Four of the 6 period deviant stars are outliers by this definition, as are 3 of the 5 of distance deviant stars. This is more evidence that they may not be members of Crater II's tidal streams. There are two additional stars,  which are neither period deviant nor distance deviant, that lie beyond the whiskers at about $5^{\circ}$ from the median of the $\eta$ distribution.  We believe they are more likely to field stars than stream members, and they and the period deviant and distance deviant stars are not considered in the following discussion of the streams from Crater II.  They are labelled as ``N'' in the final column of Table~\ref{tab:fieldstars}. The stars that we identify as members of the leading and trailing streams (``L'' and ``T'' in Table~\ref{tab:fieldstars}) may not be pure samples, but without additional observations, such as radial velocities, the interlopers, if any, are not easily recognized.

 \subsection{The orientation of the streams and the mass lost from Crater II}

 The leading stream stars, as defined by the direction of the proper motion vector, have on average smaller values of $d_{\odot}$ than the trailing stream (see Fig.~\ref{fig:Sky_Map_10_11}), which suggests that Crater II has passed apogalacticon in its orbit.  This is consistent with the orbit calculations mentioned above. 

From the variation of $d_{\odot}$ with $\xi$ in Fig.~\ref{fig:Dist_long_10_14}, we estimate that inclination of the streams to the line-of-sight is only $\sim 25^{\circ}$.  Consequently, some unbound stream stars are likely to be projected onto Crater II, and this needs to be considered when deriving the dynamical properties of the main body.  

The size of the Crater II streams suggest that it has lost a significant fraction of its initial mass through tidal stripping.  Assuming that the lower completeness of our survey for the more distant trailing stream is offset by its greater completeness for the nearer leading stream, we can use the ratio of the number of RRLS in the streams after the removal of likely field interlopers (34) to the total of Crater II and stream RRLS found by the survey (111) as an estimate of the stellar mass that has been lost ($\sim 30\%$).  Because the Crater II sample may include some stream stars, this {\em rough} estimate may be an underestimate. 

%\subsection{The streams in galactocentric coordinates}

\begin{table*}
\caption{Crater II RR Lyrae Stars and Anomalous Cepheids}
\label{tab:crater2stars}
\begin{tabular}{lccccccccc}
\hline
Variable ID & $\alpha$ & $\delta$ & Type & $N_{fit}$ & P & Amp & ${\rm HJD}^1_{max}$ & <V> & $A_v$  \\
& (deg) & (deg) & & & (day) & (mag) & (day) & (mag) & (mag)   \\
\hline
  V1 &  177.19560  & -18.62472 & AC   & 129 &   0.76330   &  0.87  &  55301.05078  &  20.60   &  0.09 \\
  V2  & 177.24789  & -18.16947 & ab   & 127 &   0.60428  &   0.80 & 56718.33594 &   20.96  &   0.09 \\
  V3  & 177.16150  & -18.52789 & ab   & 215 &   0.60088  &   0.85  &  57047.31641  &  20.90  &   0.09 \\
  V4  & 177.21321  & -18.27621 & c    & 165  &  0.41693  &   0.38   & 56305.31250  &  20.85  &   0.09 \\
  V5 &  177.21616  & -18.53249 & ab   & 167  &  0.59899 &    0.65  &  56738.11328  &  20.91 &    0.09 \\
  V7  & 177.12840 &  -17.69956 & ab   &  30  &  0.59119  &   0.98  &  55648.07422  &  20.72 &    0.08 \\
  V8 &  177.19585  & -18.52521 & ab   & 180  &  0.64963  &  0.45  &  56752.08203  &  20.88  &   0.09 \\
  V9  & 176.50267 &  -18.82727 & ab    & 80  &  0.59108  &   0.80  &  56772.17969  &  20.98   &  0.11 \\
  V10  & 176.96548 &  -18.18674 & ab  &  166   & 0.62121  &   0.74 &   57425.21094 &   20.81  &   0.10 \\
  V12 & 177.17084  & -18.52932 & ab   & 222  &  0.64034  &   0.56   & 57405.17188  &  20.94  &   0.09 \\
  V13 &  177.21498 &  -18.69282  & ab  &   72  &  0.64863   &  0.48   & 57010.33984  &  20.95  &   0.10 \\
  V14  & 177.16512 &  -18.15675  & ab  &   90  &  0.60915   &  0.64  &  56746.13281 &   20.96  &   0.09 \\
  V15 &  176.98763  & -18.14438 & ab  &  186  &  0.63777  &   0.54  &  56742.03906  &  20.88   &  0.10 \\
  V16  & 177.14012  & -18.07522 & ab   & 195  &  0.60235   &  0.64  &  56655.30859  &  20.88  &   0.09 \\
  V17 & 176.89240  & -18.65648  & ab   &  68  &  0.62319   &  0.59  &  56285.30859  & 20.99   &  0.12 \\
  V18  & 176.88820  & -18.10849 & ab   & 150 &   0.62814  &   0.50 &   57447.08984  &  20.93  &   0.10 \\
  V19 &  177.01732  & -18.07904 & ab  & 220  &  0.63028   &  0.60  &  56724.25000  &  21.01   &  0.09 \\
  V20  & 177.14749  & -18.51523 & ab   & 239 &   0.62322  &   0.51  &  57012.27734  &  20.87  &  0.09 \\
  V21 &  177.27402  & -18.53420 & ab   & 111  &  0.63534  &   0.30 &   57070.17578  &  21.09 &    0.09 \\
  V22  & 177.06213  & -18.25257 & ab   & 261  &  0.59807  &   0.73 &   56691.30469  &  20.91  &   0.09 \\
  V23 &  177.04428  & -18.29524  & ab   & 200 &   0.61473  &   0.63  &  57043.33594  &  21.00   &  0.09 \\
  V24  & 177.17685  & -19.17122 & ab   & 163 &   0.62163   &  0.58  &  56766.05078  &  21.03  &   0.13 \\
  V26 &  177.93806 &  -18.49433 & AC   & 201  &  0.77377  &   0.53  &  56711.37109   & 20.53  &   0.09 \\
  V27 &   177.60977 &   -17.98441 & ab  &  155 &   0.64655  &   0.52 &   56307.21484  &  20.85  &   0.09 \\
  V28  & 177.56619 &  -18.16063 & ab   & 200 &   0.62496   &  0.59 &   56669.32812  &  20.98  &   0.09 \\
  V29 & 177.34789   & -18.33577 & c   &  168 &   0.42028  &   0.39 &   56299.33984 &   20.85 &    0.09 \\
  V31 & 177.93794  & -18.12585  & ab    &139 &   0.62512   &  0.44  &  56754.06641  &  20.83  &   0.09 \\
  V32 &  177.62062  & -18.27895 & ab   & 166 &  0.60989   &  0.57 &   57433.20703   & 20.91  &   0.09 \\
  V34  & 177.68721 &  -18.58765 & c   & 127 &   0.42522   &  0.30  &  57400.32031  &  20.76 &    0.10 \\
  V35 &  177.48459  & -18.62868 & ab   &142 &  0.63094   &  0.53 &   57402.25391  &  20.77  &   0.09 \\
  V36  & 177.75523  & -18.53922 & ab   & 153  &  0.68134  &   0.32  &  57390.26953  &  20.92  &   0.10 \\
  V37  & 178.26863 &  -17.92537 & ab   &213 &   0.59350  &   0.77 &  57092.09766 &   20.86  &   0.09 \\
  V39 &  177.37064 & -18.02627 & ab  & 100 &   0.64212   &  0.48  &  56779.13672 &   20.83  &   0.09 \\
  V40 &  177.44287  & -18.69602  & ab  &  93 &   0.62881  &   0.50   & 57451.19141  &  21.00  &   0.09 \\
  V41  & 177.51491  & -18.74163  & ab  & 108   & 0.61815   &  0.55  &  57445.12109  &  21.02 &   0.10 \\
  V42 &  177.49997  & -18.09225  & ab  &  152 &   0.62580   &  0.45  &  56655.26172  &  21.02 &    0.09 \\
  V43 &  177.67096  & -18.27519  & ab   & 195  &  0.62560   &  0.55  &  57398.21875  &  20.95  &   0.09 \\
  V44  & 177.46115  & -18.75020  & ab  &  151  &  0.64849   &  0.44  &  57043.36328 &   20.99 &    0.10 \\
  V45  & 177.65083  & -17.95820  & ab &  159 &   0.61528   &  0.62  &  56681.26953  &  20.79  &   0.09 \\
  V46  & 177.40338  & -18.59077  & ab  &  145  &  0.61550  &   0.68  &  57016.26562  &  20.82 &    0.09 \\
  V50 &  177.44466  & -18.53697 & ab   & 180  &  0.58945  &   0.61  &  57437.18359  &  20.87 &    0.09 \\
  V51 & 177.55490  & -18.99419  & ab    &181  &  0.60895   &  0.63 &   56754.07422 &   20.92  &   0.12 \\
  V52 &  177.41907 &  -18.31799  & ab  &  162  &  0.62905 &    0.71  &  57396.21484  &  20.98  &   0.09 \\
  V54  & 178.07124  & -18.58136 & ab   &  81 &   0.58283   &  0.85  &  56730.16797  &  20.91  &   0.10 \\
  V55  & 177.95036  & -18.62933 & ab   &  63  &  0.55358   &  1.01  &  57092.10547  &  20.91  &   0.10 \\
  V56 & 177.43690  & -18.58406  & ab  &  159 &  0.56666   &  0.95  &  57068.26172  &  20.66  &   0.09 \\
  V57 &  177.47368 &  -18.27769 & ab   & 171  &  0.63248  &  0.46  &  57421.23438  &  21.03  &   0.09 \\
  V58  & 177.52641  & -18.76455 & ab  & 130  &  0.57392   &  0.85  &  56783.08594  &  20.95  &   0.10 \\
  V59  & 177.31120  & -18.76809 & c   &  175  &  0.44186    & 0.36  &  56305.32031 &   20.84  &   0.10 \\
  V61  & 177.29405 &  -18.01776  & ab   &  73 &   0.61830  &   0.57  &  56716.32812  &  20.85 &    0.09 \\
  V62 & 177.59517 & -18.41398  & ab   & 162 &   0.61389    & 0.65   & 56645.35156  &  20.84   &  0.09 \\
  V63 & 177.62825  & -18.35910  & ab   & 187  &  0.63748   &  0.55  &  57457.05078  &  20.82 &    0.09 \\
  V64 &  177.57773  & -18.41141  & c   &  202  &  0.42077   &  0.37  &  56669.30859  &  20.87 &    0.09 \\
  V66 & 177.44580  & -18.45585  & ab   & 155  &  0.62032   &  0.58  &  56772.15625  &  20.93  &   0.09 \\
  V67 &  177.28468  & -18.80381  & ab  &  120  &  0.57740  &   0.94  &  56691.35547  &  21.02 &    0.10 \\
  V68 &  177.30215 & -18.32689  & ab    &152  &  0.65250   &  0.37   & 56720.07422  &  20.97   &  0.08 \\
  \hline
\end{tabular}
\\(1) Heliocentric Julian Date of Maximum Light - 2,400,000.0 \hfill{}\break
\end{table*}

\begin{table*}
\contcaption{, Crater II RR Lyrae Stars and Anomalous Cepheids}
\label{tab:crater2starscont}
\begin{tabular}{lccccccccc}
\hline
Variable ID & $\alpha$ & $\delta$ & Type & $N_{fit}$ & P & Amp & ${\rm HJD}^1_{max}$ & <V> & $A_v$  \\
& (deg) & (deg) & & & (day) & (mag) & (day) & (mag) & (mag)   \\
\hline
  V70  & 177.30013  & -18.26403  & ab   & 158 &   0.62228   &  0.54  &  56659.32422  &  20.93  &   0.09 \\
  V71 &  177.56625  & -18.43157 & ab  &  247  &  0.60510   &  0.77  &  57429.21875  &  20.92  &   0.09 \\
  V72 &  177.20828  & -18.47437 & ab   & 163  &  0.65645   &  0.32  &  56058.04297  &  20.95  &   0.09 \\
  V74  & 177.51637 &  -18.43635  & ab  &   164 &   0.63426  &   0.45 &   56724.24219 &   21.02   &  0.09 \\
  V75 &  177.52498 &  -18.40929  & ab  &   180 &   0.60762  &   0.65 &   57405.22656 &   20.90 &    0.09 \\
  V76 &  177.26701 &  -18.15219 & ab  &   131  &  0.65156  &   0.29 &   57092.10938 &  20.93  &   0.09 \\
  V77 & 177.32239  & -18.64888  & ab  &    79 &   0.61315   &  0.68  &  56770.00781  &  20.92  &   0.09 \\
  V78  & 177.32043  & -18.56513  & ab  &   105  &  0.60193 &    0.67  &  57459.07812 &   20.89 &    0.09 \\
  V79 &  178.68275 & -17.91151  & ab  &   200 &   0.67272   &  0.58  &  56728.05469 &   20.77  &   0.09 \\
  V80 &  177.96477  & -18.51461  & c   &   192  &  0.46229  &   0.38  &  57043.36328  &  20.69 &    0.09 \\
  V81  & 176.78436  & -18.87495  & ab  &    76&   0.72695  &   0.47  &  56728.10547 &  20.94  &   0.11 \\
  V83 &  177.18942 &  -18.16633  & ab   &  127  &  0.62073  &   0.63  &  56722.33594  &  20.94  &   0.09 \\
  V84 & 176.67931  & -17.81400  & ab   &  190 &  0.61615  &   0.61  &  57376.29297 &   20.94  &   0.09 \\
  V85 &  176.50948  & -18.53670 & ab  &  168  &  0.60711 &    0.51  &  57431.33203  &  20.92  &   0.11 \\
  V86 &  177.73819  & -18.76755  & AC   &  150  &  0.41206   &  0.54  &  57031.32031  &  20.72  &   0.11 \\
  V88 &  177.44080  & -17.64594 & ab  &   162  &  0.63036  &   0.61  &  56782.96875  &  20.87  &   0.09 \\
  V89  & 177.70779  & -17.69896 & ab  &   163 &  0.62690  &   0.37  &  56764.19141  & 20.94  &   0.09 \\
  V94 &  177.32343 &  -18.10083 & ab   &  134  &   0.67079  &   0.22  &  57053.36719 &   20.95  &   0.09 \\
  V96  & 178.22525  & -19.47534 & ab   &  226 &   0.66989  &   0.58  &  56795.10156&    20.86  &   0.11  \\
  V99 & 177.55411  & -18.29874  & ab   &  231  &  0.66976  &   0.25  &  56307.10938 &  20.92  &   0.09 \\
 V101 &  177.86771  & -18.36618 & ab  &   126  &  0.61134  &   0.66  &  57065.22266 &   20.94  &   0.09 \\
 V103 &  177.30565  & -18.86133  & ab  &   139  &  0.65725  &   0.35 &   55653.25781 &   21.07 &    0.10 \\
 V104 &  177.32655 &  -17.78050  & c   &   145  &  0.37747  &   0.62  &  56285.29297 &   20.92 &   0.08 \\
 V105 &  177.41866 &  -17.80933 & c    &  136  &  0.40911 &    0.40 &   56780.95312  &  20.91 &    0.09 \\
 V106 &  176.61044 &  -18.17140 & c   &   210 &   0.40081  &   0.46  &  57101.08203  &  20.96 &   0.10 \\
 V107  & 177.41129 & -18.07878 & AC  &   285  &  1.02985   &  0.32  &  56677.33594  &  19.98  &   0.09 \\
 ${\rm V}108^2$ & 176.75842 &  -18.30692  & c  &   304  &  0.37031  &   0.44  &  56788.97266  &  18.89   &  0.10 \\
 V109 &  176.95432  & -18.28700  & AC  &  202  & 0.91305   &  0.66  &  56740.05469  &  20.09  &   0.09 \\
 V110 &  177.26743  & -18.72982  & AC   &  159  &  0.52217  &   0.69 &   57026.30469 &   20.22  &   0.10 \\
\hline
\end{tabular}
\\(1) Heliocentric Julian Date of Maximum Light - 2,400,000.0 \hfill{}\break
(2) Field star
\hfill{}\break
\end{table*}

\begin{table*}
\caption{Field RR Lyrae Stars and Anomalous Cepheids}
\label{tab:fieldstars}
\begin{tabular}{lccccccccccccc}
\hline
Field ID & $\alpha$ & $\delta$ & Type & $N_{fit}$ & P & Amp & ${\rm HJD}^1_{max}$ & <V> & $A_v$ & $d_{\odot}$ &  $\xi$ & $\eta$ & ${\rm Stream?}^2$\\
& (deg) & (deg) & & & (day) & (mag) & (day) & (mag) & (mag) & (kpc) & (deg) & (deg) \\
\hline
f1  & 170.05624  & -15.81666  & ab   & 260  & 0.54964  &  0.92  & 57375.26953  & 20.54  &  0.15  &  98.6  &  -5.2594   & 5.0122  &   N \\
f2  & 170.17073  & -24.93658  & ab   & 264  & 0.60674  &  0.29  & 57045.26562  & 20.88  &  0.12  & 116.7  &  -9.3337   & -3.1478  &   N  \\
f3  & 170.45433  & -24.66067  & AC   & 288  & 1.23683  &  1.35  & 56279.29688  & 20.05  &  0.11  & 144.3  &  -8.9552  &  -3.0324   &  T   \\
f4  & 172.33728  & -20.16157  & ab   & 258  & 0.48590  &  0.67  & 56752.15625  & 20.73  &  0.12  & 109.0  &  -5.2211  &  0.1051   &  N  \\
f5  & 172.93295  & -21.18871  &  c   & 171  & 0.42439  &  0.33  & 56402.20703  & 21.00  &  0.13  & 122.9  &  -5.1619  & -1.0808   &  T  \\
f6  & 173.13455  & -19.34159  & ab   & 268  & 0.52305  &  0.47  & 57152.96094  & 20.49  &  0.12  &  97.5  &  -4.1369  &    0.4690   &  N  \\
f7  & 173.39830  & -17.91318  & ab   & 179  & 0.59843  &  0.82  & 56780.97656  & 20.74  &  0.08  & 111.5  &  -3.2484    &  1.6181  &   T  \\
f8  & 173.71851  & -19.34602  & ab   & 137  & 0.67343  &  0.48  & 57010.35156  & 21.20  &  0.21  & 130.0  &  -3.6198    &  0.1977  &   T  \\
f9  & 173.85162  & -20.57696  & ab   & 182  & 0.61915  &  0.48  & 57041.21875  & 21.25  & 0.17  & 136.0   & -4.0651   &  -0.9576  &   T  \\
f10 &  174.91232 &  -20.96411 &   c  & 214  & 0.41352  &  0.34  & 56279.33203  & 21.11  & 0.14  & 128.9   & -3.2994   &  -1.7874  &   T   \\
f11 &  174.96010 &  -18.95842 &   c  & 215  & 0.42631  &  0.35  & 57045.37109  & 21.08  &  0.17  & 124.9   & -2.3385   &  -0.0263   &  T \\
f12 & 175.29012  & -18.01681  & ab  & 145   & 0.61663  &  0.69  & 57439.10547  & 20.90  &  0.10  & 119.0   & -1.6140    &  0.6597  &   T \\
f13 &  175.79668 &  -18.50458 &  ab &  144  & 0.59328  &  0.66  & 56655.29297  & 20.97  &  0.13  & 121.2   &  -1.3870  &   -0.0059   &  T \\
f14 &  175.99728 &  -16.60162 &  ab &   94  & 0.55849  &  0.87  & 57135.09766  & 21.38  &  0.10  & 148.3  &   -0.3373  &    1.5940  &  N \\
f15 &  176.20923 &  -19.40963  & ab &   91  & 0.64959  &  0.44  & 56663.23438  & 21.06  &  0.09  & 128.6  &  -1.4347  &   -0.9994  &   T \\
f16 &  176.36882 &  -11.07577  &  c &  256  & 0.34201  &  0.50  & 56729.21094  & 20.24  &  0.07  &  89.1  &   2.5233   &   6.3364  &   N \\
f17 &  176.43452 &  -19.96669  & ab &  199  & 0.60780  &  0.65  & 56740.14062  & 20.95  &  0.09  & 122.0  &  -1.4895   &  -1.5977  &   T \\
f18 &  176.84460 &  -12.53517  & ab &   88  & 0.55302  &  0.81  & 56305.21484  & 21.03  &  0.09  & 127.3  &   2.2780   &   4.8211 &   N \\
f19 & 177.05568  & -21.10849   & c  & 199   & 0.41603  &  0.34  & 56308.29688  & 20.99  &  0.12  & 123.0  &  -1.4601   &  -2.8972   &  T \\
f20 &  177.34325 &  -13.60686   & c &  196  & 0.33962  &  0.47  & 56648.22656  & 20.65  &  0.11  & 105.8  &   2.2305   &  3.6400   &  N \\
f21 &  177.86626 &  -24.94741   & ab &  137 &  0.50474 &   0.82 &  57396.23047 &  20.29 &   0.16  &  87.7 &   -2.4973  &   -6.6812  &   N \\
f22 &  177.98520 &  -17.01995   & c  & 177  & 0.41435  &  0.36  & 56795.15234  & 20.82  &  0.12  & 113.8  &   1.2384   &   0.3118  &  L \\
f23 &  178.15160 &  -16.96868  & ab &  294  & 0.68827  &  0.68  & 57065.18750  & 20.71  &  0.11  & 108.3  &   1.4098   &   0.2812  &   L \\
f24 &  178.38667 &  -17.60863  & ab &  202  & 0.65887  &  0.45  & 57435.05469  & 20.72  &  0.10  & 109.2  &   1.3258  &   -0.3953  &  L \\
f25 &  178.40797 &  -16.92769  & ab &  188  & 0.60689  & 0.61  & 56795.12500   & 20.69  &  0.14  & 106.5  &   1.6565   &   0.2003  &  L \\
f26 &  178.44644 &  -16.98862  & ab &  162  & 0.69528  &  0.67  & 57451.17188  & 20.79  &  0.14  & 110.8  &   1.6628   &   0.1285  &  L \\
f27 &  178.69778 &  -13.21142  &  c &  185  & 0.28654  &  0.19  & 55656.05859  & 20.97  &  0.12  & 122.2  &   3.6158   &   3.3713  &   N \\
f28 &  179.30569 &  -24.22535  & ab &  298  & 0.58705  &  0.88  & 56696.21094  & 20.51  &  0.17  &  96.6  &  -0.8870   &  -6.6984   &  N \\
f29 &  179.32715 & -14.20893   & ab &  123  & 0.62537   & 0.50  & 57066.27344  & 20.78  &  0.15  & 110.3  &   3.7186  &    2.1964  &   L \\
f30 &  179.65877 & -17.57593   & ab &  139  & 0.60946  &  0.87  & 55643.06250  & 20.70  &  0.12  & 107.4  &   2.4716  &  -0.9487   &  L \\
f31 &  179.87257 & -16.51783   & ab &  223  & 0.60163  &  0.67  & 57453.28516  & 20.80  &  0.18  & 109.9  &   3.1462  &   -0.1060   &  L \\
f32 &  179.94196 & -17.51188   & ab &  190  & 0.61760  &  0.70  & 57121.08594  & 20.66  &  0.13  & 105.4  &  2.7527  &  -1.0215   &  L  \\
f33 &  180.01028 & -19.03309   & c  & 132   & 0.37787  &  0.49  & 56711.26953  & 20.74  &  0.11  & 110.3  &  2.1169  &  -2.4051   &  L \\
f34 &  180.06273 &  -16.31652  & ab &  180  & 0.60395  &  0.76  & 56779.01953  & 20.68  &  0.17  & 104.0  &  3.4074  &   -0.0141  &   L \\
f35 & 180.07935  & -16.05497   & ab &  126  & 0.61617  &  0.38  & 56372.08984  & 20.61  &  0.17  & 101.3  &    3.542  &    0.2108  &   L \\
f36 &  180.08133 &  -16.29470  &  c &  212  & 0.39568  &  0.35  & 57447.12109  & 20.64  &  0.17  & 102.5  &   3.4339   &   -0.0032   &  L \\
f37 &  180.08730 &  -16.05149  &  c &  110  & 0.42135  &  0.36  & 56740.11719  & 20.71  &  0.17  & 106.0   &  3.5506   &   0.2103  &   L \\
f38 &  180.10497 &  -16.43388  & ab &  285  & 0.59700  &  0.55  & 56297.36328  & 20.77  &  0.18  & 108.3  &   3.3912   &  -0.1378   &  L \\
f39 &  180.20529 &  -16.21458  & AC &  284  & 0.72808  &  0.99  & 57016.33984  & 20.01  &  0.15  &  99.0  &    3.5808  &   0.0112  &   L \\
f40 &  181.27802 &  -15.80885  & ab &  218  & 0.63232  &  0.45  & 56682.28906  & 20.64  &  0.14  & 103.6  &   4.7203  &  -0.1193  &   L \\
f41 &  182.03732 &  -15.61743  &   c &  214  & 0.41671 &   0.25  & 57451.27734 &  20.32 &   0.16 & 88.9   & 5.4830   & -0.2968   &  L \\
f42 &  182.83061 &  -15.32346  & ab  & 193  & 0.61784  &  0.60  & 56285.28125  & 20.59  &  0.19  &  99.1  & 6.3228   &  -0.3987   &  L \\
f43 &  182.91818 &  -14.59507  & ab  & 202  & 0.65789  &  0.20  & 57067.22266  & 20.60  &  0.19  &  99.8  & 6.7342   &  0.2087  &  L \\
f44 &  183.61346 &  -11.41523  & ab  & 101  & 0.55194  &  0.83  & 55299.10156  & 21.08  &  0.14  & 126.8  & 8.8083   &   2.7173  &   N \\
f45 &  184.74527 &  -13.42233  & ab  & 215  & 0.59700   & 0.99  & 56004.23438  & 20.61  &  0.14  & 102.3  & 8.8954   &   0.4147 &   L \\
f46 &  185.68770 &  -19.12055  & ab  & 237  & 0.67284   & 0.66  & 57101.12403  & 20.64  &  0.14  & 103.6  & 7.1241   &  -5.086 &   N \\
f47 &  186.04571 &  -12.00381  &  c  & 248  & 0.35443   & 0.44  &           56752.08832  & 20.08  &  0.12  &  81.0  & 10.7011  &   1.0803 &   L \\
f48 &  186.56663 &  -11.72141  & ab  & 128  & 0.55864   & 0.58  &           55660.29473  & 20.10  &  0.12  &  81.9  & 11.2846  &   1.0974 &   L \\
f49 &  189.00499 &  -14.82940  &  c  &  168 & 0.35409   & 0.46  &           56307.23572  & 20.55  &  0.15  &  98.9  & 12.0380  &   -2.7867 &  L \\
\hline
\end{tabular}
\\(1) Heliocentric Julian Date of Maximum Light - 2,400,000.0 \hfill{}\break
(2) Stream Membership: L = leading, T = trailing, N=no, membership unlikely \hfill{}
\end{table*}

\section{Red Giant Branch Stars in the Crater II Region}

If our interpretation of the RRLS distribution near Crater II is correct, i.e., that we are seeing tidally stripped material from Crater II, then the RRLS are simply the ``tip of the iceberg'' in terms of the stellar populations found in the streams, and the stream  
stellar populations should resemble those of Crater II itself.  A definitive test of the stream hypothesis will be possible only when deep imaging, comparable to that obtained by \citet{walker19} in Crater II, becomes available for the streams and can reach the main sequence, where stars are much more numerous.  A partial test  can be carried out now, however, by searching for rare stars that stand out above the background of unrelated foreground stars.  Because blue stars are relatively rare in the halo, a logical choice would be to search for blue HB stars, but Crater II, and presumably the streams, contain few of these stars \citep{monelli18}. 
Auxiliary data, such as proper motion, are also limited for these stars because of their faintness (${\rm V} \sim 21$ at the distance of Crater II, which is near or beyond the Gaia detection limit). Moreover, if we plot the overall density of stars in the region as a function of colour, we find a strong accumulation of foreground stars near the colours of the Crater II HB stars, i.e., a search for HB Crater II stream stars must also contend with a strong foreground, and our attempts to extract them from this foreground did not succeed.  Instead, we show here 
that the streams are detectable using luminous red giant branch (RGB) stars.  In the Crater II colour-magnitude diagram (CMD), these stars also define a tight locus, e.g., see \citet{walker19} and Appendix, that can be used to select them, and while luminous RGB stars represent an even smaller fraction of the overall stellar population of Crater II, the portion of colour space they inhabit has a significantly lower intrinsic foreground contamination rate than that of the HB stars in this part of the sky. Luminous RGB stars have one further important advantage over the typical HB stars.They are sufficiently bright that their proper motions are usefully constrained by Gaia. We outline below the major steps in our analysis and the results.  More of the details can be found in the Appendix. Because of the expected weakness of the signal and because the stream density likely decreases with distance down a stream, note that we have restricted our analysis here to a tighter $15^\circ\times15^\circ$ region, roughly centered on Crater II.

The photometry used to isolate the candidate red giants was taken from DELVE DR2. We only considered objects that had the DELVE catalog parameter  $extended\_class\_g=0,$ which strongly rejects galaxies and any other extended objects.  We adopted the interstellar extinction values reported in the DR2 catalogue  and applied the related reddening corrections to the stellar colours.  We then used the DELVE extinction-corrected magnitudes and colours to create $g$ vs. $g-r$ and $i$ vs. $g-i$ CMDs, as in \citet{walker19} but now for the entire Crater II region.  As shown in the Appendix, we next used the locations of the Crater II member stars identified by the spectroscopic survey of \citet{ji21} as a guide to define RGB emperical extraction `boxes' (regions) in the CMDs. We require our RGB candidate stars to fall inside the boxes for {\it both} diagrams. Using two CMDs rather than just one to select RGB candidates reduces significantly the number of foreground interlopers that survive our RGB CMD selection cuts as most foreground stars do not have the correct combination of spectral type and distance to end up in the RGBs of both diagrams (see Appendix). 
The distribution of candidate RGB stars selected in this way, using two CMDs, shows a significant spatial excess in the location of Crater II itself but not along the streams. In hindsight, this is because the streams likely contain only a fraction of the stars in Crater II ($\sim 30\%$ based on the RRLS), and these stars are spread out over a much larger area of sky: the signal we are searching for is very weak. 

To further boost our selectivity for true Crater II RGB stars, we rely on the fact that the density of stars in the halo drops rapidly with distance. In other words, we expect almost all the contaminants to the RGB selection to be foreground stars, i.e., stars with typically larger proper motions than the stream stars since they are closer.  We can thus use Gaia DR3 data \citep{Gaiadr3_sum_2022} to reduce the foreground contamination by imposing a stellar proper motion cut. This will work (at the cost of completeness) even if Gaia does not actually significantly detect the proper motion of the individual RGB stars.
Because Crater II has a non-zero proper motion comparable in size to some of the proper motion (PM) constraints considered below, it is important that the proper motion size constraint be $relative$ to the proper motion of Crater II, i.e., the constraint $PM < X$ that we actually use below is $\sqrt{(pmRA-pmRA_0)^2+(pmDE-pmDE_0)^2} < X$ where $pmRA_0=-0.073$ mas/year and $pmDE_0=-0.123$ mas/year are the components of the Crater II proper motion reported in \citet{ji21}.  Within the errors, note that the proper motion measured by \citet{ji21} agrees with that from other recent determinations \citep{pace22}.  

There is one further important point regarding a proper motion cut. Gaia's sky coverage has improved considerably from DR2 to DR3, but it is still not completely uniform. Using just a proper motion cut and loosening it from ${\rm PM_{cut}}<0.1$ to ${\rm PM_{cut}}<3,$ we have verified that that the stars passing the cut (except near Crater II, of course) do indeed appear to be distributed quite uniformly across the $15^\circ\times15^\circ$ Crater II 
region -- just as stars selected only using an extinction-corrected CMD cut appear to be. (We have also checked this in other $15^\circ\times15^\circ$ fields immediately adjacent to the Crater II one.) In other words, away from Crater II and a possible stream, the stars passing the proper motion and cut CMD cuts appear as a $\sim$ uniform background whose density decreases with
the tightness of the proper motion cut. There is no evidence for systematics associated with proper motion cuts alone or combined with CMD cuts that can match the spatial properties of the signal we report below.  Indeed, assuming that the foreground contamination inside Crater II and the RRLS stream is also uniform and the same as outside of those areas leads to predicted RGB star counts in those areas that match well what we find below.  

\begin{figure}
    \centering
    \includegraphics[width=\columnwidth]{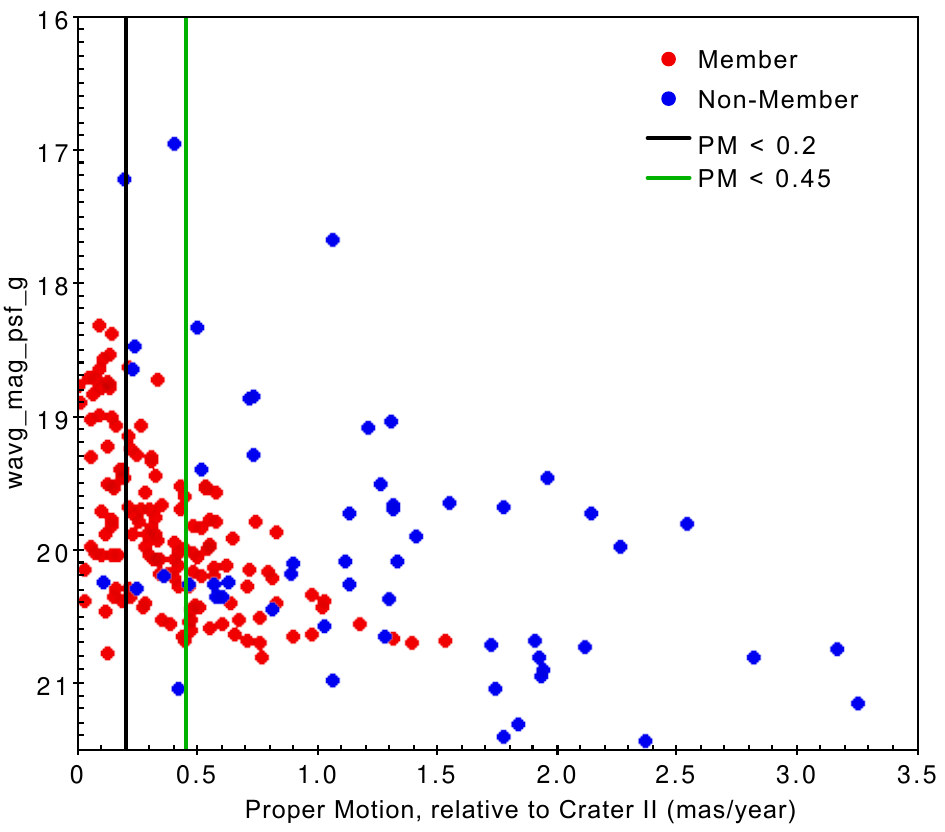}
    \caption{Weighted-average $g$ magnitude from DELVE DR2 vs. total proper motion (PM) from Gaia DR3 for spectroscopically confirmed members ($blue$ dots)  and non-confirmed stellar members ($red$ dots) of Crater II. The list of members and non-members is from \citet{ji21}, who took spectra over the entire $3\sigma$ elliptical sky region containing the RRLS in the Crater II galaxy.  The vertical lines illustrate the effects of two proper motion cuts; the tighter the cut, the less relative contamination from foreground stars although the number of members that pass the cut also drops. Note that proper motions are measured {\it relative} to the proper motion vector reported in \citet{ji21}. }
    \label{fig:jipmcuts}
\end{figure}

The efficiency of the proper motion cuts is illustrated in Fig.\ref{fig:jipmcuts}, where the member and non-member stars identified by \citet{ji21} are plotted against proper motion relative to Crater II's motion.  Fig.\ref{fig:jipmcuts} shows that the member stars are concentrated at the smallest proper motions, while the non-member stars are scattered over a wide range. Note that the proper motion contrast between member and non-member stars decreases as the stars become fainter and Gaia measurement errors increase. In addition to a cut on absolute proper motion magnitude, we thus impose an additional cut on the brightness of the RGB stars we consider on the CMD: $g<19.5, i<18.4$ for stars at the distance of Crater II (see Appendix). 

There is one final complication we must deal with in order to pull stream RGB stars out of the background. The bulk of the stream RRLS are {\it not} actually at the distance of Crater II, and presumably the same is true of stream RGB stars.
To take into account the spread in $d_{\odot}$ of the streams (see Fig.~\ref{fig:Dist_long_10_14}), we thus shifted the selection boxes of the CMDs in steps of 0.1 mag from -0.4 to +0.4 with respect to the CMDs of Crater II and combined the selection results.  The selection box for the faintest shift (+0.4) has a faint limit of $g<19.9$, which one can see from Fig.\ref{fig:jipmcuts} is roughly where many of the member stars start to have a measured $PM > 0.5,$ i.e., Gaia measurement errors become large.  To keep the CMD selection boxes the same size, we raised our faint magnitude limits in lockstep with the shifts in magnitude.      

\begin{figure}
    \centering
    \includegraphics[width=\columnwidth]{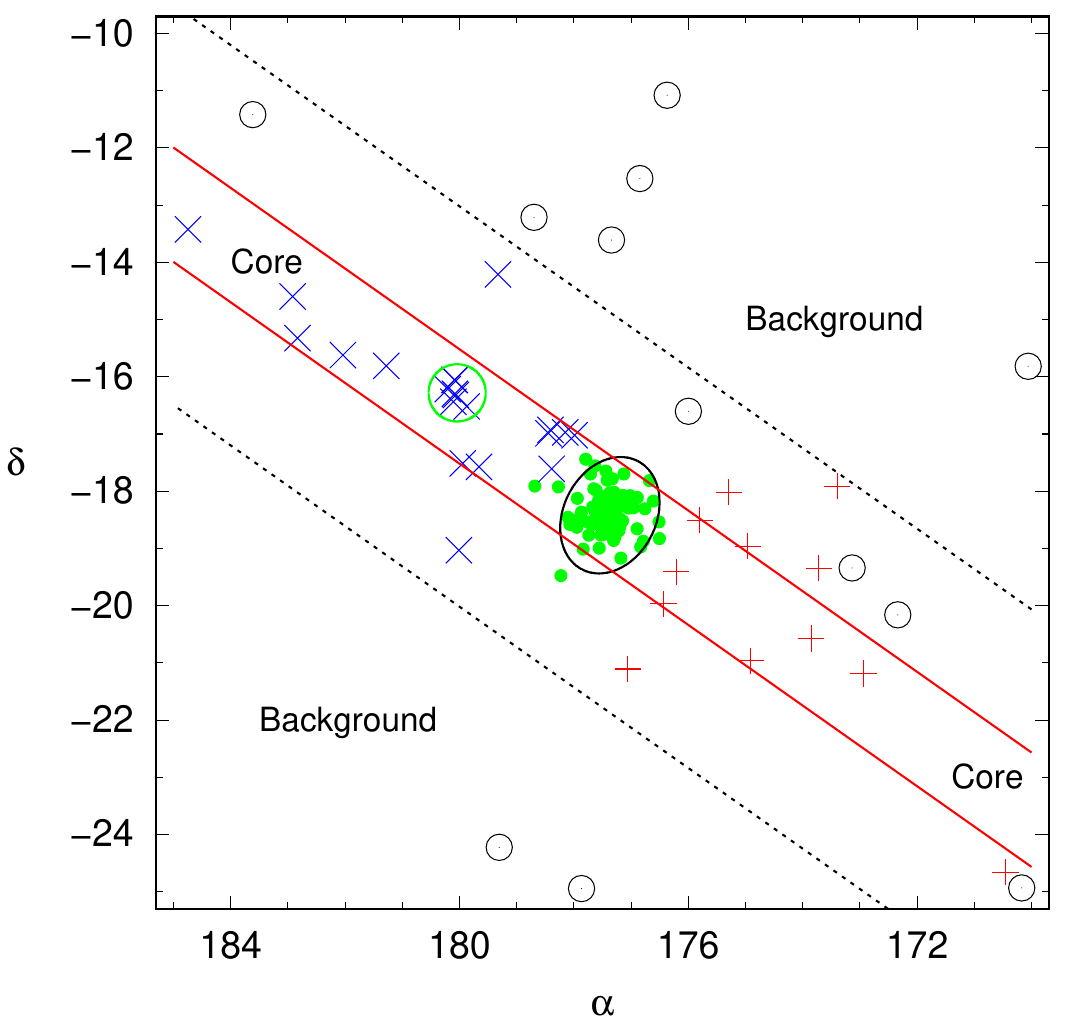}
    \caption{This diagram of the sky shows the positions of the RRLS and ACs and the regions used to estimate the S/N of the stream detections by the variable stars and the RGB candidates.  The ellipse, which is the same as that plotted in Fig.~\ref{fig:Sky_Map_10_11}, is adopted as Crater II proper.  The `core' of the streams is defined as the region between the solid red lines, not including the part covered by the ellipse.  The `background' region, which we assume contains no stream RRLS, is between the dashed lines and the boundaries of the figure.  
    The {\it green points} depict our sample of Crater II RRLS.  The {\it blue X's} and the {\it red crosses} are RRLS in the leading and trailing streams, respectively.  The {\it open circles} are the RRLS rejected as stream members.  The small {\it green circle} encompasses the most significant clump of stars in the streams.}
    \label{fig:rrls_sky_with_regions}
\end{figure}

To aid in the interpretation of our search for stream RGB stars, we have plotted again in Fig. \ref{fig:rrls_sky_with_regions} the positions of RRLS and ACs around Crater II, but now with different regions of the sky delineated. In the following, we will consider the stars falling within the ellipse (plotted also in Fig.~\ref{fig:Sky_Map_10_11}) as potentially Crater II stars, and the ones outside the ellipse as potentially stream stars.  The green circle in the figure encircles the ``clump" of 6 RRLS and one AC.  The red solid lines are parallel to a least-squares fit of a line to the stream stars that was constrained to pass through the center of Crater II, and they are offset from this line by $\pm 1^{\circ}$ in $\delta$.  Within the region between these lines, which we call the `core', are 25 of the 33 RRLS and ACs in this region of the sky that we identified as members of the Crater II streams in table \ref{tab:fieldstars}.  The remaining 8 stream stars, $24\%$ of the total, are scattered between the red solid lines and the dashed lines, which are offset from the red lines by $\pm 2\overset{\circ}{.}5$.  We call this region the `wings' of the streams, which also contains 5 RRLS that we rejected as possible stream members.  There are no stream members between the dashed lines and the boundaries of the $15^\circ \times 15^\circ$ Crater II region shown in Fig. \ref{fig:rrls_sky_with_regions}, and the 7 RRLS in this region thus provide an estimate of the RRLS background.  Since this region encompasses $\sim 123$ sq. deg., our estimate for the  `background' of RRLS with $d_{\odot} > 80$ kpc is $0.06\pm 0.02$ per sq. deg. This number does indeed appear to be the background, for it equals the density we found in a $\sim 300~{\rm deg
^2}$ region away from Crater II (see above).  Assuming RGB stars in the stream track the RRLS + ACs, we will use the same procedure and background region to provide an estimate of the background density for our candidate RGB stars.  Since the excess of RGB candidates in the diffuse wings of the streams does not stand out much above the background, in the following we concentrate on the detection of the core, which does.  While marginally significant, note that there {\it is} an excess of RGB stars in the wings, and it is consistent with that expected from the fraction of RRLS found in the wings.   

\begin{figure*}
    \centering
  \includegraphics[width=\columnwidth]{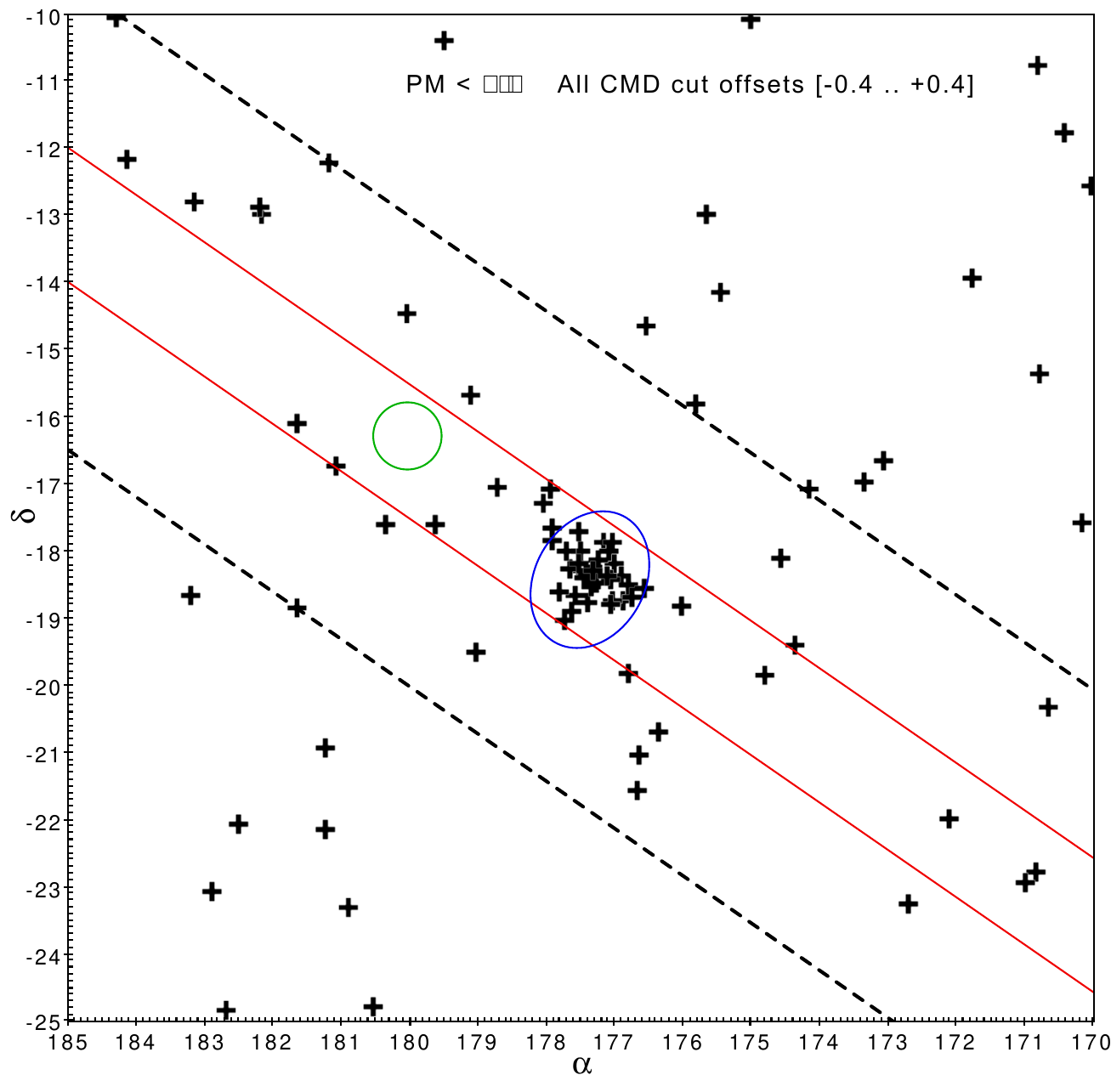}
     \includegraphics[width=\columnwidth]{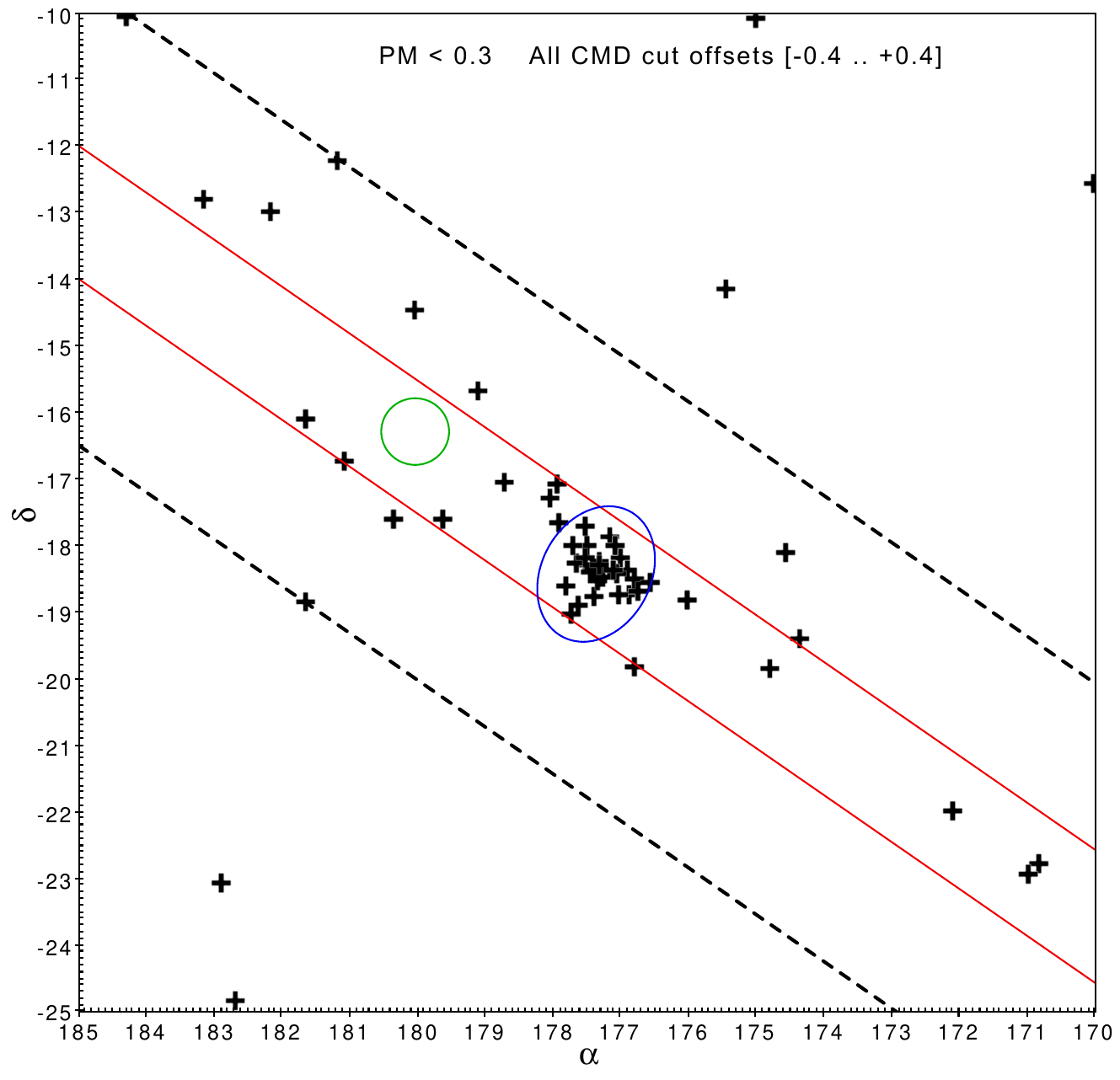}
  \includegraphics[width=\columnwidth]{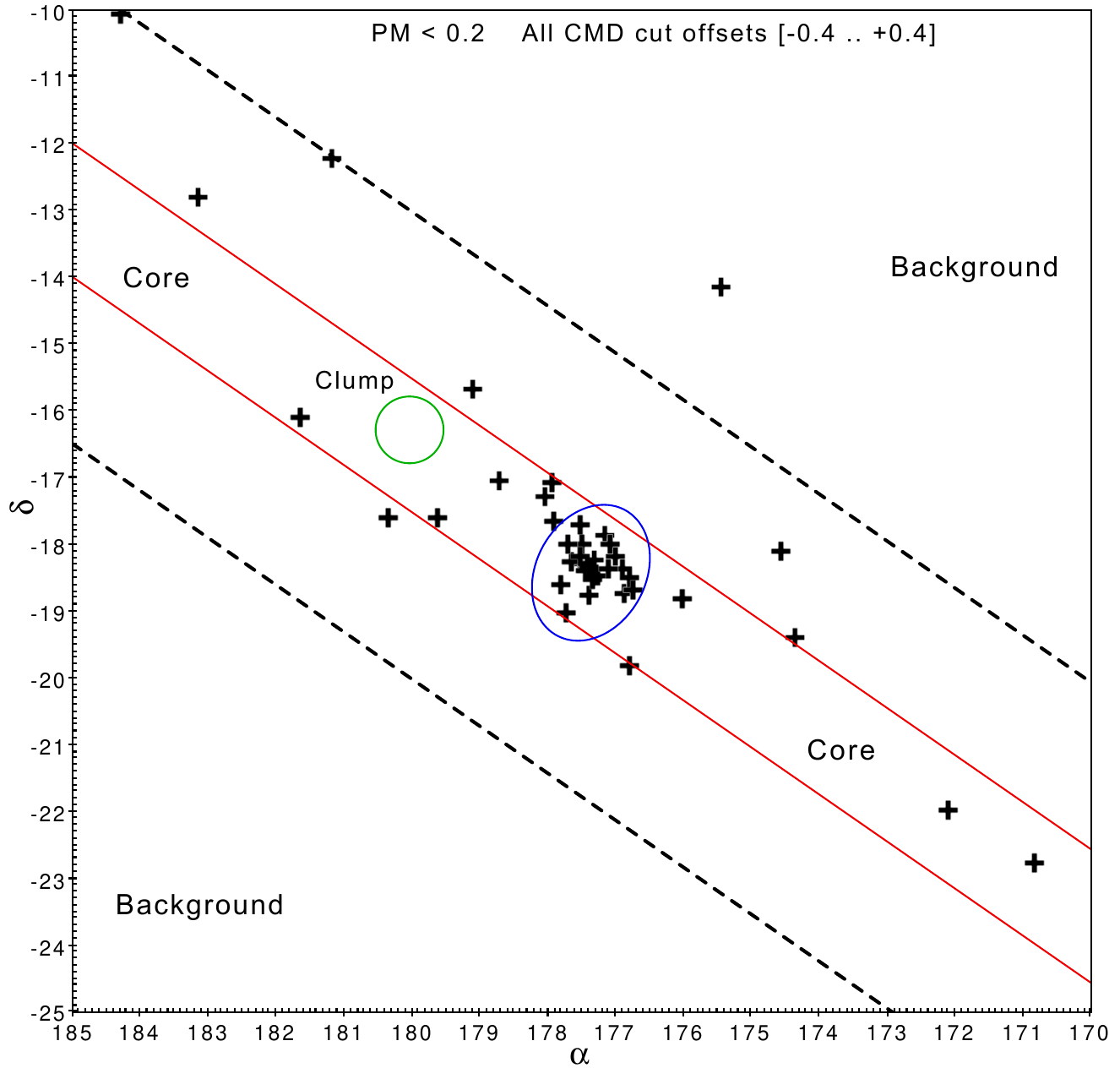}
 \includegraphics[width=\columnwidth]{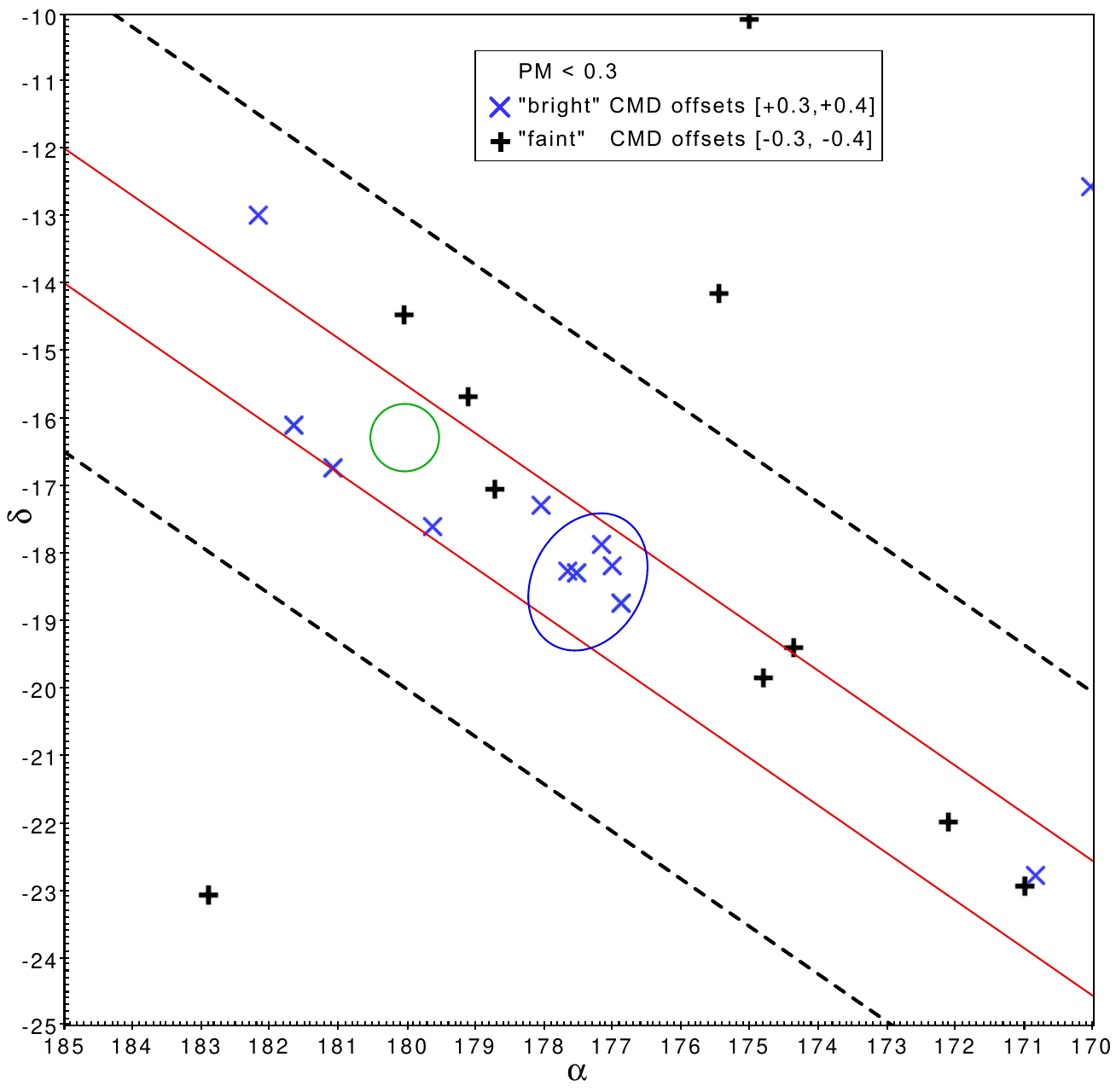}
    \caption{Spatial distribution of RGB stream candidates ({\it black crosses}) selected by varying the CMD cut offsets from -0.4 to +0.4 for a given proper motion cut. The proper motion cut value decreases as we go clockwise from {\it top left} ($PM < 0.5$) to {\it bottom left} ($ PM < 0.2$). The {\it bottom left} panel also labels the spatial regions discussed in the text, which are the same as in Fig. \ref{fig:rrls_sky_with_regions}. The green circle indicates the location of the RRLS clump. The {\it bottom right panel} attempts to check if the RGB candidates show a similar trend in brightness (distance) as the stream RRLS by plotting the positions of RGB candidates selected using only the two brightest (+0.3, +0.4) CMD offsets ({\it blue Xs}) and the two faintest (-0.3, -0.4) offsets ({\it black crosses}). The proper motion cut used is the same as in the panel above it ($PM<0.3$) In the core stream region, the ``bright'' RGB candidates (for which the proper motion cuts work best) do show a 4:1 asymmetry in the same sense as the RRLS (where brighter objects are at $\alpha \gtrsim 178$). The ``faint'' candidates show a 3:1 asymmetry, also in the correct sense as the RRLS (with fainter objects at $\alpha \lesssim 176$ }
    \label{fig:cmdscan}
\end{figure*}

Fig. \ref{fig:cmdscan} and Table \ref{tab:pmcut_counts} summarize the results of our search for candidate RGB stars using a Gaia DR3 proper motion cut and the two (shifted) CMD selection boxes. For a proper motion cut of $PM<0.5,$ which in Fig. \ref{fig:jipmcuts} lets in quite a few non-member stars, we see that the background regions contains many candidates, and except for the concentration of candidates inside the Crater II ellipse, there is no strong evidence for inhomogeneity in the spatial distribution of the candidates. As we tighten up the proper motion cuts, though, the number of candidate stars in the background regions drops as expected (as we remove foreground contamination) but the number of candidates in the stream region, in particular inside the ``core", {\it does not} until we get to $PM \lesssim 0.3.$ For proper motion cuts tighter than this, note that the number of candidates found inside the Crater II ellipse also begins to decrease. As all the candidates in the ellipse are in fact known \citet{ji21} member stars, 
this means that further tightening of the proper motion cuts is no longer just reducing the background but also throwing out signal too, i.e., decreasing our RGB selection completeness. We do show in Fig. \ref{fig:cmdscan}, however, the case for a proper motion cut of $PM<0.2.$ Most of the member stars in the ellipse ($\sim 22/30 \sim 75\%$) survive this cut but now there is only {\it one} star found in the background regions. The expected number of foreground contaminants in the 
core stream region is thus $\sim 0$ yet we still find 9 candidates there, i.e., the distribution of candidate stars outside the ellipse is clearly not spatially homogeneous and happens to align with the direction of the RRLS stream. 

\begin{table}
    \caption{Number of RGB candidates in different spatial regions {\it vs.} proper motion cut. See text for definition of regions and the calculation of the expected counts of foreground interloper stars in the core region along with the 3 $\sigma$-equivalent confidence intervals (enclosed by the parentheses) for those counts.\hfill{} 
 \break}
    \centering
    \begin{tabular}{lccccc}
    \hline
     ${\rm PM_{cut}}$ & stream core & core interlopers & ellipse & background  \\
           ({\rm mas/year})         &   (counts) & (counts) & (counts) & (counts) \\
     \hline
      0.1   & 4 & .2 (0,4) &  8 & 1 \\
      0.2   & 9 & .2 (0,4) &  22 & 1  \\
      0.3   & 12 & 1. (0,6) & 27 & 5 \\
      0.4   & 12 & 3. (0,10) &   31 &  13   \\
      0.5   & 12 & 5. (0,14) &  33 & 23  \\
      1     & 23 & 14 (3,27) &  35  & 60  \\
    \hline
    \end{tabular}
    \label{tab:pmcut_counts}
\end{table}

To put things on a more quantitative footing, we report in Table \ref{tab:pmcut_counts} the number of candidates found in the various spatial regions of Fig. \ref{fig:rrls_sky_with_regions} as well as the predicted number of foreground interlopers for the core stream region 
based on assuming that foreground interlopers are distributed uniformly on the sky, i.e., that the number of interlopers found in a given spatial region is proportional to the area of the region. The ratio of the areas of the regions is $A_{back}:A_{wings}:A_{core}:A_{ell}=1:0.56:0.23:0.024.$ The number of expected foreground interlopers in a given region is then $N_{fore}\approx (A_{region}/A_{back})\times N_{back}$ where $N_{back}$ is the observed number of candidates in the background region for a given $PM_{cut}$. In particular, note that for the core region, the interloper counts $N_{fore} \approx 0.23 N_{back} \ll N_{back},$ while the observed number counts in the core are always greater than $N_{fore}$ for all proper motion cuts shown. 

As a first rough estimate for the significance of the observed excess counts in the core region, let us assume the interloper background is uniform and obeys Poisson statistics, with distribution ${\bf Pois}(k,\lambda)$ where $k$ is the observed number of counts in a region and $\lambda,$ the mean, is set to the expected number of interloper counts, $N_{fore}$. The integers enclosed by the parentheses in Table \ref{tab:pmcut_counts} then give the minimum and maximum count values $k$ such that the cumulative probabilities ${\bf Pois}(k\le k_{min},\lambda)$ and ${\bf Pois}(k\ge k{max},\lambda)$ are less then $1.35\times10^{-3},$ the probability contained in one of the $>3\sigma$ wings of a Gaussian. Note that for large $k,$ the Poisson distribution converges to a Gaussian, and the quantities in the parentheses thus correspond to the standard 3$\sigma$ confidence interval limits. For small $k,$ the Poisson distribution is of course not symmetric and given that $k\ge 0,$ $k_{min}$ may not exist. In that case, we simply report 
$k_{min}=0$ as the lower bound of the confidence interval.  Comparing the observed core counts in Table \ref{tab:pmcut_counts} to the confidence intervals in parenthesis, we see that the observed counts are on the high side of the confidence interval for all proper motion cuts. For $PM_{cut}=0.2, 0.3,$ the counts actually significantly exceed the $3 \sigma$ equivalent upper limits.

The main flaw in this estimate for the significance of the core count excess is the assumption that $\lambda_{core}=0.23 N_{back}.$  Because the observed values of $N_{back}$ are small and thus subject to significant Poisson fluctuations for $PM_{cut} \lesssim 0.4,$ there is significant uncertainty in what the actual mean background density is in these cases. It could be, for example, that for $PM_{cut}=0.2,$ the observed value of $N_{back}=1$ is actually a large ($\sim 3 \sigma$) negative deviation from a true expected mean of $\sim 4,$ which would make the detection of 9 counts in the core significantly more likely (though still rather unlikely overall). The fact that for all proper motion cuts, the observed $N_{back}$ always has to be on the low side of the true expected value actually argues against this being the explanation for an apparent excess of counts in the core. Let us ignore this, however, and instead use Bayesian statistical reasoning to marginalize (integrate) over an unknown mean source density value. 

Following, e.g., \cite{GregoryLoredo1992}, let us compute the Bayesian odds ratio $O_{21}$ comparing the following two models: (1) the mean source density is the same in the core and background regions and simply due to foreground interlopers, and (2) the source densities in the background and core regions are independent and different, e.g., due to the presence of stream stars in the core region. If we have no prior preference for one of these two models, then the odds ratio is just the Bayes factor, which is the ratio of the global likelihoods (evidence) for the models,  $P(D|M_2)/P(D|M_1).$ Here $D$ is the set of counts data to be explained by a model, e.g., $N_{back}=N_1=1$, $N_{core}=N_2=9$ for $PM_{cut}=0.2,$ and we assume that for both models that the probability of producing a certain number of counts $k$ in a region is ${\bf Pois}(k,\lambda)$ where the value of the mean, $\lambda,$ is now allowed to vary. The difference between model (1) and model (2), then, is that $\lambda_{core}$ is not free in model (1) and is instead tied to $\lambda_{back}$ by the ratio of the areas of the spatial regions, ie., $\lambda_{core}=0.23\lambda_{back}.$ In order to carry out a Bayesian model comparison involving models with continuously variable parameters (in this case, $\lambda_{core}$ and  $\lambda_{back}$), we must further specify the prior probability distribution for those parameters, i.e., the ranges over which we believe those parameters can vary. The larger the allowed parameter range, the bigger the ``Occam's razor" penalty factor a model with an added parameter must overcome by explaining the data better. The appropriate range is somewhat arbitrary (to the chagrin of frequentists) but it cannot be infinite for an underlying probability distribution with compact support like the Poisson one. Here, we will assume that the range of allowed $\lambda_i$ for region $i$ is such that the observed counts, $N_i,$ must always be less than a $\sim 5\sigma$ fluctuation away from the expected value, $\lambda_i.$  Specifically, we require $\lambda_{min,i}<\lambda_i<\lambda_{max,i},$ where ${\bf Pois}(N_i,\lambda_{min/max,i}) = 3\times 10^{-7}.$ Assuming no preferred value for $\lambda_i$ over this range, the corresponding prior for $\lambda_i$ is then $P(\lambda_i)=1/\Delta \lambda_i,$ where $\Delta \lambda_i=\lambda_{max,i}-\lambda_{min,i}.$ For example, for the case $PM_{cut}=0.3$ where $N_1=N_{back}=5,$ then $\lambda_{min,1} = 0.134,$ $\lambda_{max,1} = 26.6,$ and $\Delta \lambda_1 = 26.4.$ For the same $PM_{cut},$ if we consider the density in the core region to be independent of that in the background (Model 2), then $N_2=N_{core}=12,$
$\lambda_{min,2} = 1.75,$ $\lambda_{max,2} = 39.0,$ and $\Delta \lambda_2 = 37.3.$ If we denote the background region as region $i=1,$ and the core region as region $i=2,$ then we have, 
\begin{equation} 
    P(D|M_1) = \frac{1}{\Delta \lambda_1}\int_{\lambda_{min,1}}^{\lambda_{max,1}} d\lambda_1 {\bf Pois}(N_2,0.23\lambda_1)\times{\bf Pois}(N_1,\lambda_1).
\end{equation}
Similarly, we have, 

\begin{multline}
     P(D|M_2) = \frac{1}{\Delta \lambda_1 \Delta \lambda_2} \int_{\lambda_{min,1}}^{\lambda_{max,1}}\int_{\lambda_{min,2}}^{\lambda_{max,2}} d\lambda_1 d\lambda_2 {\bf Pois}(N_2,\lambda_2)\\\
    \times{\bf Pois}(N_1,\lambda_1).        
\end{multline}

Evaluating these expressions for the counts data in Table \ref{tab:pmcut_counts}, we then obtain the following odds ratios favoring Model (2) over Model (1): 10.0, 
16400, 8240, 51.2, 2.53, 1.88 for $PM_{cut} = 0.1, 0.2, 0.3, 0.4, 0.5, 1.0$ respectively. As we estimated above, the hypothesis that the source density in the core region is different from that of the background is favored for all proper motion cuts, even with our quite generous priors on the range of allowed density values. The only strong evidence favoring Model 2 over Model 1, however, comes (as above) from the cases with $PM_{cut}=0.2$ and $0.3.$ As a sanity check, we also computed the odds ratio for the case $PM_{cut}=0.3,$ but with the number of counts in the core region ($N_2$) set to 1, the value expected if the mean source density was the same in the core as in the background region, and with a mean such that the expected background counts matched the observed ones, i.e.,  $\lambda_1=N_1=5.$ The odds ratio in this case was only $0.17,$ indicating that allowing the density in the core to be different from that of background region was indeed not justified.  In sum, based on more detailed statistical considerations, we think that there is strong evidence for an excess of Crater II-like RGB stars outside of Crater II. One can repeat this statistical analysis but now shifting, shrinking, and rotating what we call the "core" region. The fact that the signal for an RGB excess is maximized when one considers a strip of sky of approximately the width of Crater II, passing through Crater II,  and aligned with Crater II's proper motion vector -- exactly the same as for the RRLS -- is again further, independent evidence that we are likely seeing streams associated with Crater II.

Assuming the streams exist and do contain RGB stars, it is instructive to try to model the observed counts to see if we can estimate the true numbers of RGB candidate stars, corrected for completeness and foreground contamination. We start with the number of candidate stars inside the Crater II ellipse, where we have direct confirmation of Crater II membership, e.g., \cite{ji21}, and the foreground contamination is likely negligible until $PM_{cut}\gtrsim 0.5.$ A good guess for the intrinsic number appears to 
be $N_{ell}^0 \approx 31.$ An estimate for the completeness of the candidates that survive the tighter proper motion cuts is then the ratio of the observed counts inside the ellipse, $N_{ell},$ to $N_{ell}^0.$  For $PM_{cut}=0.1,0.2, 0.3,$ this gives estimates of our cut completeness of, respectively, $8/31=0.26,$ $22/31=0.71,$ and $27/31=0.87.$ If the asymptotic/intrinsic number of candidates in the core is $N_{core}^0 \approx 12$ (the value for $PM_{cut}=0.5,$ by which the ellipse counts seem to level off), then we predict core counts of $3.1, 8.5,$ and $10.4$ for $PM_{cut}=0.1,0.2,0.3,$ respectively. Given the small number statistics, this in reasonable agreement with the observed counts, $N_{core} = 4, 9,$ and $12.$ Using similar logic, guessing $N_{wings}^0 \approx 5$ leads to a reasonable match to the counts in the stream wings regions. 

With these estimates for the background-free (intrinsic) RGB candidate counts, 
we can then compare the ratios of the numbers in the different regions to see if the RGB 
and RRLS spatial distributions are, at least to zeroth order, compatible.
First, the ratio of wing to core RGB stars is $\approx 5/12 = 0.42,$ while for the RRLS it is $\approx 7/26=0.27$   
Given the small numbers, especially for the RGB stars, these numbers are roughly consistent. The key point here is there is likely an excess of RGB candidates in the stream wings, which is roughly consistent with the number expected from the RRLS distribution. This reinforces the argument, based on the $\sim 7$ RRLS found outside the core stream region, that the stream has a dense core, with width $\sim$ that of the Crater II galaxy, surrounded by a much less dense, broader envelope. The length of the stream (which may actually extend to right ascensions  $<170^\circ$ given our incompleteness for the distant stream) and this stream density profile are relevant for Crater II orbit/tidal stripping calculations. In particular, to explain the stream wings may require multiple orbital passes and precession. Second, we can come up with an RGB-based estimate for the fraction of mass stripped from Crater II by taking the ratio of the number of RGB candidates found outside the Crater II ellipse to the 
total number, including those found inside the ellipse, i.e., 
$f_{disrupt} \sim (N_{core}+N_{wings})/(N_{ell}+N_{core}+N_{wings})  \approx 17/41 = 0.35,$ which is close to the corresponding RRLS-based estimate of $\approx 0.3,$ i.e., both the RGB candidates and RRLS tell the same story that Crater II has undergone significant stripping.  Again, because of projection effects that likely put some stream stars inside the Crater II ellipse, this estimate may in fact be an {\it underestimate}.

Interestingly, one area where the RGB vs. RRLS counts do not appear to agree well is in the region containing the clump of 6 RRLS and 1 AC (green circle in Fig. \ref{fig:rrls_sky_with_regions}). If the clump of RRLS points to a bound structure, e.g., a globular cluster such as Pal 3, or a true overdensity in 3 dimensions (vs. a projected one in 2 dimensions) that has the same ratio of RGB stars to RRLS as Crater II, then in both cases we should also have found a few ($\sim 3-6$) RGB candidates in the RRLS clump region. We do not see any, and the overall CMD near the clump (after applying proper motion cuts to remove backgrounds) seems similar, again subject to small number statistics, to what is found by looking at other similarly sized control regions in the stream. In other words, from the point of view of RGB stars, nothing special seems to be going on in the clump. We may just be seeing chance fluctuations in two very sparse samples of the same underlying stream distribution, although the number of RRLS involved in the clump fluctuation is quite large and not replicated elsewhere in the stream. This bears further investigation.

The only way to conclusively check if the candidate stars we selected are compatible with the RGB of Crater II, and if their radial velocities are compatible with being in the
stream, is to obtain spectroscopy for those stars. (Note that a few of our candidates could actually be on the AGB because of the imprecision of our CMD cuts.) We list their coordinates and $g$ and $i$ magnitudes in Table \ref{tab:core_stream_rgb} and Table \ref{tab:outer_stream_rgb}. The candidates are selected using a proper motion cut $PM < 0.3,$ which should give good completeness at the cost of $\sim 1$ interloper in the stream core and $\sim 2-3$ in the stream wings. The likely presence of RGB stars in the stream strengthens the case that the remainder  of the stripped Crater II stellar population (e.g., its main sequence) should also be present in the stream. This should show up with deep imaging comparable to that already available for Crater II itself.

\begin{table}
    \caption{Coordinates and magnitudes (from DELVE DR2) for candidate RGB stars in the ``core" streams.}
    \centering
    \begin{tabular}{lccccc}
    \hline
     ID & $\alpha$ & $\delta$ & g & r & i \\
     \# & deg & deg & mag & mag & mag \\
     \hline
CS1 & 170.848061	&-22.77174028	&18.35	&17.22	&16.80 \\
CS2 & 171.002862	&-22.92557487	&18.97	&17.91	&17.52\\
CS3 & 172.103052	&-21.97855757	&19.34	&18.43	&18.13\\
CS4 & 174.357550	&-19.39302575	&19.32	&18.23	&17.85\\
CS5 & 174.795559	&-19.85049436	&19.72	&18.84	&18.49\\
CS6 & 176.024999	&-18.81781961	&18.82	&17.84	&17.49\\
CS7 & 176.813035	&-19.83220698	&18.80	&17.84	&17.48\\
CS8 & 177.913881	&-17.66181371	&18.89	&17.95 &17.57\\
CS9 & 179.633722	&-17.59651485	&18.95	&18.07	&17.72\\
CS10 & 178.055730	&-17.28305147	&18.36	&17.31	&16.91\\
CS11 & 177.933934	&-17.0759734	&18.94	&18.06  &17.70\\
CS12 & 178.707060	&-17.04233558	&19.16	&18.10	&17.84\\
CS13 & 181.082231	&-16.72914585	&18.95	&18.05	&17.71\\
CS14 & 181.638775	&-16.10120047	&18.82	&17.90	&17.55\\
 \hline
    \end{tabular}
    \label{tab:core_stream_rgb}
\end{table}

\begin{table}
    \caption{Coordinates and magnitudes (from DELVE DR2) for candidate RGB stars in the ``wings'' of the streams.}
    \centering
    \begin{tabular}{lccccc}
    \hline
     ID & $\alpha$ & $\delta$ & g & r & i \\
     \# & deg & deg & mag & mag & mag \\
     \hline
WS1 174.5832240	& -18.10150950	&19.28	&18.46	&18.16 \\
WS2 179.0963181	& -15.67693691	&19.87	&19.03	&18.72 \\
WS3 181.6353074	& -18.85721807	&19.51	&18.62	&18.33  \\
WS4 180.3541692	& -17.61190255	&18.93	&17.97 &17.61 \\
WS5 180.0464870	& -14.46247758	&18.94	&17.83	&17.46 \\
WS6 182.1510874	& -12.99774679	&18.87	&17.86 &17.46 \\
WS7 183.1370640	& -12.79596808	&19.61	&18.75	&18.46 \\
WS8 181.1864106	& -12.22119653	&19.51	&18.68	&18.35 \\
WS9 184.2760686	& -10.07081370	&18.90	&17.88	&17.55  \\
    \hline
    \end{tabular}
    \label{tab:outer_stream_rgb}
\end{table}

\section{Conclusions}

\begin{figure*}
	\includegraphics[width=7in]{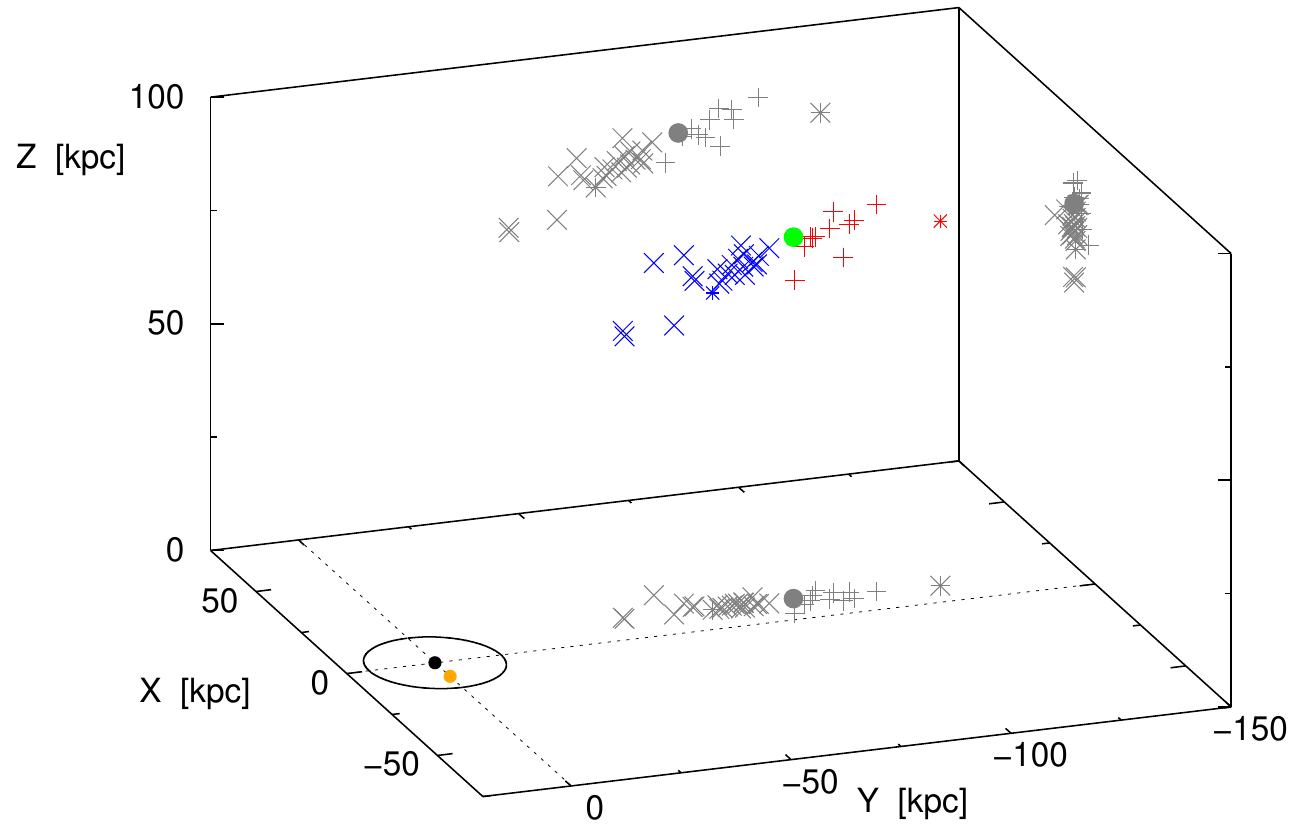}
    \caption{A three-dimensional view of the Crater II streams. Rectangular galactic coordinates X, Y and Z (with origin X=Y=Z=0 at the Galactic Centre) are 
    used to plot the positions of RRLs in the leading stream
({\it blue X’s}), the trailing stream ({\it red crosses}), and at the center of Crater II ({\it large green dot}).
Plotted in the XY plane is a cartoon of the Galaxy, showing the positions of the Galactic Centre
and the Sun ({\it large black} and {\it large orange dots}, respectively), and the galactic disk ({\it black circle}). To help
visualize the geometries of the streams, the projections of the plotted points onto the XY, YZ, and XZ
planes are shown in grey. Especially from the projection in the YZ plane, it is evident that depending on viewing angle, projection effects may cause
apparent overlap of the streams with the Crater II galaxy.  Membership of stars in Crater II thus needs to be evaluated carefully.   }
    \label{fig:Crater3D}
\end{figure*}

Our major result is the detection of tidal streams from the Crater II galaxy, which have been suspected on the basis of Crater II's orbit, low velocity dispersion, and deviation from the luminosity-metallicity relation of dwarf galaxies  \citep{fritz18, fattahi18, sanders18, fu19,ji21, borukhovetskays22, battaglia22}, but until now not observationally confirmed.  To summarize, the streams, which are inclined by $\sim 25^{\circ}$ to the line-of-sight, span at least $\sim 15^{\circ}$ on the sky, and $\sim 50$ kpc in $d_{\odot}$.  Their detection would not have been possible without the major improvements in the photometric calibration of the LSQ RRLS survey.  

Perhaps the best way to visualize and understand our results is to plot the position of the Crater II stream stars in rectangular Galactic coordinates, as in
 Fig.~\ref{fig:Crater3D}. 
In this coordinate system, the origin of the system is located at the Galactic Centre. 
 The X axis is the line between the Sun and the Galactic Centre, with positive X
in the direction opposite to the Sun. The Y axis is positive in the direction of galactic rotation,
and the Z axis is perpendicular to the galactic plane. To give a sense of the scale, we have also
plotted a cartoon representation of the Galaxy showing the locations of the Sun [at coordinates (-8.3, 0, 0) kpc ]  and
the Galactic Centre. The circle in the figure with radius 15 kpc represents the Galactic Disc.
The total lengths from the end of the trailing stream to the opposite end of the leading stream,
not including the very distant AC, are $\sim 12, \sim 57,$ and $\sim 23$ kpc in X, Y, and Z, respectively, and
$\sim 57$ kpc in $R_{gc}$. These variations are perhaps best visualized by examining the projections of
the points on the XY, YZ, and XZ planes, which are depicted in grey in Fig.~\ref{fig:Crater3D}. As the cartoon
illustrates, the large length of the streams in Y dwarfs the familiar inner regions of the Galaxy. The projections also illustrate how when viewed on the sky, the streams may overlap the main body of Crater II and ``contaminate'' its stellar population.  
The positions of the streams indicate that Crater II has passed apogalacticon in its orbit, which
is consistent with the orbit calculations \citep[e.g.,][]{ji21,borukhovetskays22}. The decrease in Z along the streams
suggests that Crater II is starting to plunge towards the Galactic Plane and the inner Galaxy. We note that because of our survey incompleteness at the faint end, the distant stream may in fact extend further than indicated here.

To check on the stream interpretation of the RRLS distribution, we made a first attempt at searching for other stellar population components that should be in the stream by looking for luminous stars on the Red Giant Branch of the Crater II CMD, which are bright enough to have their proper motions constrained by Gaia even at the distance of Crater II. We found $\approx$ 17 candidate stars, whose spatial distribution and number is consistent (given the relatively small number statistics) with what would expect from the RRLS. 

While the RRLS and RGB stars detected here provide a glimpse of the overall dimensions of the streams, these are very rare objects. The streams would be better defined by utilizing more numerous tracers such as subgiants and main-sequence turnoff stars.  Because such stars are very faint at the distances of the Crater II and its streams ($22.5\lesssim i \lesssim 24.5$, see \citet{walker19}), and because the streams occupy a large areas of the sky, this may require the Rubin telescope.  

There is no doubt that Crater II has lost a significant amount of its stellar mass. From the numbers of stream and Crater II RRLS, we estimate the fraction of mass lost is $\gtrsim 30\%$. Our attempt to find RGB stream stars agrees with this estimate and suggests it could be even higher. The streams may help to define better Crater II's orbit, which is an essential part of modelling the stripping process and the removal of both stars and dark matter.  It may be possible then to answer the some of the questions about the form of dark matter and its density profile prior to stripping, which have been raised by previous investigations (see section 1).  The mystery of its kinematically-cold, dark matter halo, first recognized by \citet{caldwell17}, may then be solved. We caution again that because the streams are oriented close to our line of sight, when viewed in projection on the sky, they will overlap the Crater II galaxy itself and may contaminate it with unbound objects. Additionally, various selection cuts of possible Crater II members show a non-relaxed morphology for Crater II, e.g., with two density peaks as in \citet{walker19}. This may be solely due to the inclusion of spurious stream members, but it may also reflect that fact that the galaxy itself may have been significantly disrupted. Any dynamical conclusions, e.g., from velocity dispersions, should thus be treated with corresponding caution. 

The Crater II streams may ultimately provide a good probe of the form of the Milky Way's potential and the influence the LMC has on the halo, because they are far from the galactic center and extend over a large range of $R_{gc}$  The identification of additional red giants in the streams, whose proper motions and radial velocities may be measurable, is therefore desirable. A first list of possible candidates for spectroscopy is provided here.   

\section*{Acknowledgements}

We thank the referee for their careful reading of our paper. Their comments improved it significantly. 

PC, DR, and TG acknowledge essential support from National Science Foundation grant AST-1909101, which enabled the re-reduction and preservation of the La Silla QUEST survey data. The construction of the RRLs catalog used here would not have been possible without the computing support provided by the grant. The construction and execution of the La Silla QUEST survey itself was supported by the Department of Energy and Yale University.

Our work makes use of data from SDSS DR17 and SDSS-IV. Funding for the Sloan Digital Sky Survey IV has been provided by the Alfred P. Sloan Foundation, the U.S. Department of Energy Office of Science, and the Participating Institutions. SDSS acknowledges support and resources from the Center for High-Performance Computing at the University of Utah. The SDSS web site is www.sdss4.org.  SDSS is managed by the Astrophysical Research Consortium for the Participating Institutions of the SDSS Collaboration including the Brazilian Participation Group, the Carnegie Institution for Science, Carnegie Mellon University, Center for Astrophysics.  Harvard \& Smithsonian (CfA), the Chilean Participation Group, the French Participation Group, Instituto de Astrofísica de Canarias, The Johns Hopkins University, Kavli Institute for the Physics and Mathematics of the Universe (IPMU) / University of Tokyo, the Korean Participation Group, Lawrence Berkeley National Laboratory, Leibniz Institut für Astrophysik Potsdam (AIP), Max-Planck-Institut für Astronomie (MPIA Heidelberg), Max-Planck-Institut für Astrophysik (MPA Garching), Max-Planck-Institut für Extraterrestrische Physik (MPE), National Astronomical Observatories of China, New Mexico State University, New York University, University of Notre Dame, Observatório Nacional / MCTI, The Ohio State University, Pennsylvania State University, Shanghai Astronomical Observatory, United Kingdom Participation Group, Universidad Nacional Autónoma de México, University of Arizona, University of Colorado Boulder, University of Oxford, University of Portsmouth, University of Utah, University of Virginia, University of Washington, University of Wisconsin, Vanderbilt University, and Yale University.

Our work also makes use of data from the PanSTARRS PS1 survey, DR 2. The Pan-STARRS1 Surveys (PS1) and the PS1 public science archive have been made possible through contributions by the Institute for Astronomy, the University of Hawaii, the Pan-STARRS Project Office, the Max-Planck Society and its participating institutes, the Max Planck Institute for Astronomy, Heidelberg and the Max Planck Institute for Extraterrestrial Physics, Garching, The Johns Hopkins University, Durham University, the University of Edinburgh, the Queen's University Belfast, the Harvard-Smithsonian Center for Astrophysics, the Las Cumbres Observatory Global Telescope Network Incorporated, the National Central University of Taiwan, the Space Telescope Science Institute, the National Aeronautics and Space Administration under Grant No. NNX08AR22G issued through the Planetary Science Division of the NASA Science Mission Directorate, the National Science Foundation Grant No. AST-1238877, the University of Maryland, Eotvos Lorand University (ELTE), the Los Alamos National Laboratory, and the Gordon and Betty Moore Foundation.

This work also makes use of data from the DELVE survey, DR2. The DECam Local Volume Exploration Survey (DELVE; NOAO Proposal ID 2019A-0305, PI: Drlica-Wagner) is partially supported by Fermilab LDRD project L2019-011 and the NASA Fermi Guest Investigator Program Cycle 9 No. 91201.
This project used data obtained with the Dark Energy Camera (DECam), which was constructed by the Dark Energy Survey (DES) collaboration. Funding for the DES Projects has been provided by the U.S. Department of Energy, the U.S. National Science Foundation, the Ministry of Science and Education of Spain, the Science and Technology Facilities Council of the United Kingdom, the Higher Education Funding Council for England, the National Center for Supercomputing Applications at the University of Illinois at Urbana–Champaign, the Kavli Institute of Cosmological Physics at the University of Chicago, the Center for Cosmology and Astro-Particle Physics at the Ohio State University, the Mitchell Institute for Fundamental Physics and Astronomy at Texas A\&M University, Financiadora de Estudos e Projetos, Fundação Carlos Chagas Filho de Amparo à Pesquisa do Estado do Rio de Janeiro, Conselho Nacional de Desenvolvimento Científico e Tecnológico and the Ministério da Ciência, Tecnologia e Inovação, the Deutsche Forschungsgemeinschaft and the Collaborating Institutions in the Dark Energy Survey.  The Collaborating Institutions are Argonne National Laboratory, the University of California at Santa Cruz, the University of Cambridge, Centro de Investigaciones Enérgeticas, Medioambientales y Tecnológicas–Madrid, the University of Chicago, University College London, the DES-Brazil Consortium, the University of Edinburgh, the Eidgenössische Technische Hochschule (ETH) Zürich, Fermi National Accelerator Laboratory, the University of Illinois at Urbana-Champaign, the Institut de Ciències de l'Espai (IEEC/CSIC), the Institut de Física d'Altes Energies, Lawrence Berkeley National Laboratory, the Ludwig-Maximilians Universität München and the associated Excellence Cluster Universe, the University of Michigan, the National Optical Astronomy Observatory, the University of Nottingham, the Ohio State University, the OzDES Membership Consortium, the University of Pennsylvania, the University of Portsmouth, SLAC National Accelerator Laboratory, Stanford University, the University of Sussex, and Texas A\&M University.

Finally, our work has made use of data from the European Space Agency (ESA) mission
{\it Gaia} (\url{https://www.cosmos.esa.int/gaia}), processed by the {\it Gaia}
Data Processing and Analysis Consortium (DPAC,
\url{https://www.cosmos.esa.int/web/gaia/dpac/consortium}). Funding for the DPAC
has been provided by national institutions, in particular the institutions
participating in the {\it Gaia} Multilateral Agreement.

%%%%%%%%%%%%%%%%%%%%%%%%%%%%%%%%%%%%%%%%%%%%%%%%%%
\section*{Data Availability}

The photometric data from the Sloan Digital Sky Survey is available at https://www.sdss4.org/dr17. The PanSTARRS DR2 data is available at https://outerspace.stsci.edu/display/PANSTARRS. The DELVE DR2 data is available at https://datalab.noirlab.edu/delve. The Gaia DR3 data is available at https://www.cosmos.esa.int/web/gaia/dr3.
The remaining data underlying this article, including the LSQ lightcurves for the Crater II region RRLS, are available in the article and in its online supplementary material.



\bibliographystyle{mnras}
\bibliography{crater2} % if your bibtex file is called example.bib

\begin{thebibliography}{}
\makeatletter
\relax
\def\mn@urlcharsother{\let\do\@makeother \do\$\do\&\do\#\do\^\do\_\do\%\do\~}
\def\mn@doi{\begingroup\mn@urlcharsother \@ifnextchar [ {\mn@doi@}
  {\mn@doi@[]}}
\def\mn@doi@[#1]#2{\def\@tempa{#1}\ifx\@tempa\@empty \href
  {http://dx.doi.org/#2} {doi:#2}\else \href {http://dx.doi.org/#2} {#1}\fi
  \endgroup}
\def\mn@eprint#1#2{\mn@eprint@#1:#2::\@nil}
\def\mn@eprint@arXiv#1{\href {http://arxiv.org/abs/#1} {{\tt arXiv:#1}}}
\def\mn@eprint@dblp#1{\href {http://dblp.uni-trier.de/rec/bibtex/#1.xml}
  {dblp:#1}}
\def\mn@eprint@#1:#2:#3:#4\@nil{\def\@tempa {#1}\def\@tempb {#2}\def\@tempc
  {#3}\ifx \@tempc \@empty \let \@tempc \@tempb \let \@tempb \@tempa \fi \ifx
  \@tempb \@empty \def\@tempb {arXiv}\fi \@ifundefined
  {mn@eprint@\@tempb}{\@tempb:\@tempc}{\expandafter \expandafter \csname
  mn@eprint@\@tempb\endcsname \expandafter{\@tempc}}}

\bibitem[\protect\citeauthoryear{{Abdurro'uf} et~al.,}{{Abdurro'uf}
  et~al.}{2022}]{sdss_dr17_2022}
{Abdurro'uf} et~al., 2022, \mn@doi [\apjs] {10.3847/1538-4365/ac4414}, \href
  {https://ui.adsabs.harvard.edu/abs/2022ApJS..259...35A} {259, 35}

\bibitem[\protect\citeauthoryear{{Akritas}, {Murphy}  \& {Lavalley}}{{Akritas}
  et~al.}{1995}]{akritas95}
{Akritas} M.~G.,  {Murphy} S.~A.,   {Lavalley} M.~P.,  1995, \mn@doi [Journal
  of the American Statistical Association] {10.1080/01621459.1995.10476499},
  90, 170

\bibitem[\protect\citeauthoryear{{Amorisco}}{{Amorisco}}{2019}]{amorisco19}
{Amorisco} N.~C.,  2019, \mn@doi [\mnras] {10.1093/mnrasl/slz121}, \href
  {https://ui.adsabs.harvard.edu/abs/2019MNRAS.489L..22A} {489, L22}

\bibitem[\protect\citeauthoryear{{Applebaum}, {Brooks}, {Christensen},
  {Munshi}, {Quinn}, {Shen}  \& {Tremmel}}{{Applebaum}
  et~al.}{2021}]{applebaum21}
{Applebaum} E.,  {Brooks} A.~M.,  {Christensen} C.~R.,  {Munshi} F.,  {Quinn}
  T.~R.,  {Shen} S.,   {Tremmel} M.,  2021, \mn@doi [\apj]
  {10.3847/1538-4357/abcafa}, \href
  {https://ui.adsabs.harvard.edu/abs/2021ApJ...906...96A} {906, 96}

\bibitem[\protect\citeauthoryear{{Baltay} et~al.,}{{Baltay}
  et~al.}{2007}]{baltay07}
{Baltay} C.,  et~al., 2007, \mn@doi [\pasp] {10.1086/523899}, \href
  {https://ui.adsabs.harvard.edu/abs/2007PASP..119.1278B} {119, 1278}

\bibitem[\protect\citeauthoryear{{Baltay} et~al.,}{{Baltay}
  et~al.}{2013}]{Lsq_description_2013}
{Baltay} C.,  et~al., 2013, \mn@doi [\pasp] {10.1086/671198}, \href
  {https://ui.adsabs.harvard.edu/abs/2013PASP..125..683B} {125, 683}

\bibitem[\protect\citeauthoryear{{Battaglia}, {Taibi}, {Thomas}  \&
  {Fritz}}{{Battaglia} et~al.}{2022}]{battaglia22}
{Battaglia} G.,  {Taibi} S.,  {Thomas} G.~F.,   {Fritz} T.~K.,  2022, \mn@doi
  [\aap] {10.1051/0004-6361/202141528}, \href
  {https://ui.adsabs.harvard.edu/abs/2022A&A...657A..54B} {657, A54}

\bibitem[\protect\citeauthoryear{{Bersier} \& {Wood}}{{Bersier} \&
  {Wood}}{2002}]{fornaxvariables2002}
{Bersier} D.,  {Wood} P.~R.,  2002, \mn@doi [\aj] {10.1086/338315}, \href
  {https://ui.adsabs.harvard.edu/abs/2002AJ....123..840B} {123, 840}

\bibitem[\protect\citeauthoryear{{Borukhovetskaya}, {Navarro}, {Errani}  \&
  {Fattahi}}{{Borukhovetskaya} et~al.}{2022}]{borukhovetskays22}
{Borukhovetskaya} A.,  {Navarro} J.~F.,  {Errani} R.,   {Fattahi} A.,  2022,
  \mn@doi [\mnras] {10.1093/mnras/stac653}, \href
  {https://ui.adsabs.harvard.edu/abs/2022MNRAS.512.5247B} {512, 5247}

\bibitem[\protect\citeauthoryear{{Caldwell} et~al.,}{{Caldwell}
  et~al.}{2017}]{caldwell17}
{Caldwell} N.,  et~al., 2017, \mn@doi [\apj] {10.3847/1538-4357/aa688e}, \href
  {https://ui.adsabs.harvard.edu/abs/2017ApJ...839...20C} {839, 20}

\bibitem[\protect\citeauthoryear{{Cartier} et~al.,}{{Cartier}
  et~al.}{2015}]{cartier2015}
{Cartier} R.,  et~al., 2015, \mn@doi [\apj] {10.1088/0004-637X/810/2/164},
  \href {https://ui.adsabs.harvard.edu/abs/2015ApJ...810..164C} {810, 164}

\bibitem[\protect\citeauthoryear{{Deason}, {Belokurov}  \& {Evans}}{{Deason}
  et~al.}{2011}]{deason11}
{Deason} A.~J.,  {Belokurov} V.,   {Evans} N.~W.,  2011, \mn@doi [\mnras]
  {10.1111/j.1365-2966.2011.19237.x}, \href
  {https://ui.adsabs.harvard.edu/abs/2011MNRAS.416.2903D} {416, 2903}

\bibitem[\protect\citeauthoryear{{Deason}, {Belokurov}, {Koposov}  \&
  {Lancaster}}{{Deason} et~al.}{2018}]{deason18}
{Deason} A.~J.,  {Belokurov} V.,  {Koposov} S.~E.,   {Lancaster} L.,  2018,
  \mn@doi [\apjl] {10.3847/2041-8213/aad0ee}, \href
  {https://ui.adsabs.harvard.edu/abs/2018ApJ...862L...1D} {862, L1}

\bibitem[\protect\citeauthoryear{{Di Criscienzo}, {Caputo}, {Marconi}  \&
  {Cassisi}}{{Di Criscienzo} et~al.}{2007}]{DiCriscienzo2007}
{Di Criscienzo} M.,  {Caputo} F.,  {Marconi} M.,   {Cassisi} S.,  2007, \mn@doi
  [\aap] {10.1051/0004-6361:20066541}, \href
  {https://ui.adsabs.harvard.edu/abs/2007A&A...471..893D} {471, 893}

\bibitem[\protect\citeauthoryear{{Drlica-Wagner} et~al.,}{{Drlica-Wagner}
  et~al.}{2022}]{delvedr2_2022}
{Drlica-Wagner} A.,  et~al., 2022, \mn@doi [\apjs] {10.3847/1538-4365/ac78eb},
  \href {https://ui.adsabs.harvard.edu/abs/2022ApJS..261...38D} {261, 38}

\bibitem[\protect\citeauthoryear{{Errani}, {Navarro}, {Ibata}  \&
  {Pe{\~n}arrubia}}{{Errani} et~al.}{2022}]{errani22}
{Errani} R.,  {Navarro} J.~F.,  {Ibata} R.,   {Pe{\~n}arrubia} J.,  2022,
  \mn@doi [\mnras] {10.1093/mnras/stac476}, \href
  {https://ui.adsabs.harvard.edu/abs/2022MNRAS.511.6001E} {511, 6001}

\bibitem[\protect\citeauthoryear{{Fattahi}, {Navarro}, {Frenk}, {Oman},
  {Sawala}  \& {Schaller}}{{Fattahi} et~al.}{2018}]{fattahi18}
{Fattahi} A.,  {Navarro} J.~F.,  {Frenk} C.~S.,  {Oman} K.~A.,  {Sawala} T.,
  {Schaller} M.,  2018, \mn@doi [\mnras] {10.1093/mnras/sty408}, \href
  {https://ui.adsabs.harvard.edu/abs/2018MNRAS.476.3816F} {476, 3816}

\bibitem[\protect\citeauthoryear{{Flewelling} et~al.,}{{Flewelling}
  et~al.}{2020}]{ps1db2020}
{Flewelling} H.~A.,  et~al., 2020, \mn@doi [\apjs] {10.3847/1538-4365/abb82d},
  \href {https://ui.adsabs.harvard.edu/abs/2020ApJS..251....7F} {251, 7}

\bibitem[\protect\citeauthoryear{{Frings}, {Macci{\`o}}, {Buck}, {Penzo},
  {Dutton}, {Blank}  \& {Obreja}}{{Frings} et~al.}{2017}]{frings17}
{Frings} J.,  {Macci{\`o}} A.,  {Buck} T.,  {Penzo} C.,  {Dutton} A.,  {Blank}
  M.,   {Obreja} A.,  2017, \mn@doi [\mnras] {10.1093/mnras/stx2171}, \href
  {https://ui.adsabs.harvard.edu/abs/2017MNRAS.472.3378F} {472, 3378}

\bibitem[\protect\citeauthoryear{{Fritz}, {Battaglia}, {Pawlowski},
  {Kallivayalil}, {van der Marel}, {Sohn}, {Brook}  \& {Besla}}{{Fritz}
  et~al.}{2018}]{fritz18}
{Fritz} T.~K.,  {Battaglia} G.,  {Pawlowski} M.~S.,  {Kallivayalil} N.,  {van
  der Marel} R.,  {Sohn} S.~T.,  {Brook} C.,   {Besla} G.,  2018, \mn@doi
  [\aap] {10.1051/0004-6361/201833343}, \href
  {https://ui.adsabs.harvard.edu/abs/2018A&A...619A.103F} {619, A103}

\bibitem[\protect\citeauthoryear{{Fu}, {Simon}  \& {Alarc{\'o}n Jara}}{{Fu}
  et~al.}{2019}]{fu19}
{Fu} S.~W.,  {Simon} J.~D.,   {Alarc{\'o}n Jara} A.~G.,  2019, \mn@doi [\apj]
  {10.3847/1538-4357/ab3658}, \href
  {https://ui.adsabs.harvard.edu/abs/2019ApJ...883...11F} {883, 11}

\bibitem[\protect\citeauthoryear{{Gaia Collaboration} et~al.,}{{Gaia
  Collaboration} et~al.}{2022}]{Gaiadr3_sum_2022}
{Gaia Collaboration} et~al., 2022, \mn@doi [arXiv e-prints]
  {10.48550/arXiv.2208.00211}, \href
  {https://ui.adsabs.harvard.edu/abs/2022arXiv220800211G} {p. arXiv:2208.00211}

\bibitem[\protect\citeauthoryear{{Garofalo}, {Delgado}, {Sarro}, {Clementini},
  {Muraveva}, {Marconi}  \& {Ripepi}}{{Garofalo} et~al.}{2022}]{garofalo22}
{Garofalo} A.,  {Delgado} H.~E.,  {Sarro} L.~M.,  {Clementini} G.,  {Muraveva}
  T.,  {Marconi} M.,   {Ripepi} V.,  2022, \mn@doi [\mnras]
  {10.1093/mnras/stac735}, \href
  {https://ui.adsabs.harvard.edu/abs/2022MNRAS.513..788G} {513, 788}

\bibitem[\protect\citeauthoryear{{Gregory} \& {Loredo}}{{Gregory} \&
  {Loredo}}{1992}]{GregoryLoredo1992}
{Gregory} P.~C.,  {Loredo} T.~J.,  1992, \mn@doi [\apj] {10.1086/171844}, \href
  {https://ui.adsabs.harvard.edu/abs/1992ApJ...398..146G} {398, 146}

\bibitem[\protect\citeauthoryear{{Han} et~al.,}{{Han} et~al.}{2022}]{han22}
{Han} J.~J.,  et~al., 2022, \mn@doi [\aj] {10.3847/1538-3881/ac97e9}, \href
  {https://ui.adsabs.harvard.edu/abs/2022AJ....164..249H} {164, 249}

\bibitem[\protect\citeauthoryear{{Honeycutt}}{{Honeycutt}}{1992}]{Honeycutt1992}
{Honeycutt} R.~K.,  1992, \mn@doi [\pasp] {10.1086/133015}, \href
  {https://ui.adsabs.harvard.edu/abs/1992PASP..104..435H} {104, 435}

\bibitem[\protect\citeauthoryear{{Ivezi{\'c}} et~al.,}{{Ivezi{\'c}}
  et~al.}{2007}]{Ivezic2007}
{Ivezi{\'c}} {\v{Z}}.,  et~al., 2007, \mn@doi [\aj] {10.1086/519976}, \href
  {https://ui.adsabs.harvard.edu/abs/2007AJ....134..973I} {134, 973}

\bibitem[\protect\citeauthoryear{James}{James}{1994}]{minuit94}
James F.,  1994, CERN Program Library Long Writeup, D506

\bibitem[\protect\citeauthoryear{{Ji} et~al.,}{{Ji} et~al.}{2021}]{ji21}
{Ji} A.~P.,  et~al., 2021, \mn@doi [\apj] {10.3847/1538-4357/ac1869}, \href
  {https://ui.adsabs.harvard.edu/abs/2021ApJ...921...32J} {921, 32}

\bibitem[\protect\citeauthoryear{{Joo} et~al.,}{{Joo} et~al.}{2018}]{joo18}
{Joo} S.-J.,  et~al., 2018, \mn@doi [\apj] {10.3847/1538-4357/aac4a3}, \href
  {https://ui.adsabs.harvard.edu/abs/2018ApJ...861...23J} {861, 23}

\bibitem[\protect\citeauthoryear{{Jurcsik} \& {Kovacs}}{{Jurcsik} \&
  {Kovacs}}{1996}]{jurcsik96}
{Jurcsik} J.,  {Kovacs} G.,  1996, \aap, \href
  {https://ui.adsabs.harvard.edu/abs/1996A&A...312..111J} {312, 111}

\bibitem[\protect\citeauthoryear{{Kallivayalil} et~al.,}{{Kallivayalil}
  et~al.}{2018}]{kallivayalil18}
{Kallivayalil} N.,  et~al., 2018, \mn@doi [\apj] {10.3847/1538-4357/aadfee},
  \href {https://ui.adsabs.harvard.edu/abs/2018ApJ...867...19K} {867, 19}

\bibitem[\protect\citeauthoryear{{Lafler} \& {Kinman}}{{Lafler} \&
  {Kinman}}{1965}]{lafler_kinman1965}
{Lafler} J.,  {Kinman} T.~D.,  1965, \mn@doi [\apjs] {10.1086/190116}, \href
  {https://ui.adsabs.harvard.edu/abs/1965ApJS...11..216L} {11, 216}

\bibitem[\protect\citeauthoryear{{Layden}}{{Layden}}{1998}]{laydentemplates1998}
{Layden} A.~C.,  1998, \mn@doi [\aj] {10.1086/300195}, \href
  {https://ui.adsabs.harvard.edu/abs/1998AJ....115..193L} {115, 193}

\bibitem[\protect\citeauthoryear{{Lomb}}{{Lomb}}{1976}]{lombscargle1976}
{Lomb} N.~R.,  1976, \mn@doi [\apss] {10.1007/BF00648343}, \href
  {https://ui.adsabs.harvard.edu/abs/1976Ap&SS..39..447L} {39, 447}

\bibitem[\protect\citeauthoryear{{Mart{\'\i}nez-V{\'a}zquez}
  et~al.,}{{Mart{\'\i}nez-V{\'a}zquez} et~al.}{2016}]{martinezvasquez2016}
{Mart{\'\i}nez-V{\'a}zquez} C.~E.,  et~al., 2016, \mn@doi [\mnras]
  {10.1093/mnras/stw1895}, \href
  {https://ui.adsabs.harvard.edu/abs/2016MNRAS.462.4349M} {462, 4349}

\bibitem[\protect\citeauthoryear{{Matsunaga} et~al.,}{{Matsunaga}
  et~al.}{2006}]{Matsunaga2006}
{Matsunaga} N.,  et~al., 2006, \mn@doi [\mnras]
  {10.1111/j.1365-2966.2006.10620.x}, \href
  {https://ui.adsabs.harvard.edu/abs/2006MNRAS.370.1979M} {370, 1979}

\bibitem[\protect\citeauthoryear{{McCarthy} \& {Nemec}}{{McCarthy} \&
  {Nemec}}{1997}]{McCarthy1997}
{McCarthy} J.~K.,  {Nemec} J.~M.,  1997, \mn@doi [\apj] {10.1086/304118}, \href
  {https://ui.adsabs.harvard.edu/abs/1997ApJ...482..203M} {482, 203}

\bibitem[\protect\citeauthoryear{{McConnachie} \& {Venn}}{{McConnachie} \&
  {Venn}}{2020}]{mcconnachie20}
{McConnachie} A.~W.,  {Venn} K.~A.,  2020, \mn@doi [Research Notes of the
  American Astronomical Society] {10.3847/2515-5172/abd18b}, \href
  {https://ui.adsabs.harvard.edu/abs/2020RNAAS...4..229M} {4, 229}

\bibitem[\protect\citeauthoryear{{McGaugh}}{{McGaugh}}{2016}]{mcgaugh16}
{McGaugh} S.~S.,  2016, \mn@doi [\apjl] {10.3847/2041-8205/832/1/L8}, \href
  {https://ui.adsabs.harvard.edu/abs/2016ApJ...832L...8M} {832, L8}

\bibitem[\protect\citeauthoryear{{Monelli} et~al.,}{{Monelli}
  et~al.}{2018}]{monelli18}
{Monelli} M.,  et~al., 2018, \mn@doi [\mnras] {10.1093/mnras/sty1645}, \href
  {https://ui.adsabs.harvard.edu/abs/2018MNRAS.479.4279M} {479, 4279}

\bibitem[\protect\citeauthoryear{{Ngeow}, {Bhardwaj}, {Graham}, {Groom},
  {Masci}  \& {Riddle}}{{Ngeow} et~al.}{2022}]{ngeow22}
{Ngeow} C.-C.,  {Bhardwaj} A.,  {Graham} M.~J.,  {Groom} S.~L.,  {Masci} F.~J.,
    {Riddle} R.,  2022, \mn@doi [\aj] {10.3847/1538-3881/ac8df2}, \href
  {https://ui.adsabs.harvard.edu/abs/2022AJ....164..191N} {164, 191}

\bibitem[\protect\citeauthoryear{{Odenkirchen} et~al.,}{{Odenkirchen}
  et~al.}{2003}]{odenkirchen03}
{Odenkirchen} M.,  et~al., 2003, \mn@doi [\aj] {10.1086/378601}, \href
  {https://ui.adsabs.harvard.edu/abs/2003AJ....126.2385O} {126, 2385}

\bibitem[\protect\citeauthoryear{{Osborn}, {Kopacki}  \& {Haberstroh}}{{Osborn}
  et~al.}{2012}]{Osborn2012}
{Osborn} W.,  {Kopacki} G.,   {Haberstroh} J.,  2012, \actaa, \href
  {https://ui.adsabs.harvard.edu/abs/2012AcA....62..377O} {62, 377}

\bibitem[\protect\citeauthoryear{{Pace}, {Erkal}  \& {Li}}{{Pace}
  et~al.}{2022}]{pace22}
{Pace} A.~B.,  {Erkal} D.,   {Li} T.~S.,  2022, arXiv e-prints, \href
  {https://ui.adsabs.harvard.edu/abs/2022arXiv220505699P} {p. arXiv:2205.05699}

\bibitem[\protect\citeauthoryear{{Padmanabhan} et~al.,}{{Padmanabhan}
  et~al.}{2008}]{ubercal2008}
{Padmanabhan} N.,  et~al., 2008, \mn@doi [\apj] {10.1086/524677}, \href
  {https://ui.adsabs.harvard.edu/abs/2008ApJ...674.1217P} {674, 1217}

\bibitem[\protect\citeauthoryear{{Pozo}, {Broadhurst}, {Emami}  \&
  {Smoot}}{{Pozo} et~al.}{2022}]{pozo22}
{Pozo} A.,  {Broadhurst} T.,  {Emami} R.,   {Smoot} G.,  2022, \mn@doi [\mnras]
  {10.1093/mnras/stac1862}, \href
  {https://ui.adsabs.harvard.edu/abs/2022MNRAS.515.2624P} {515, 2624}

\bibitem[\protect\citeauthoryear{{Press} \& {Rybicki}}{{Press} \&
  {Rybicki}}{1989}]{press_rybicki1989}
{Press} W.~H.,  {Rybicki} G.~B.,  1989, \mn@doi [\apj] {10.1086/167197}, \href
  {https://ui.adsabs.harvard.edu/abs/1989ApJ...338..277P} {338, 277}

\bibitem[\protect\citeauthoryear{{Press}, {Teukolsky}, {Vetterling}  \&
  {Flannery}}{{Press} et~al.}{1992}]{press92}
{Press} W.~H.,  {Teukolsky} S.~A.,  {Vetterling} W.~T.,   {Flannery} B.~P.,
  1992, {Numerical recipes in FORTRAN. The art of scientific computing}.
Cambridge University Press

\bibitem[\protect\citeauthoryear{{Ripepi} et~al.,}{{Ripepi}
  et~al.}{2014}]{ripepi14}
{Ripepi} V.,  et~al., 2014, \mn@doi [\mnras] {10.1093/mnras/stt2047}, \href
  {https://ui.adsabs.harvard.edu/abs/2014MNRAS.437.2307R} {437, 2307}

\bibitem[\protect\citeauthoryear{{Sanders}, {Evans}  \& {Dehnen}}{{Sanders}
  et~al.}{2018}]{sanders18}
{Sanders} J.~L.,  {Evans} N.~W.,   {Dehnen} W.,  2018, \mn@doi [\mnras]
  {10.1093/mnras/sty1278}, \href
  {https://ui.adsabs.harvard.edu/abs/2018MNRAS.478.3879S} {478, 3879}

\bibitem[\protect\citeauthoryear{{Schlafly} \& {Finkbeiner}}{{Schlafly} \&
  {Finkbeiner}}{2011}]{schlafly11}
{Schlafly} E.~F.,  {Finkbeiner} D.~P.,  2011, \mn@doi [\apj]
  {10.1088/0004-637X/737/2/103}, \href
  {https://ui.adsabs.harvard.edu/abs/2011ApJ...737..103S} {737, 103}

\bibitem[\protect\citeauthoryear{{Schlegel}, {Finkbeiner}  \&
  {Davis}}{{Schlegel} et~al.}{1998}]{schlegel98}
{Schlegel} D.~J.,  {Finkbeiner} D.~P.,   {Davis} M.,  1998, \mn@doi [\apj]
  {10.1086/305772}, \href
  {https://ui.adsabs.harvard.edu/abs/1998ApJ...500..525S} {500, 525}

\bibitem[\protect\citeauthoryear{{Schwarzenberg-Czerny}}{{Schwarzenberg-Czerny}}{1996}]{schwarzenbergczernyAov1996}
{Schwarzenberg-Czerny} A.,  1996, \mn@doi [\apjl] {10.1086/309985}, \href
  {https://ui.adsabs.harvard.edu/abs/1996ApJ...460L.107S} {460, L107}

\bibitem[\protect\citeauthoryear{{Soszy{\'n}ski} et~al.,}{{Soszy{\'n}ski}
  et~al.}{2015}]{soszynski15}
{Soszy{\'n}ski} I.,  et~al., 2015, \actaa, \href
  {https://ui.adsabs.harvard.edu/abs/2015AcA....65..233S} {65, 233}

\bibitem[\protect\citeauthoryear{{Stetson}}{{Stetson}}{2000}]{stetson2000}
{Stetson} P.~B.,  2000, \mn@doi [\pasp] {10.1086/316595}, \href
  {https://ui.adsabs.harvard.edu/abs/2000PASP..112..925S} {112, 925}

\bibitem[\protect\citeauthoryear{{Stetson}}{{Stetson}}{2005}]{stetson2005}
{Stetson} P.~B.,  2005, \mn@doi [\pasp] {10.1086/430281}, \href
  {https://ui.adsabs.harvard.edu/abs/2005PASP..117..563S} {117, 563}

\bibitem[\protect\citeauthoryear{{Stringer} et~al.,}{{Stringer}
  et~al.}{2021}]{StringerDESRRL2021}
{Stringer} K.~M.,  et~al., 2021, \mn@doi [\apj] {10.3847/1538-4357/abe873},
  \href {https://ui.adsabs.harvard.edu/abs/2021ApJ...911..109S} {911, 109}

\bibitem[\protect\citeauthoryear{{Tachibana} \& {Miller}}{{Tachibana} \&
  {Miller}}{2018}]{tachibana2018}
{Tachibana} Y.,  {Miller} A.~A.,  2018, \mn@doi [\pasp]
  {10.1088/1538-3873/aae3d9}, \href
  {https://ui.adsabs.harvard.edu/abs/2018PASP..130l8001T} {130, 128001}

\bibitem[\protect\citeauthoryear{{Tonry} et~al.,}{{Tonry}
  et~al.}{2012}]{tonry2012}
{Tonry} J.~L.,  et~al., 2012, \mn@doi [\apj] {10.1088/0004-637X/750/2/99},
  \href {https://ui.adsabs.harvard.edu/abs/2012ApJ...750...99T} {750, 99}

\bibitem[\protect\citeauthoryear{{Torrealba}, {Koposov}, {Belokurov}  \&
  {Irwin}}{{Torrealba} et~al.}{2016}]{torrealba16}
{Torrealba} G.,  {Koposov} S.~E.,  {Belokurov} V.,   {Irwin} M.,  2016, \mn@doi
  [\mnras] {10.1093/mnras/stw733}, \href
  {https://ui.adsabs.harvard.edu/abs/2016MNRAS.459.2370T} {459, 2370}

\bibitem[\protect\citeauthoryear{{Vivas}, {Alonso-Garc{\'\i}a}, {Mateo},
  {Walker}  \& {Howard}}{{Vivas} et~al.}{2019}]{vivas19}
{Vivas} A.~K.,  {Alonso-Garc{\'\i}a} J.,  {Mateo} M.,  {Walker} A.,   {Howard}
  B.,  2019, \mn@doi [\aj] {10.3847/1538-3881/aaf4f3}, \href
  {https://ui.adsabs.harvard.edu/abs/2019AJ....157...35V} {157, 35}

\bibitem[\protect\citeauthoryear{{Vivas} et~al.,}{{Vivas}
  et~al.}{2020}]{vivas20}
{Vivas} A.~K.,  et~al., 2020, \mn@doi [\mnras] {10.1093/mnras/stz3393}, \href
  {https://ui.adsabs.harvard.edu/abs/2020MNRAS.492.1061V} {492, 1061}

\bibitem[\protect\citeauthoryear{{Walker} et~al.,}{{Walker}
  et~al.}{2019}]{walker19}
{Walker} A.~R.,  et~al., 2019, \mn@doi [\mnras] {10.1093/mnras/stz2826}, \href
  {https://ui.adsabs.harvard.edu/abs/2019MNRAS.490.4121W} {490, 4121}

\bibitem[\protect\citeauthoryear{{Zinn} \& {Dahn}}{{Zinn} \&
  {Dahn}}{1976}]{Zinn1976}
{Zinn} R.,  {Dahn} C.~C.,  1976, \mn@doi [\aj] {10.1086/111916}, \href
  {https://ui.adsabs.harvard.edu/abs/1976AJ.....81..527Z} {81, 527}

\bibitem[\protect\citeauthoryear{{Zinn}, {Horowitz}, {Vivas}, {Baltay},
  {Ellman}, {Hadjiyska}, {Rabinowitz}  \& {Miller}}{{Zinn}
  et~al.}{2014}]{zinn14}
{Zinn} R.,  {Horowitz} B.,  {Vivas} A.~K.,  {Baltay} C.,  {Ellman} N.,
  {Hadjiyska} E.,  {Rabinowitz} D.,   {Miller} L.,  2014, \mn@doi [\apj]
  {10.1088/0004-637X/781/1/22}, \href
  {https://ui.adsabs.harvard.edu/abs/2014ApJ...781...22Z} {781, 22}

\makeatother
\end{thebibliography}

%\begin{thebibliography}{99}
%\end{thebibliography}

%%%%%%%%%%%%%%%%%%%%%%%%%%%%%%%%%%%%%%%%%%%%%%%%%%

%%%%%%%%%%%%%%%%% APPENDICES %%%%%%%%%%%%%%%%%%%%%

\appendix

\section{Photometry Pipeline Description}

As in paper I, the photometry pipeline for this paper is designed to monitor the variability of point sources in non-crowded, high galactic latitude ($|b| > 15$) fields. The QUEST collaboration has a separate difference imaging pipeline, see \cite{Lsq_description_2013} to look for variables in extended objects,e.g., supernovae. Even though we are primarily interested in point sources, the best way to photometer them is not immediately obvious. LSQ is an example of a modern, large area survey based on a wide field-of-view camera where point spread function (PSF) variations across the field of view can often be significant. Also, especially at the faint end, misphotometering of quasi-point sources, like an AGN nucleus surrounded by lower surface brightness extended brightness emission from the host galaxy, can create significant spurious variability as the seeing varies if one is not careful. After some experimentation and consideration of the computational expense vs. accuracy vs. robustness for various schemes, we decided to not attempt to do PSF fitting. Rather, as in Paper I, we rely on simple aperture photometry applied to the target and comparison stars chosen to be near the target, so as to minimize the uncorrected effects of spatial variations, including those in 
the PSF. While careful PSF-based photometry can be definitely be more accurate (albeit with increased computational cost), errors in modeling and properly centering the PSF can often cause more problems than a PSF-based approach solves, and when applied to moderately extended sources such as faint galaxies misidentified as stars, the results can actually be worse than using aperture photometry, e.g., if the aperture happens to cover most of the source.  Having decided to do aperture photometry, however, one must still decide what aperture size to use.  To maximize the signal to noise, we decided to use a 4.8\arcsec (6 pixel) diameter aperture for target sources brighter than V$\sim 18.3$ and 2.4\arcsec (3 pixels) for sources fainter than this. The target brightness we use to decide whether we apply the large or small aperture is derived, as we discuss below, by taking the median of the  first pass photometry (chip-wide) lightcurve. Once the appropriate aperture size is chosen for the target, it remains fixed during the remainder of the photometric processing, and the that same aperture size is always used to measure the comparison star fluxes. As can be seen in Fig. \ref{fig:photcomp_stetv}, there is no major glitch in the photometric accuracy near the source magnitude (V $\sim 18.3$) where we switch aperture sizes. 

Our revamped photometry pipeline is designed to easily work with a variety of comparison star and input target catalogs.  Here we will specialize to the case relevant to this paper, where we used PanSTARRS DR2 as both our comparison and target catalog. As a first step, we apply quality cuts to the PanSTARRS DR2 mean photometry catalog and remove blended objects, objects that are extended (galaxies), or objects where the PanSTARRS pipeline sets a photometry warning flag. This creates the ``clean star" catalog from which we draw our initial comparison stars. Note that the set of comparison  stars selected this way turn may still turn out to contain objects that should not be used for a given target object, and as described below, we implement on-the-fly checks out to weed them out.  As a general comment on the importance of having a good comparison star catalog, we compared the performance of using SDSS Stripe 82 stars cleaned in a similar way to that obtained using stars from the Stripe 82 standard star catalog of \cite{Ivezic2007}, which includes further checks such as removing low-amplitude variable objects. The performance using the Stripe 82 standard stars was uniformly better, at the ~$0.5-1\%$ level, presumably because the SDSS catalog did a better job of excluding variable stars or high proper motion objects than our on-the-fly checks. Note that the photometric accuracy of the comparison star catalog also matters, especially at the faint end, as we demonstrate below by comparing the performance obtained using a catalog based on PanSTARRS DR2 to that obtained using DELVE DR2, which goes much deeper.

After determining which of our initial ``clean'' comparison stars fall on chips used to observe a given target object, we load them in and match them (using a 1" search radius) to the Sextractor-detected sources on the chip. Once we have identified the comparison stars on the chip for a given exposure, we make a first guess for their true (pre-atmosphere and telescope) instrumental magnitudes using the values of the $g$ and $r$ magnitudes provided by the external catalog. This is possible to a suprisingly good degree of accuracy because LSQ observations are made in a single filter, ``Q," centered on V (5450 \AA ), that is essentially the sum of the SDSS $g$ and $r$ filters,  
To within $\sim 1-2\%$ for $0<g_{s} - r_{s}<2$ where the subscript $s$ indicates that the $g$ and $r$ filters are in the SDSS system, we have
\begin{equation}
 {\rm Q} \approx g_{s} -2.5\log[1.0+ 10^{0.4(g_{s}-r_{s})}].
 \label{eqn:gandrtoq}
 \end{equation}
 Similar, though more complicated expressions could be derived to transform from the $g$ and $r$ values in the PanSTARRS and DELVE catalogs to instrumental Q magnitude. For this paper, however, we simply transform all $g$ and $r$ values in the external catalogs into the SDSS system using the transformation equations provided by the respective collaborations and then use the SDSS system values.
 Once we have estimates for the true comparison star magnitudes, we obtain a first-pass correction to the observed Q (instrumental) magnitudes
 in a given exposure by using the median magnitude offset between the predicted and observed values of the comparison stars. To maximize the numbers of comparison stars and thus maximize the chances for obtaining a reasonable photometric solution, we initially consider all the comparison stars that fall on the same chip as the target; hence, we call it a ``chip-wide" correction. Because  of the non-linearities present in the chips, the offset turns out to have a significant magnitude dependence. If we consider all comparisons stars on the chip, there are usually enough of them that we can follow \citep{cartier2015} (but now for a single exposure) to obtain a reasonably well-constrained linear fit to that magnitude dependence. Using this fit, we then apply a magnitude dependent correction to each comparison star's observed magnitude in that exposure. We then use these magnitude offsets and corrected comparison star magnitudes to make an educated first guess for those values in the next phase of the pipeline. If the next phase fails, e.g., due to there not being enough good comparison stars located near the target, then the pipeline returns a backup target lightcurve derived by applying these ``chip-wide'' magnitude offsets to the observed instrumental target magnitudes.  Because this first-pass photometric correction for a given exposure does not depend on what happens in other exposures (this is not true for the next step in the photometry pipeline), we also the use our first-pass photometry to apply some initial cuts on the overall photometric quality of an exposure.  While we strive to keep as many exposures as possible to maximize the number points in a lightcurve, we found that there are some exposures that are simply so corrupted by external factors, e.g., patchy cirrus clouds, that they
 seriously skew the subsequent photometric analysis. These ``bad'' exposures are thus removed by the pipeline at this stage, e.g., by dropping exposures where the number of successful matches to comparison stars is too low,   or the derived zero point correction or magnitude limit for a given exposure is discrepant from the median values of these quantities across all exposures. Note that the first pass photometry uses only the large aperture flux measurements, with the goal being to anchor the relative offsets of the exposures as accurately possible at the bright end, where the photon statistics are best.

 Unfortunately, because of intrachip variations (especially in the charge transfer inefficiency and associated non-linearity corrections), the accuracy of the first pass,``chip-wide'' photometry often turns out to be no better than $\sim 3-10\%$ for fainter sources where the nonlinearity corrections are important.  
 As the next step to improve our photometric accuracy, we thus restrict ourselves to using comparison stars located close to the target objects, both spatially  and in magnitude, to minimize the effects of nonlinearities and bad chip cosmetics as well as atmospheric variations. Spatially, we initially only consider stars within 5' of the target. To decide what magnitude range to consider, we use the chip-wide corrections to provide a first-pass calibrated lightcurve of the target object and then take the median of it to estimate the typical instrumental magnitude of the target, $Q_{targ}$. We then only use those comparison stars that have instrumental magnitudes predicted from the external ``clean star" catalog that fall within half a magnitude of $Q_{targ}.$ Using a single value of the target magnitude, $Q_{targ},$ is clearly not optimal for extremely variable objects like transients but it saves 
 considerable computing time as we only have to apply the magnitude cut once. As the magnitude dependence of the detector non-linearities 
 is usually not that strong, the recipe just 
 presented works well-enough in practice (to percent level) for ab-type RR Lyrae and objects that vary with amplitudes $\Delta Q \lesssim 1,$ i.e., most common persistent objects. 
 To further improve the photometric precision (at the $\sim$ percent level), it turns out to be important to also restrict the $g-r$ colours of the comparison stars since equation \ref{eqn:gandrtoq} is only an approximation, the variable chip non-linearities likely add additional (variable) colour terms, and the photometry used to generate the external comparison star catalogs typically also contains uncorrected colour-dependent effects.  For this paper, we tune our photometry for RRLS by considering only bluer comparison stars, with $0<g_{s}-r_{s} < 0.7.$ 
 
Once we have a local set of comparison stars that are well-matched to the target object, we then follow the procedure of \cite{Honeycutt1992}. 
We perform a global, iterative least squares fit, using all exposures that survive the first-pass quality cuts, to obtain an improved estimate for the true (corrected) instrumental magnitudes of our comparison stars and the implied exposure-to-exposure zero point magnitude offsets. The overall goal is to minimize the scatter between the measurements of the comparison stars that are on the same chip as the target on a given exposure, further correcting for external effects such as extinction and internal ones such as spatially dependent sensitivity variations. Once we have our best guesses for the zero point offsets of the exposures, we then use them to construct corrected lightcurves for all comparison stars. Those comparison stars whose zero-point corrected lightcurves exhibit too much residual scatter or whose corrected median lightcurve magnitudes differ significantly from the ones predicted using the external catalog are then thrown out. We then repeat the least squares fit (omitting the rejected stars) and check the comparison star lightcurves again, throwing out stars that show issues.
We keep doing this until no further stars are removed or we no longer have enough comparison stars to work with. This comparison star ``cleaning" stage turned to be essential as the photometry in the external catalog can sometimes be bad, the comparison stars may consistently land inside an extended bad region of a chip,  or the comparison stars be intrinsically variable themselves.  If enough stars survive the cleaning process, then we are done, and the final zero point offsets are applied to the observed target instrumental magnitudes to produce a calibrated lightcurve. If we do not have enough good comparison stars, we double the search radius to bring in more comparison star candidates, and repeat the entire photometry process, doubling the companion star search radius as needed until we are basically covering the entire chip. At that point, if we still do not have enough good comparison stars, we declare the restricted ensemble photometry analysis
 a failure and revert to the chip-wide solution. The number of good comparison stars that we require for our ensemble photometry to succeed is a function of magnitude: at least 4 for bright objects with V $\lesssim 16$, and more than 20 for objects with V $\gtrsim 20.$ We need more comparison stars at fainter magnitudes because the accuracy of the comparison star photometry in the external comparison catalog also usually starts to drop significantly at faint magnitudes.     
 On good chips and for bright non-variable stars (V $\lesssim 16$) where systematic errors dominate the photometry, our (relative) ensemble photometry is routinely good to $\sim 5$ mmag, comparable to what has been achieved on other Schmidt telescopes. 
 
 The  measurement error we report for a given exposure is the rms scatter between the observed, zero point-corrected comparison star magnitudes in that exposure and the lightcurve medianed magnitudes of those comparison stars. Note that this error estimate does not rely on the external catalog and assumes that the median magnitude is a much more accurate estimate of the true magnitude (usually the case for lightcurves of non-variable objects with more than $\sim 10$ points). Although it can often work surprisingly well, this error estimate is not perfect. For catastrophic photometry failures, this recipe sometimes underestimates the magnitude of the deviations. On the other hand, the error can be overestimated if the target usually falls on a good quality CCD chip but some of the comparison stars are mostly observed using bad chips. (When a chip-wide photometric solution must be used, the error obtained in this way is also often an overestimate due to the sensitivity variations across the chip.) Problems with correct error estimation are unfortunately common in other wide area surveys using large CCD arrays, e.g., ZTF. The distribution of measured magnitudes for a star known to be non-variable can also show significant non-Gaussian tails, often caused by charge transfer inefficiencies in the detector that become apparent when the sky background is high. (We are working on corrections for this problem, but they are not ready yet for this paper.) Luckily, since the transfer inefficiencies conserve charge, the over-predicted magnitudes seem to balance out the under-predicted magnitudes, and the median of the overall lightcurve appears quite robust and converges to the correct value. We will demonstrate this below. 
 
 Up to this point, our ensemble photometry is only correct in a relative sense. To obtain an absolute calibration and tie the lightcurves to an external system, we compute the medians of all the comparison star lightcurves and then obtain the average offset between these median values and those predicted from the external catalog. This is the offset used to tie the target object to the external reference system. For our RRLS, we then use the following conversion to go from the absolutely calibrated instrumental Q magnitudes to the 
 Johnson V band values presented in the tables: 
 \begin{equation}
  {\rm V} \approx  Q + 2.5[1.0+ 10^{0.4<g-r>}] - 0.5784<g-r> - 0.0038, 
  \label{eqn:qtov}
  \end{equation}
 where $<g-r> = 0.1$ is the typical phase-averaged (SDSS) color for RRLs.   As a rough rule of thumb (good to within $\sim 1\%$ 
 for blue objects like RRLs), $ {\rm V} \approx Q + 0.75 .$

\section{Tests of Photometry Pipeline Accuracy}
        
\begin{figure}
    \centering
    \includegraphics[width=2in,height=2in]{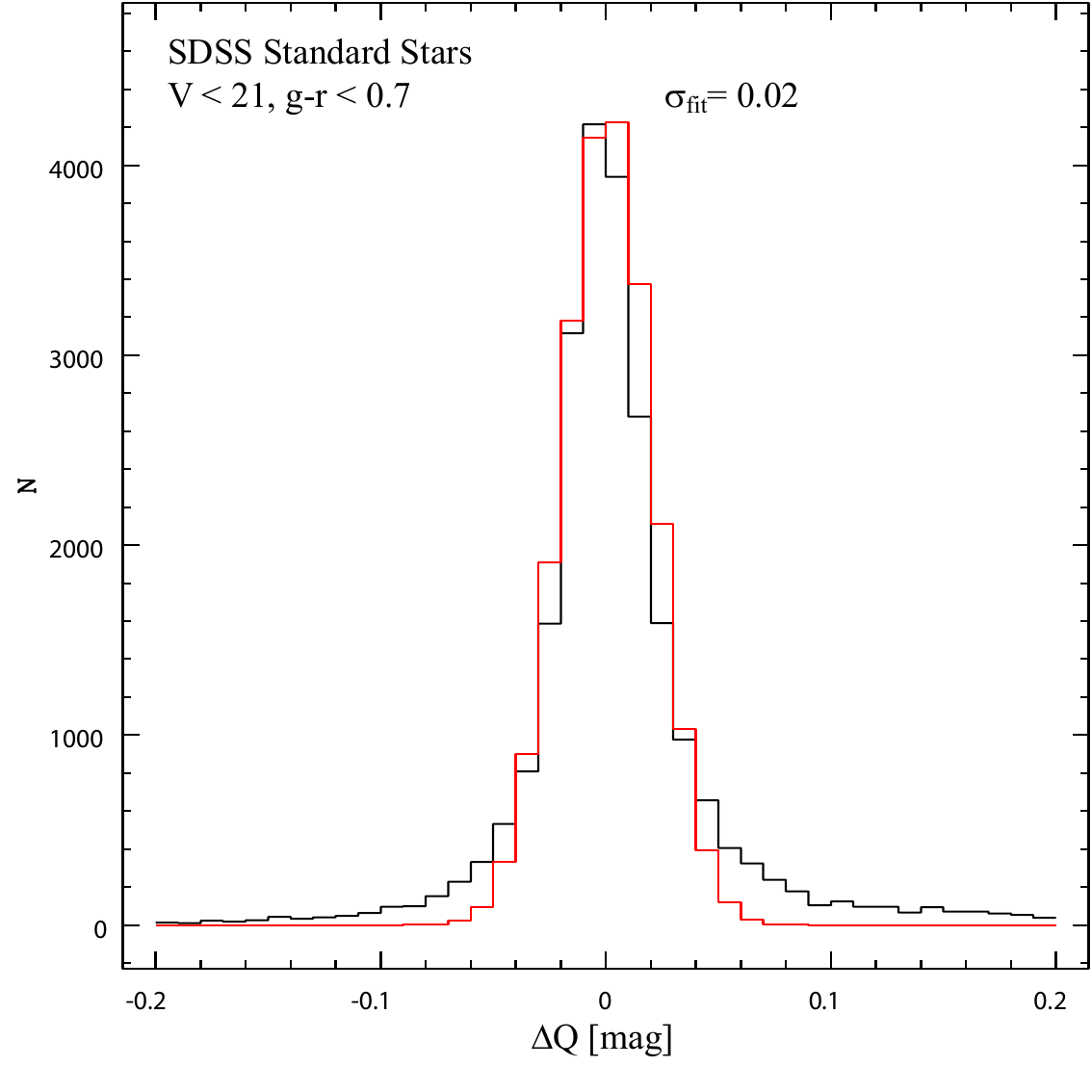}
    \includegraphics[width=2in,height=2in]{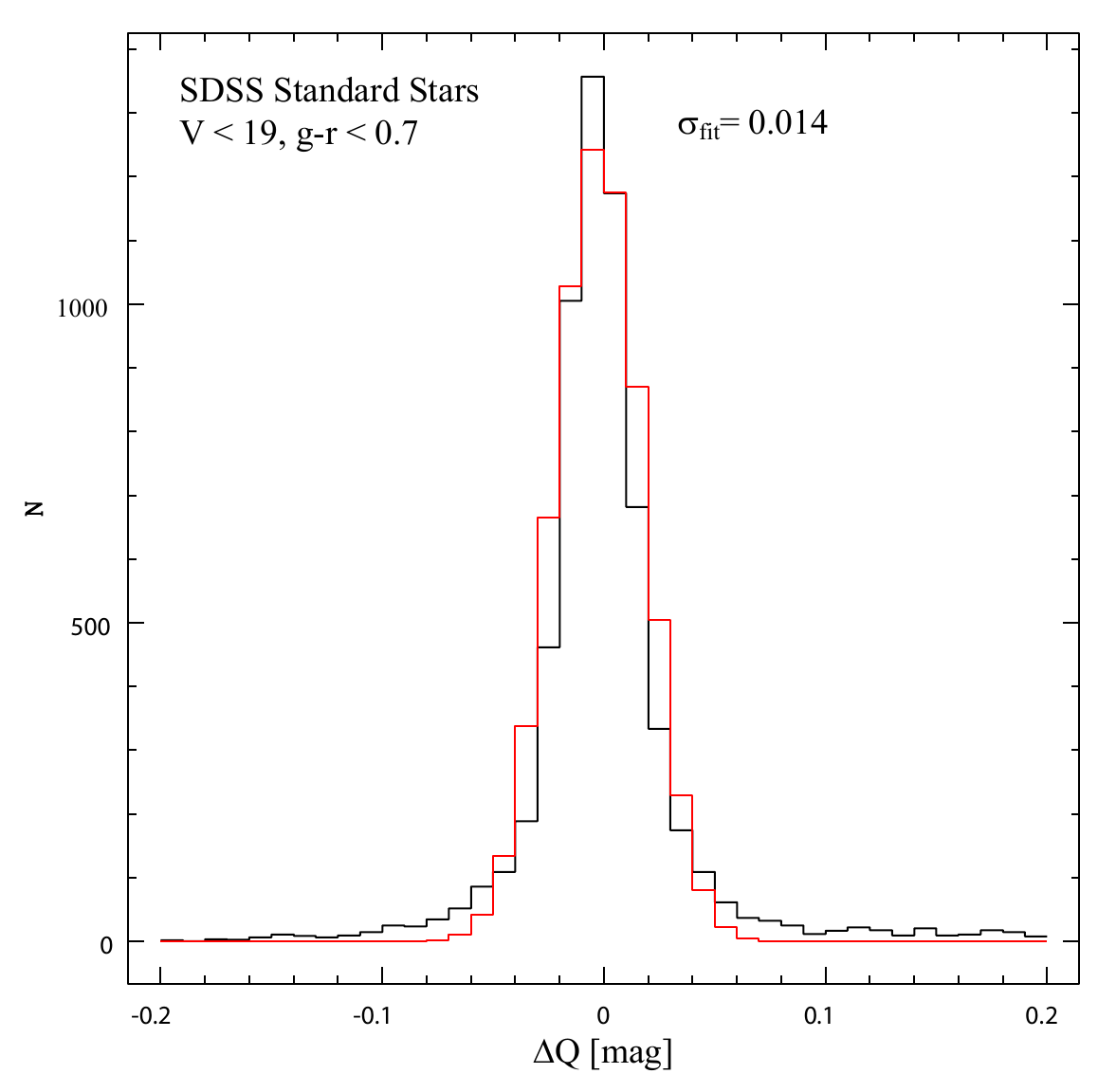}
    \caption{Distributions of $\Delta Q,$ the difference in predicted vs. measured median LSQ instrumental magnitude, for objects in the \citet{Ivezic2007} SDSS Stripe 82 standard star catalog. The {\it top} panel is for catalog objects with magnitude V < 21. The {\it bottom} panel
    is for brighter catalog objects with V < 19.  The {\it black} histograms show the data and the {\it red} histogram are Gaussian fits. }
    \label{fig:s82compall}
\end{figure}

LSQ photometry will improve in the future as we use the ``first'' pass photometry of this paper to better identify problematic comparison stars and bad chip regions. To characterize where we stand now, we first look at the distributions of the differences between the predicted instrumental magnitudes (from equation \ref{eqn:gandrtoq}) and the measured, lightcurve {\it median} instrumental magnitudes,  $\Delta {\rm Q} = {\rm Q}_{pred}-{\rm Q}_{meas}^{median}$, for objects in the SDSS Stripe 82 standard star catalog \citep{Ivezic2007}. These are shown in Fig. \ref{fig:s82compall} for two brightness cuts. The $g-r$ colours used to predict the instrumental 
 magnitudes are the SDSS ones, computed from
 the values in the standard star catalog. There are usually enough points in the lightcurves for these stars that the medianing process should remove the statistical errors, leaving us with the uncorrected systematic errors. 
 The non-Gaussian tails in Fig. \ref{fig:s82compall} appear to be mostly due to blends with an object within 10\arcsec of the target.The
 distributions show  only a weak dependence on the brightness limit, and the distributions are centered on $\Delta Q \sim 0,$ indicating the median process does work well. The relatively small width, $\sim 2\%$, of the distributions together with the fact that SDSS data itself is only calibrated to $\sim 1-2\%$ indicates that the intrinsic scatter of the LSQ measurements must be comparable to or smaller than the SDSS one. 
 
 The $50\%$ completeness detection limit (5$\sigma$) for decent LSQ chips and a 60 second exposure is typically ${\rm V}\sim 21 - 21.5.$  To work with RR Lyrae at distances up to $\sim 140$ kpc (corresponding to $<{\rm V}>_{RRL}\sim 21.5$), as we hope to here, this means working in the regime where biases due
 to Poisson fluctuations are important and the CCD chip nonlinearities are at their worst. In particular, the chips have charge traps that effectively eat photoelectrons.  For very faint sources, the fraction of photoelectrons from a source that make it to the readout depends on how filled those charge traps are, which in turn depends on quantities like the level of the variable sky background, i.e., the fraction of photoelectrons that are read out varies with each exposure. 
 
\begin{figure*}
	\includegraphics[width=2.5in]{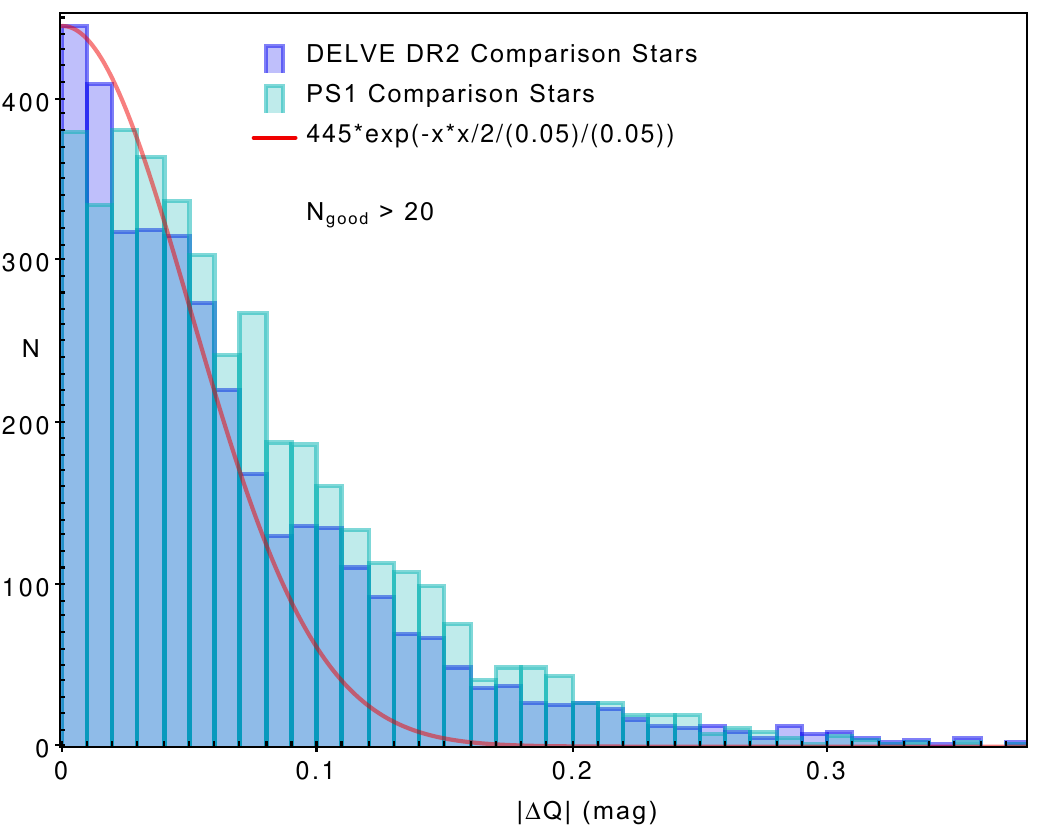}
	\includegraphics[width=2.5in]{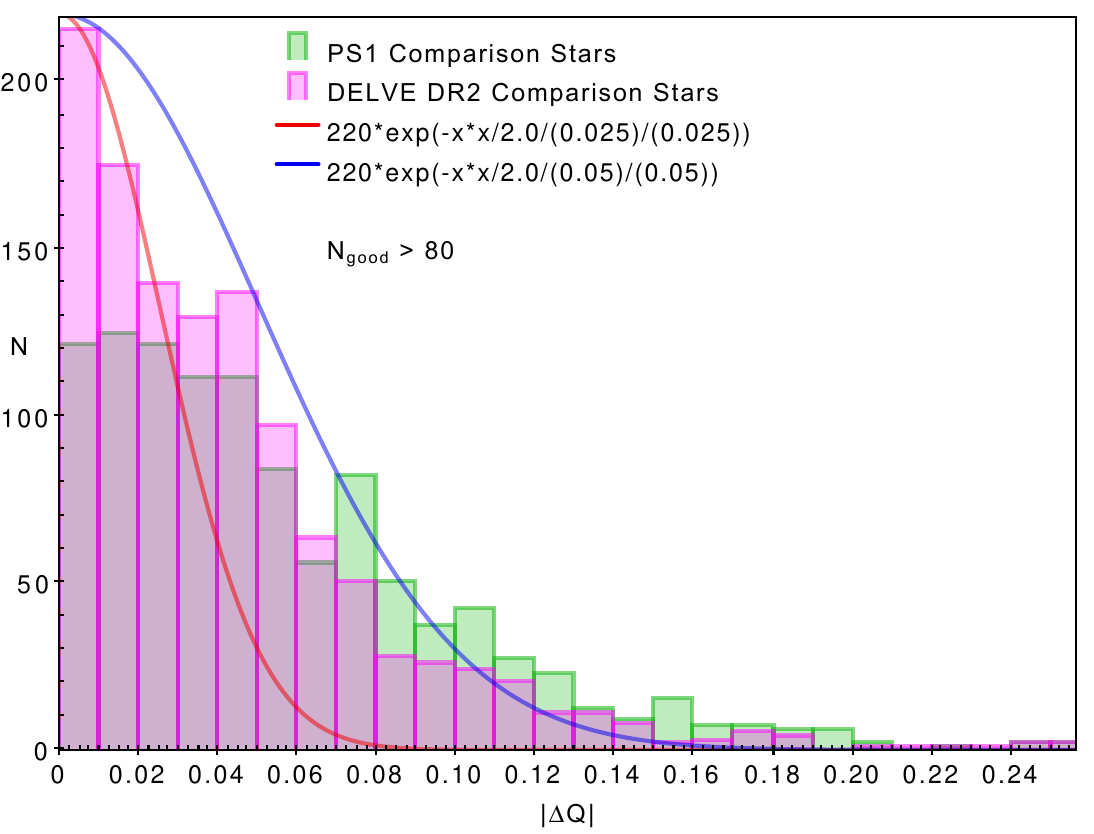}
    \includegraphics[width=1.9in]{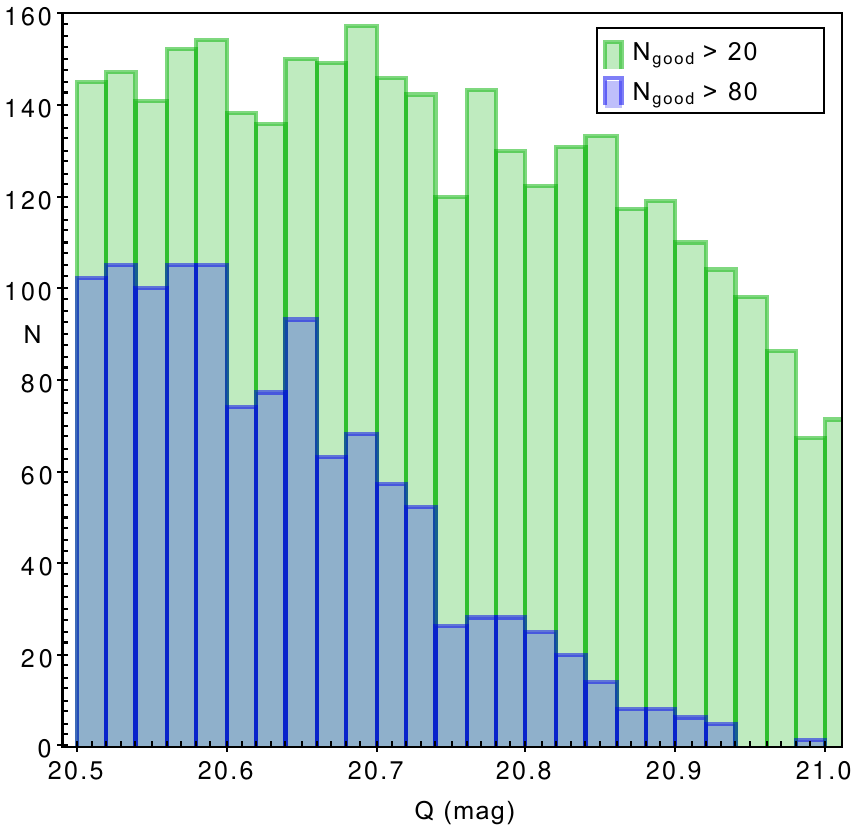}
    \caption{The {\it left} two panels show the distribution of $|\Delta Q|,$ the absolute value of the difference between the median of the calibrated LSQ lightcurve and the value predicted from the external catalog, for faint (V>21.3), stellar and likely non-variable objects in the Crater II region. The {\it leftmost} panel shows the distribution for sources where the number of lightcurve points, $N_{good},$ exceeds 20. The {\it middle} panel shows the distribution for the subset of these that have $N_{good}>80.$ A clear dependence is visible on the set of comparison star magnitudes used in the last step of the ensemble photometry pipeline, with values taken from the DELVE DR2 catalog generally performing better than those taken from the PanSTARRS PS1 DR2 catalog. This is not unexpected as DELVE DR2 goes significantly deeper than PanSTARRS.  Note that at these faint magnitudes, the Poisson 
    measurement errors per observation are large ($\sigma \sim 0.1-0.2$) and still contribute significantly to the scatter in lightcurve medians. Going from
    $N_{good}>20$ to $N_{good}>80$ points in a lightcurve, we thus expect (to first order) the scatter in the medians to drop by a factor 
    $\approx \sqrt{80/20} = 2,$ which is indeed the case as can be seen by comparing to the Gaussian curves plotted in the left two panels. The {\it rightmost} panel shows the distribution of median LSQ lightcurve magnitudes, Q, computed by the LSQ pipeline using PanSTARRS comparison stars for the sources used to make the left two panels. The number of sources with $N_{good}>80$ drops rapidly with magnitude, showing that we are starting to hit the 50\% completeness limit of LSQ at around ${\rm Q}\sim 20.7$ ($V\sim 21.5$).
    }
    \label{fig:qdiff_faint_ngood}
\end{figure*}

 To test how well our pipeline corrects for these effects, and to see how badly we are dominated by systematic errors, we now consider the distribution of $|\Delta Q|$ for a set of very faint sources (${\rm Q}>20.5, {\rm V\gtrsim 21.3}$) in the Crater II region. The sources are selected to be stellar (with parameter ${\it extended\_class\_g}=0$ in the DELVE DR2 catalog) and with a probability of being variable of less than $<0.5$ (as computed from their LSQ lightcurves).  The magnitude distribution of the sources is shown in Fig. \ref{fig:qdiff_faint_ngood}, for sources
 that have more than 20 observations and the subset of these that have more than 80 observations (as determined by the pipeline used here for RRLS detection, that relies on PanSTARRS comparison stars). If the detection limit were not an issue, the average number of exposures in this region should be $\gtrsim 100.$ The number of sources with $N_{good} > 80$ observations should thus not be very different from 
 those with 
 $N_{good} > 20$ observations since most objects in reality have $ > 100$ observations. We see that this is roughly true up to $Q \sim 20.65,$ at which point the number of sources with $N_{good}>80$ plummets, i.e., this is where the inability to distinguish sources from
 the background really sets in and means that the measurements of the $N_{good}>20$ sources fainter than $Q>20.7$ are dominated by the increasingly rare Poisson fluctuations. Nonetheless, in Fig. \ref{fig:qdiff_faint_ngood} we see that for $N_{good}>20$ (the upper panel),
 we can still carry out some useful measurements. Regardless of the set of comparison stars used, the core of the $|\Delta Q|$ distribution is approximately a Gaussian of width $\sigma \sim 0.05$ magnitude, which is still usable for RRLS-based distance measurements (which have an intrinsic scatter $\sim 3-7 \%$ per object). The distribution based on PS1 comparison stars does show evidence for 
 systematic error issues at the $2-3\%$ level as it does not peak at $|\Delta Q|\sim 0.$ However, while this paper was being completed, the 
 DELVE DR2 catalog was released, and we were able to run the pipeline using comparison stars with DELVE DR2 photometry. The performance using the DELVE stars is clearly better, with the $|\Delta Q|$ distribution now peaking at zero and less scatter overall. This is not 
 surprising since DELVE photometry (using a 4m telescope) goes much deeper, and this demonstrates the importance of choosing the set of comparison stars properly. Going forwards, we will switch to DELVE comparison stars when possible, but this paper will rely on results based on PanSTARRS comparison star photometry. We have done some checks, and the key results of this paper do not depend on the comparison star set used.  Going to the case of $N_{good}>80$ observations (which is dominated by objects with $Q<20.7$), we see further evidence that we are not yet systematics limited, particularly for the case of DELVE comparison stars. Quadrupling the number of observations should halve the statistical error contribution to $\Delta Q,$ and the core of the $|\Delta Q|$ distribution indeed now looks like a Gaussian of width $\sigma \sim 0.025$ (vs. $\sim 0.05$ for $N_{good}>20$). There is still a significant non-Gaussian tail from the fainter objects, but it now lies almost entirely under the Gaussian with $\sigma \sim 0.05$ (as opposed to the $N_{good}>20$ case). Indeed, experiments comparing the sources found in stacked LSQ exposures to those found in very deep catalogs indicate that we are not systematics-limited at the faint end until we hit $N_{good} > 100$ (reaching a limiting co-add magnitude $r \sim 23-24$). As we will see, for RRLS that fall on good chips, this means
 we can potentially measure phase-resolved RRLS lightcurve shapes to $\sim$ few percent accuracy, even for $<V> \sim 20.5-21.$ Note that in Fig. \ref{fig:qdiff_faint_ngood}, the total number of sources measured using PS1-based comparison stars is {\it lower} because the ensemble photometry routine (used to produce the measurements) fails more often using PS1 vs. DELVE DR2 comparison stars, another indication of the importance of having a good, consistent set of comparison standards.

\begin{figure*}
	\includegraphics[width=3in,height=2in]{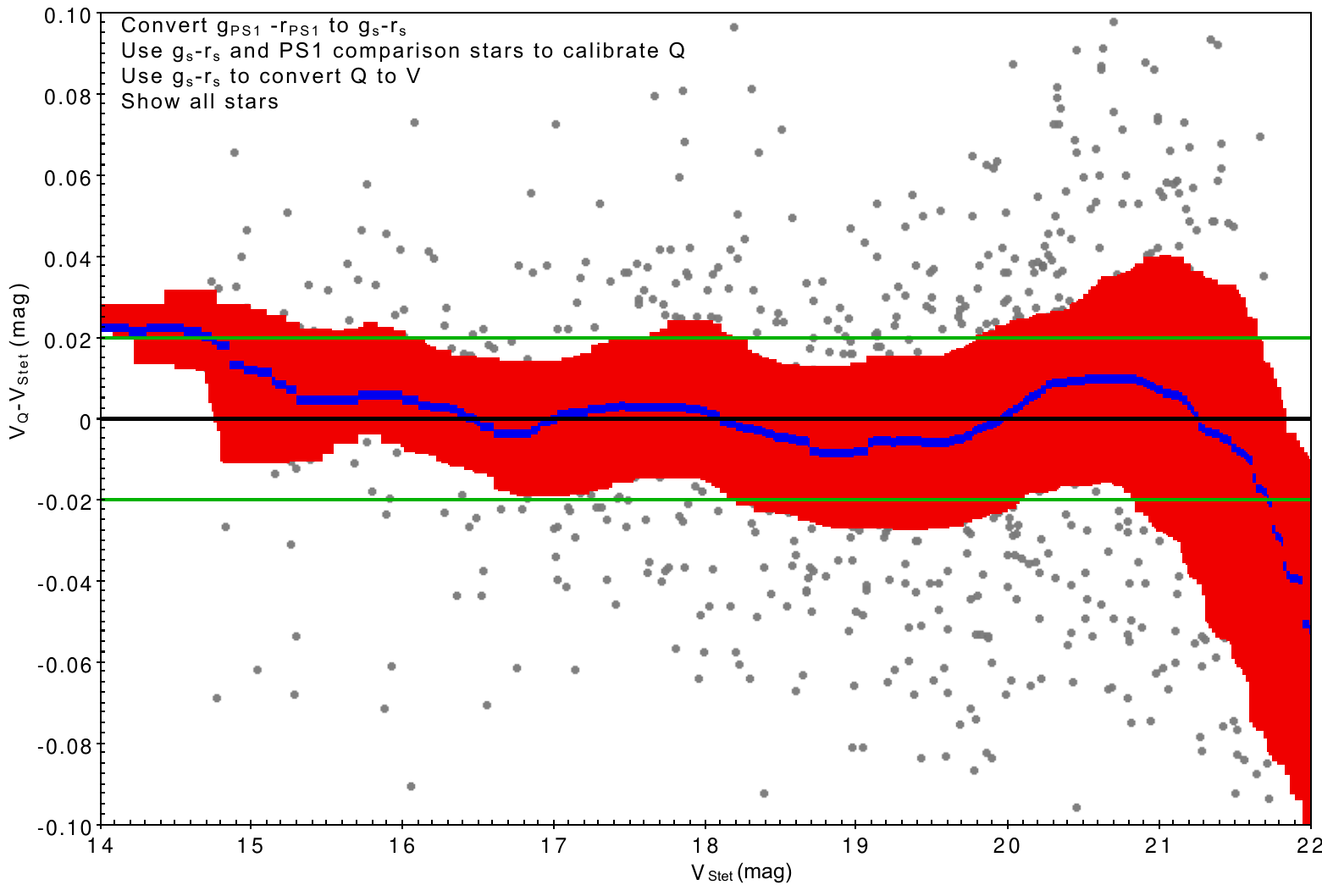}
	\includegraphics[width=3in,height=2in]{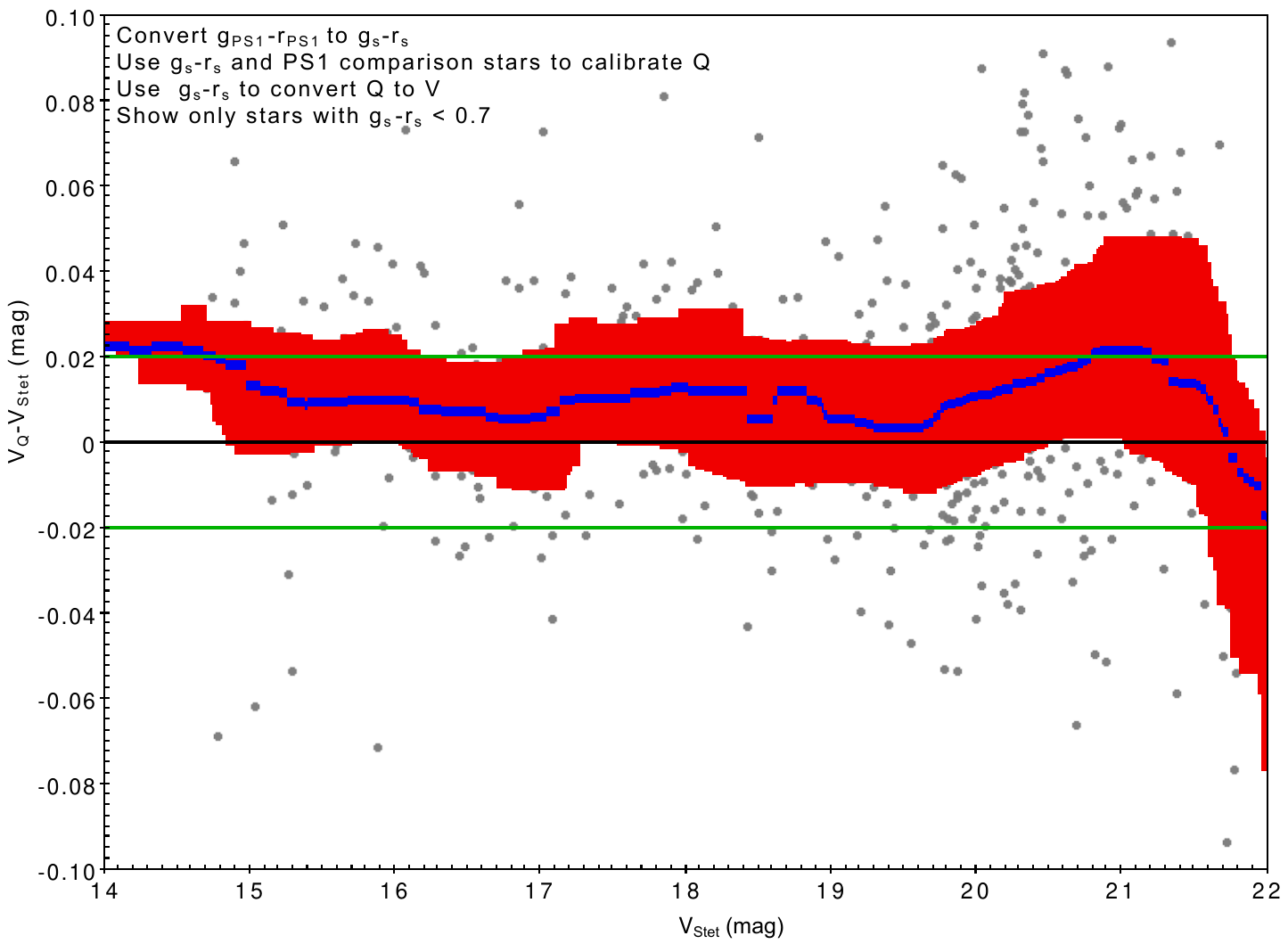}
	\includegraphics[width=3in,height=2in]{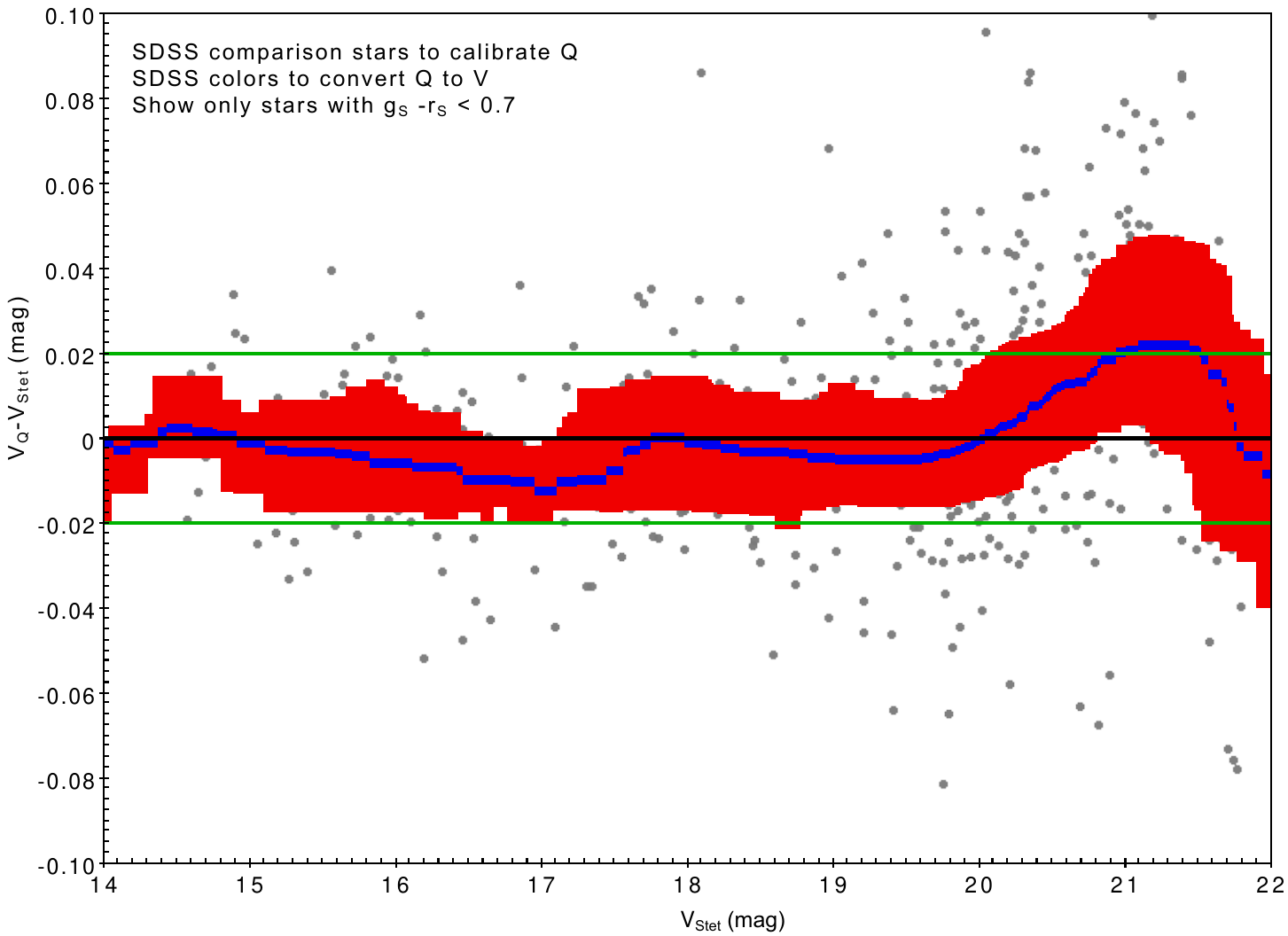}
    \includegraphics[width=3in,height=2in]{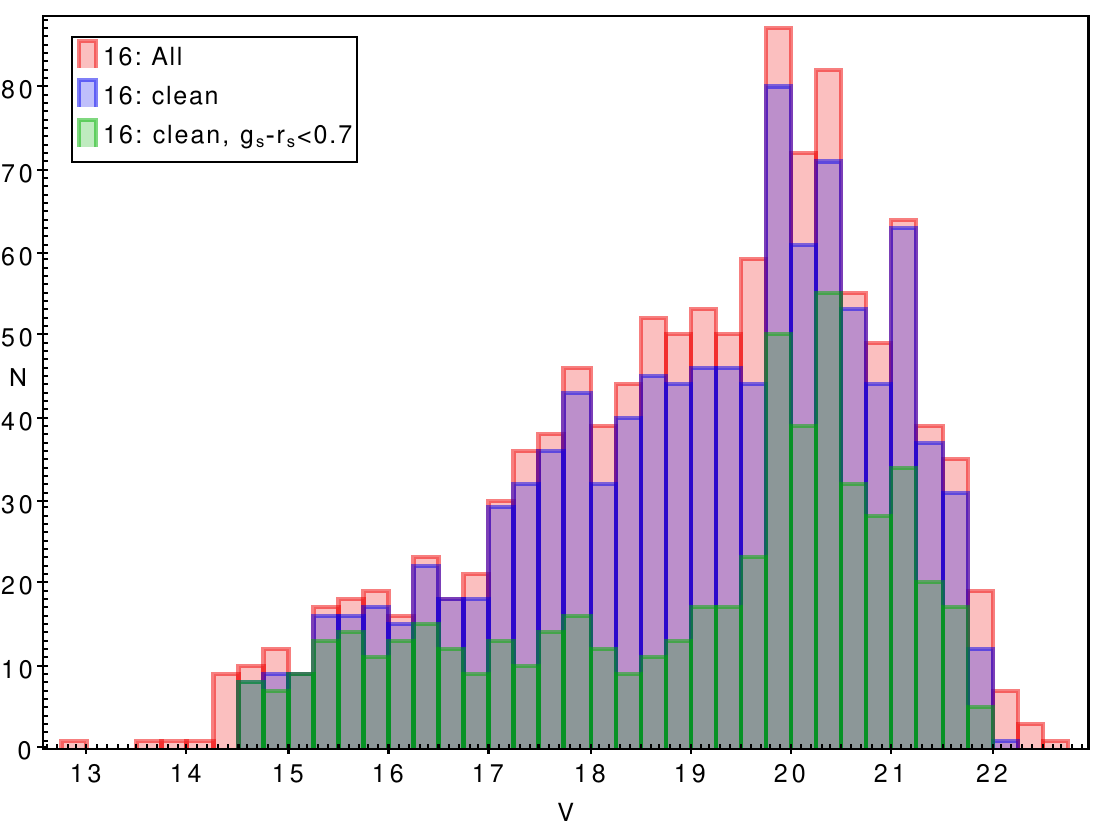}
    \caption{The {\it upper} two and {\it bottom left} plots show the difference between the V magnitude measured by LSQ (converted from Q to V using 
    Eqn. \ref{eqn:qtov}) for a set of Stetson standard stars in Sextans. The {\it thick blue} lines in the plots are the running median of the differences as a function of the V magnitude measured by Stetson. The {\it red shaded} regions are quantile plots showing the spread between the 25th and 75th percentile of the distribution of differences as a function of magnitude. The {\it grey} dots are the measurement differences for the individual stars that fall outside of the 25th-75th percentiles. The {\it bottom right} plot shows the magnitude distribution of the standard stars. The {\it red} bars show the distribution for all stars. The {\it blue/purple}  (``clean") bars show the distribution for the stars that have a Stetson measurement rms $<0.01,$ magnitude $V>14.5,$ and a successful LSQ ensemble photometry measurement. The {\it green} bars show the magnitude distribution for the subset of the ``clean" stars that has color $g_s-r_s<0.7.$}
    \label{fig:photcomp_stetv}
\end{figure*}

As a final end-to-end test, which will also demonstrate the photometry differences in current external catalogs, we run the LSQ photometry pipeline on the Stetson standard stars in Sextans \citep{stetson2000, stetson2005}, where LSQ has good lightcurve coverage ($N_{good}>100-200$). We then compare the inferred LSQ V magnitudes (from Eqn. \ref{eqn:qtov}) to those reported by Stetson. To minimize contamination from variable objects, we select stars with measurement rms values, as reported by Stetson, less than $0.01$ magnitudes. Of these 1053 objects, we select a further subset of 1043 objects with ${\rm V}>14.5$ for which LSQ CCD saturation is not an issue 
and for which the LSQ  restricted (second pass) ensemble photometry converged. (Overall, the second pass photometry had a $\approx 98\%$ success rate for Stetson stars with ${\rm V} > 14$.)  Figure \ref{fig:photcomp_stetv} shows the differences between the estimated LSQ V magnitudes (obtained by taking the median of the LSQ lightcurves again) and the corresponding Stetson V magnitudes as a function of Stetson $V$ magnitude. Because the stars are well-observed and mostly bright, the results should mostly reflect systematic offsets 
and color term issues. The conversion from $g_s$ and $r_s$ to Q (\ref{eqn:gandrtoq}), for example, is not perfect and has small but non-zero higher order color term corrections. The $g_{PS1}$ filter response is also quite different from the $g_s$ (SDSS) one, and the conversion from $g_{PS1}$ to $g_{s}$ is also not perfect. As an example of what these differences mean, the Stetson stars cover the range $-0.33<g_s-r_s<1.83$ if $g_s$ and $r_s$ values from the SDSS DR17 catalog are used. If we instead use $g_{PS1}$ and $r_{PS1}$ values from the PS1 DR2 catalog and convert them to $g_s$ and $r_s$ using the relations in \citet{tonry2012}, the same stars now cover a range $-0.42<g_s-r_s<1.61.$ Since stellar color is an issue, we consider the results for stars of all colors and for stars with $g_s-r_s< 0.7$ As explained above, the ensemble photometry routine is optimized for blue RRLS and only uses comparison stars with $0<g_s-r_s<0.7.$ Not surprisingly, the scatter between LSQ and Stetson V magnitudes is least when considering stars in the restricted range $g_s-r_s<0.7.$  To look for trends and estimate the amount of scatter, we also plot in Fig. \ref{fig:photcomp_stetv} the median difference between LSQ-measured V and Stetson V for each star as function of V as well as the 25th to 75th percentile range of the differences in a given magnitude bin. We also show the magnitude (V) distribution of the stars. Note that the number of stars at the bright and faint end are very small, so that the results at the bright and faint ends should be down-weighted accordingly.  There are three main conclusions from Fig. \ref{fig:photcomp_stetv}. First, there are definitely magnitude dependent offsets and uncorrected color terms (depending primarily on the typical color of objects in a given magnitude range and reflecting uncertainties in the external catalogs). The choice of external reference catalog, e.g., SDSS vs PS1, matters. Second, while offsets exist, they are not huge. For stars restricted to $g_s-r_s<0.7,$ the 25th and 75th percentiles of the differences typically lie within $\sim \pm 1.5\%$ of the median, and for all stellar colors, within $\sim \pm 2\%$ (except at $V\sim21$ where Poisson errors may be coming in; using DELVE DR2 stars seems to improve performance there). Second, the median difference values are also typically within $\sim 1-2\%$ of zero. In other words, while there are always some outliers (caused by bad CCD chip cosmetics and objects with unusual spectra), ``typical" absolute photometry calibration should be good to $\sim 1-2\%,$ which is not much larger than what is obtained in comparisons of SDSS and PanSTARRS to Stetson. Note that our absolute calibration is only applied at the last step of the photometry. In other words, the absolute calibration scatter seen here {\it does not} reflect the {\it relative} scatter inside a given lightcurve.  For example, as noted above, the {\it internal} rms scatter in measured values for an isolated, non-variable star at $V\sim 16$ can routinely be 5 mmag for stars falling mainly on good chips. Our final point is that the changes in the median offset between LSQ and Stetson
appear to be relatively {\it smooth}. (Remember also that the vertical scale in Fig. \ref{fig:photcomp_stetv} is measured in percents.) In other words, for RRLS distance measurements, the changes in the median offset do not increase the scatter at a given distance but rather expand and compress the distance scale slightly; structures as a function of distance should not be significantly be washed out due to sudden discontinuities or rapid large oscillations in the median photometry as a function of magnitude. (Measurements of individual RRLS mean magnitudes can of course sometimes deviate significantly from the median value expected from a structure localized in distance, e.g., due the bad chip systematics discussed here, and one has to have enough stars detected in the structure to be able to construct a meaningful median in the first place.) 
 
\section{Effects of Varying LSQ Survey Coverage in the Crater II Region}
\begin{figure}
\includegraphics[width=\columnwidth]{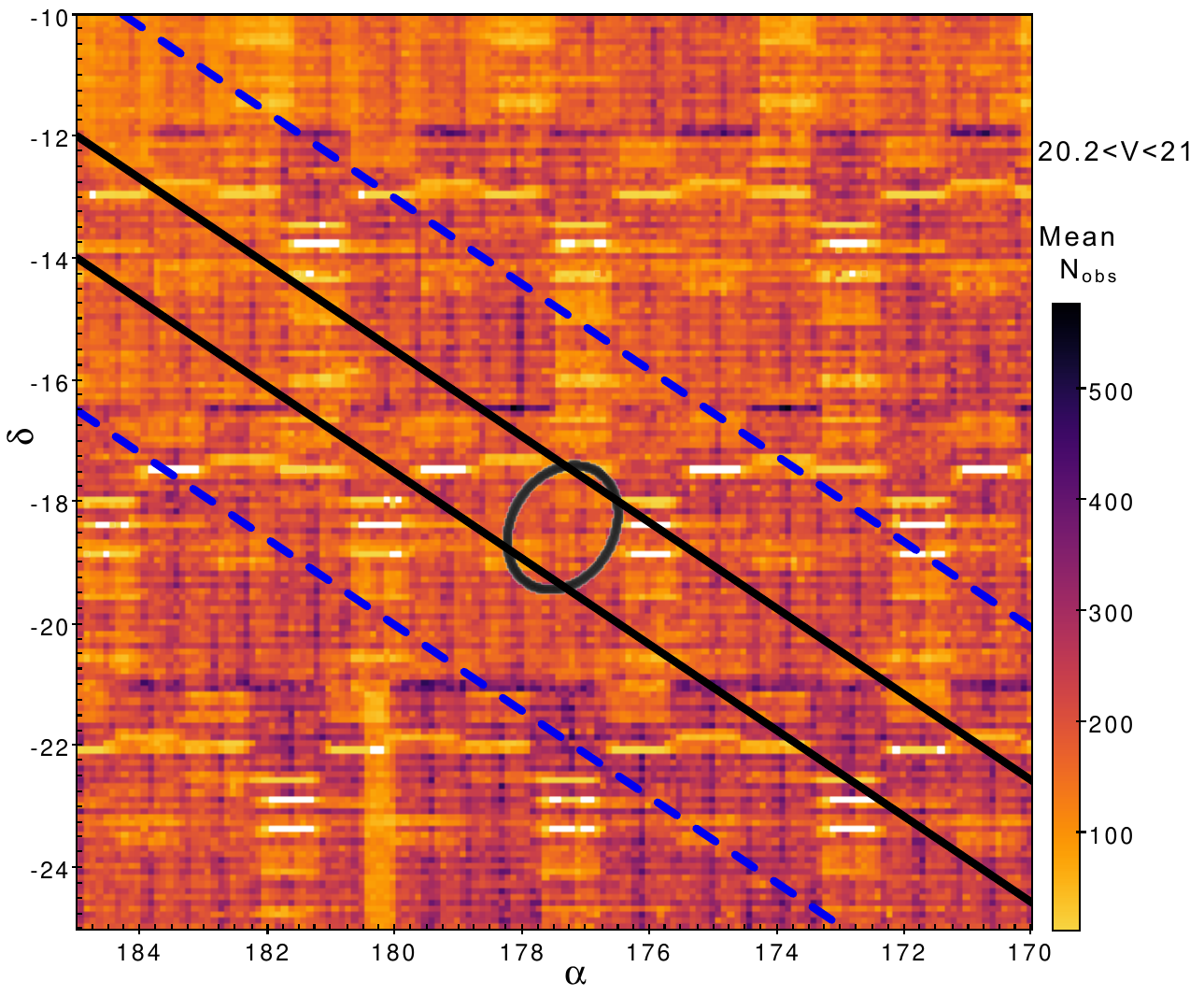}
\includegraphics[width=3.2in]{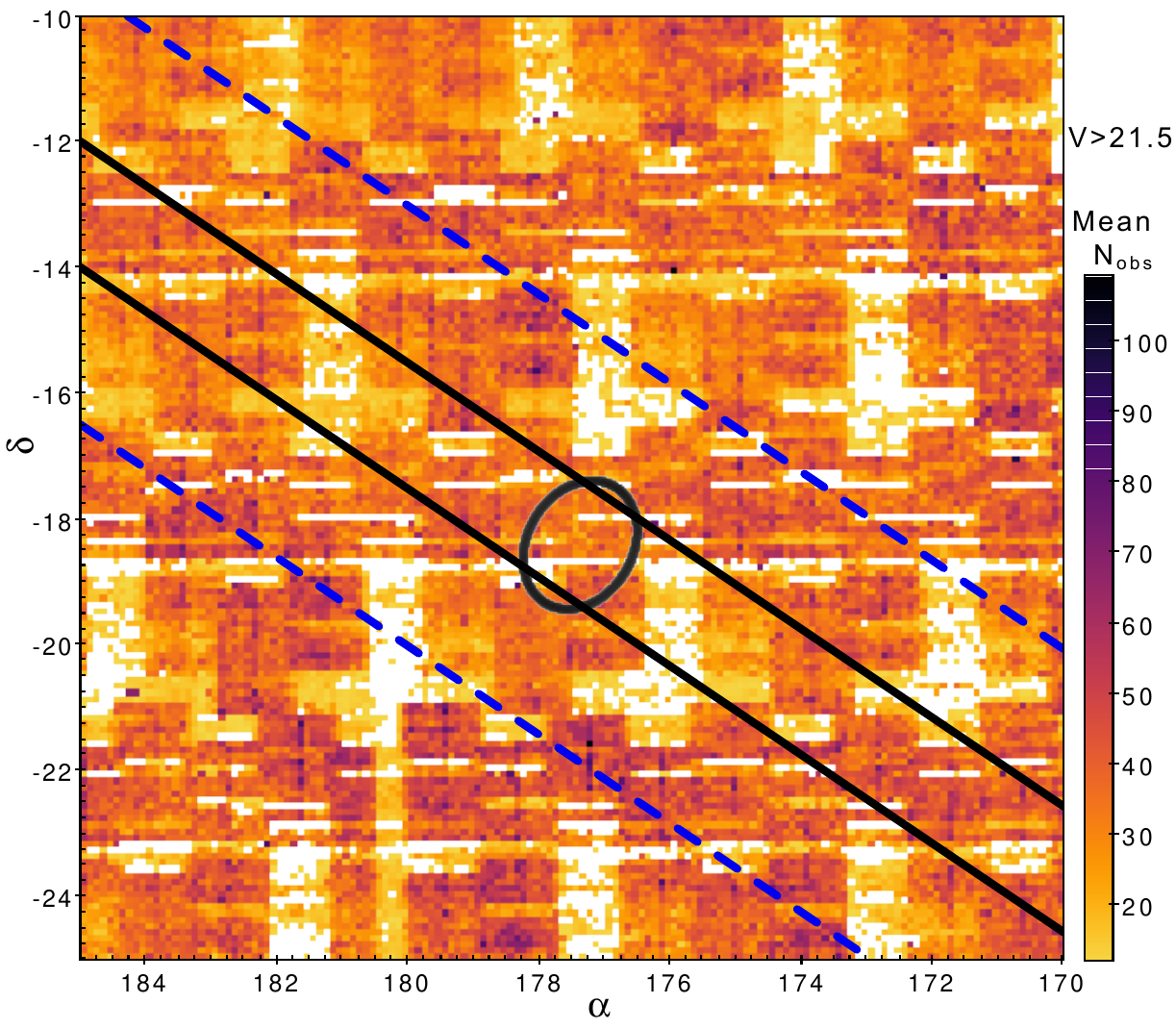}
\caption{A ``heat" map, binned in $0.1^\circ \times 0.1^\circ$ pixels, where pixel color shows the average number of ``good'' observations ($N_{obs}$) for the sources found inside the pixel. For a given source, an observation is only counted as ``good" if the source is detected significantly in that observation and the overall observation passes photometric quality checks. Pixels where sources consistently fall on dead chips or are much fainter than the exposure magnitude limit, i.e., have zero good observations, show up as {\it white} areas in the maps.
The lines and the ellipse are the same as in Fig. \ref{fig:rrls_sky_with_regions} and delineate the stream, background, and Crater II galaxy regions respectively. The {\it top} panel shows the map for sources in the magnitude range $\sim V~20.2 - 21$, corresponding to RRLS distances $80-110$ kpc. The {\it bottom} panel shows the map for sources fainter than $V\sim 21.5,$ the rough magnitude limit for the LSQ survey. In both panels, there is no obvious spatial correlation between observation number and the stream region or Crater II ellipse. }
\label{fig:expo_maps}
\end{figure}

The LSQ camera unfortunately does not provide uniform sky coverage over its field of view. Its 112 CCD chips are arranged in four rows separated by $0.5$ degree gaps. More importantly, several chips in the array are completely dead, and others operate a with a higher background and/or reduced quantum efficiency that reduces the magnitude limit they can reach for a given exposure time. These sensitivity variations, in particular the gaps between chips, are partially compensated for by dithering or shifting successive exposures. (The gaps between chips are filled in by shifting alternate exposures by $0.5$ degrees, which also results in 0.5 degree wide stripes with twice the typical number of source visits, visible as the dark vertical stripes in Fig. \ref{fig:expo_maps}.) Because combining observations from chips with very different characteristics can be problematic, however, the sky was divided into a set of fields, and exposure shifts within a given field were usually kept small, i.e., less than the width of one chip.  ``Dead" regions, where there are few or no exposures by a good CCD, thus remain in the LSQ survey. Depending on the angular size of the structures one is looking for on the sky, these regions can lead to spurious results. For example, one of these dead regions falls on the core of the Sculptor dwarf galaxy, limiting what LSQ can say about RRLS in the interior of Sculptor. Fig. \ref{fig:expo_maps} attempts to quantify and map these variations in survey coverage by spatially binning the sources detected in a given magnitude range into $0.1^\circ\times0.1^\circ$ pixels and then assigning a color to each pixel that corresponds to the average number of ``good'' observations for the sources found inside the pixel. The number of good observations for a source, key in determining how well we can tell if it is periodic, is a function of the total number of exposures that contained the source, the fraction of times the source is bright enough to be detected, and the photometric quality of individual exposures (e.g., we drop exposures where the photometric scatter compared to a reference catalog is too large or the exposure's magnitude limit is anomalously low).

For sources in the magnitude range $V \sim 20.2-21.0$ (corresponding to RRLS distances $\sim 80-110$ kpc,
the ``bright" side of the stream and  Crater II), we see in Fig. \ref{fig:expo_maps} that the survey coverage is quite uniform, with most of the sky covered by $\gtrsim 100$ observations.
This explains the relatively high completeness (especially for RRLS ab) we find above with respect to other surveys.  It {\it cannot} explain the anisotropic distribution of stream stars that we find here.  For fainter magnitudes, we begin to hit the sensitivity limit of the chips for a 60 second exposure.
For $V>21.5,$ corresponding to RRLS distance $\gtrsim 140$ kpc, we see in Fig. \ref{fig:expo_maps} that the coverage is quite patchy, with some significant areas with no observations.
As noted above, then, our survey may be missing stars in the faint (distant) side of the stream, especially if the stream extends beyond $\sim 140$ kpc. Looking on scales of order the size of the stream, however, we see that the pattern of patchiness in coverage is similar to that in the background areas (and the entire $15^\circ\times15^\circ$ Crater II region). There is nothing that would single out the stream region as special. In sum, variations in LSQ survey coverage cannot explain the spatial distribution of RRLS we find near Crater II. That distribution must be intrinsically stream-like. 

Note that our search for RGB stars uses the DELVE DR2 and Gaia DR3 surveys, which are entirely independent of the LSQ survey. The candidate stars we are considering are nowhere near the magnitude of the limit of DELVE DR2 survey, which is very uniform for brighter sources in this region of the sky.  Most of our RGB candidate stars (because of the proper motion cut) are also well above the Gaia DR3 magnitude limit, and that survey is also quite uniform for brighter objects in this part of the sky. Variations in survey coverage thus cannot explain the spatial distribution of our RGB candidates, which largely coincides with that of the RRLS. 

\section{Colour-Magnitude Diagram (CMD) Selection Cuts for RGB Stars}

\begin{figure}
    \centering
    \includegraphics[width=3.22in]{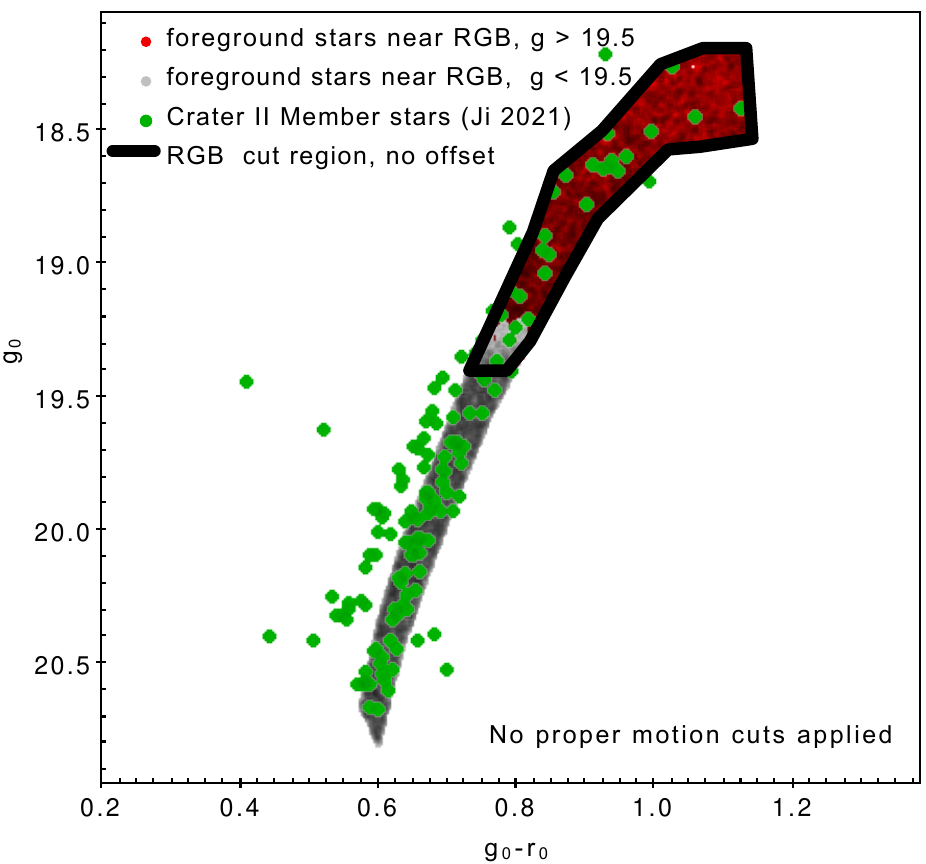}
    \includegraphics[width=3.2in]{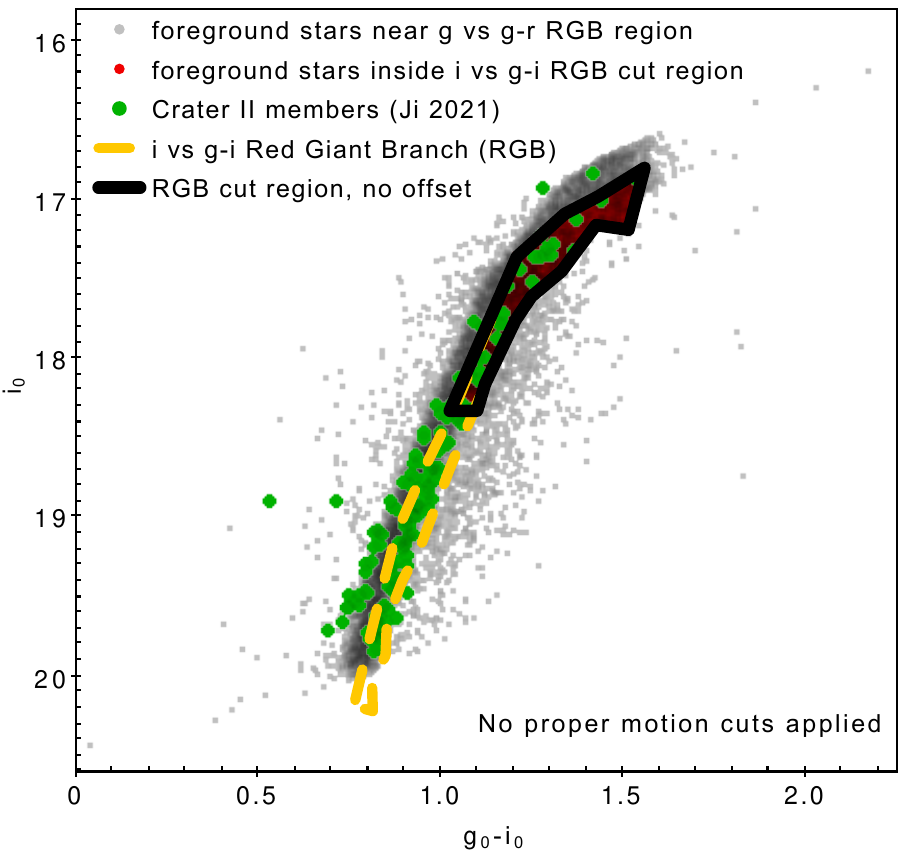}
    \caption{Locations of the CMD cut regions used for the RGB candidate search. The {\it top} and {\it bottom} panels respectively show the $g$ vs. $g-r$ and $i$ vs. $g-i$ color-magnitude diagrams for stars in the Crater II region. The ({\it green X's}) show the CMD positions of the stars determined by \citet{ji21} to be member stars of Crater II. Note that the positions of the stars in the CMD are {\it corrected for extinction} using the values given in the DELVE DR2 catalog, while the brightness cuts applied to exclude faint objects with poor Gaia proper motions are {\it not} corrected for extinction because Gaia signal-to-noise is based on observed (apparent) magnitude.  Using the locations of the \citet{ji21} stars as a guide, we then empirically define polygons ({\it heavy black outlines}) that contain the stars we think are on the RGB branch. The $grey$ points in the bottom panel are foreground stars from the 
        top panel which were selected to lie on the RGB in the $g$ vs. $g-r$ CMD (the $red$ and $grey$ points in the top panel). Note that most of these foreground stars do not end up on the $i$ vs. $g-i$ RGB. These cut regions are shifted up and down
        in brightness to scan for stream stars not located at the distance of Crater II.}
    \label{fig:cmdcuts}
\end{figure}

\begin{figure}
    \centering
    \includegraphics[width=\columnwidth,height=2.7in]{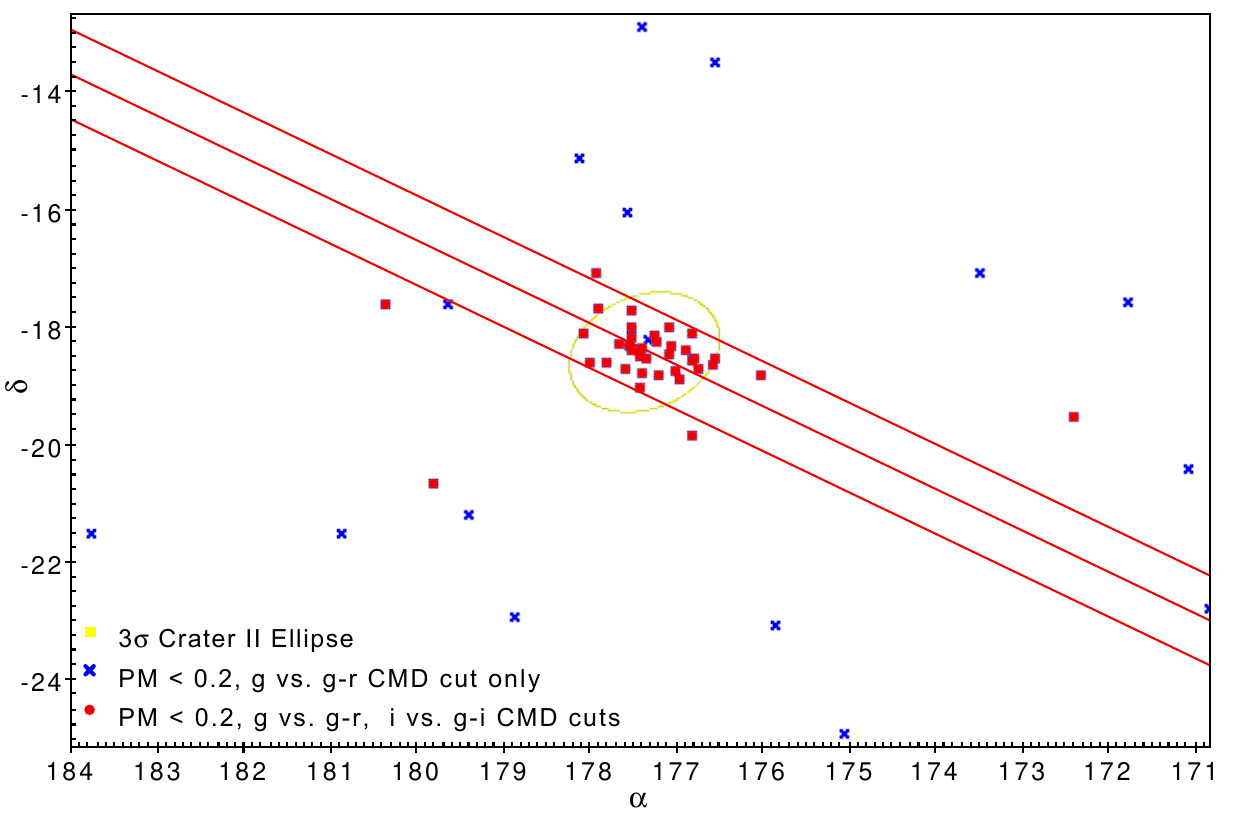}
    \caption{The {\it blue crosses} show the spatial distribution of stars selected by a Gaia DR3 proper motion cut of $PM<0.2$ and a single ($g$ vs. $g-r$) CMD cut. The middle red line is the best line fit to the {\it RRLS} stream distribution (see Fig. \ref{fig:Sky_Map_10_11}). The two {\it outer red lines} are from Fig. \ref{fig:rrls_sky_with_regions} and delineate the core stream region.  The ellipse made of {\it faint yellow triangles} is the $3\sigma$ ellipse fit to the overall stellar distribution of Crater II. No brightness (magnitude) cuts have been applied. When an additional ($i$ vs. $g-i$) RGB CMD cut is applied, only one of the remaining stars ({\it red dots}) is not consistent with the RRL stream distribution, i.e., the second CMD cut is indeed helping to exclude foreground stars.}
    \label{fig:cmd_comp_red_lines}
\end{figure}

Given the large spatial extent of the Crater II stream, standard techniques like making a Hess/CMD diagram and subtracting a background prove problematic. Applying a  proper motion cut does significantly improve foreground rejection,  but even a very tight proper motion cut of $PM<0.2$ is unfortunately still not enough given the weakness of the expected signal (e.g., see the blue crosses in  Fig. \ref{fig:cmd_comp_red_lines}). One of the remaining issues that limits what we can do is the presence of foreground stars that fall close to or on the Crater II RGB in a given CMD. These cannot be immediately rejected because the RGB of Crater II definitely has a finite width, e.g., due to age and metallicty effects. These interloper stars are not real red giants with the correct luminosities, and they significantly weaken the signal from the real Crater II red giants. One can ameliorate the problem by noting that if one selects stars that appear to lie near or on the RGB in the $g$ vs $g-r$ CMD, for example, one will find that they have a significantly increased scatter relative to the RGB when plotted on an another CMD, e.g.,  the $i$ vs. $g-i$ CMD. This is demonstrated explicitly in Fig. \ref{fig:cmdcuts} by the grey points, which are foreground stars selected to lie close to the RGB in the $g$ vs. $g-r$ CMD. One can remove many of the interloper foreground stars by simply requiring that they lie close to the RGB in {\it both} $g$ vs $g-r$ and 
$i$ vs. $g-i$ CMD space, i.e., if we apply {\it two} simultaneous CMD cuts as is the case for the red points shown in Fig. \ref{fig:cmd_comp_red_lines}.  The two CMD cuts are redundant for true Crater II RGB stars, but they are {\it not} for non-members of Crater II. One can presumably further improve selectivity by using information from other filters, {\it if} good enough photometry is available for those filters. Note that this method of removing foreground by applying two CMD cuts only works well if the RGB is also intrinsically narrow (as it luckily is in Crater II). In particular, defining an "RGB" CMD region that includes adjacent AGB stars
turns out  {\it not} to be a good idea even though it increases the number of stars that can pass a CMD cut, which could have increased the strength of a signal. 

As knowing the exact regions of the CMDs used for cuts is important for reproducing our specific results, we show them in Fig. \ref{fig:cmdcuts} and also give the vertices of the polygons in CMD space that define the cuts in Tables \ref{tab:rgbcut_gvgmr} and \ref{tab:rgbcut_ivgmi}. The exact shape of the cuts, however, is not too critical. The narrower they are, generally the better, but because of the intrinsically lower foreground in the red portion of the CMDs for this region of the sky, just requiring that candidate stars have $0.8 \lesssim < g-r \lesssim 1.2$ and $\lesssim g-i \lesssim 1.7$ (regardless of brightness), for example, and then applying the proper motion and brightness cuts  is enough to pull out much of the RGB stream signal. (No scanning by shifting the cut regions up and down in magnitude is required in this case.)

\begin{table}
    \caption{$r$ vs. $g-r$ CMD: Vertices of RGB Cut Polygon}
    \centering
    \begin{tabular}{cc}
    \hline
     $g_0 - r_0$ & $g_0$ \\
     mag & mag \\
     \hline
0.731 & 19.40 \\
 0.747 & 19.31  \\
0.823 &  18.88 \\
 0.853 & 18.65 \\
 0.923 & 18.50 \\
 1.008 & 18.25 \\
1.069 & 18.19 \\
1.132 & 18.19 \\
1.141 & 18.53 \\
1.065 & 18.56 \\
1.019 & 18.57 \\
0.917 & 18.83 \\ 
0.869 & 19.05 \\
0.819 & 19.29 \\
0.785 & 19.40 \\
0.731 & 19.40 \\
    \hline
    \end{tabular}
    \label{tab:rgbcut_gvgmr}
\end{table}

\begin{table} 
    \caption{$i$ vs. $g-i$ CMD: Vertices of RGB Cut Polygon}
    \centering
    \begin{tabular}{cc}
    \hline
     $g_0 -i_0$  & $i_0$ \\
     mag & mag \\
     \hline
1.025 & 18.33 \\
1.209 & 17.36  \\
1.340 & 17.09 \\
1.432 & 16.98  \\
1.559 & 16.80  \\
1.515 & 17.19  \\
1.426 & 17.16  \\
1.332 & 17.45  \\
1.249 & 17.61  \\
1.205 & 17.76  \\
1.122 & 18.16  \\
1.100 & 18.33  \\
1.025 & 18.33  \\
    \hline
    \end{tabular}
    \label{tab:rgbcut_ivgmi}
\end{table}

%%%%%%%%%%%%%%%%%%%%%%%%%%%%%%%%%%%%%%%%%%%%%%%%%%

% Don't change these lines
\bsp	% typesetting comment
\label{lastpage}
\end{document}